\documentclass[12pt]{article}
\usepackage{graphicx}
\usepackage{epstopdf}
\usepackage{amsmath}
\usepackage{amsfonts}
\usepackage{amssymb}
\usepackage{color}
\usepackage{mathrsfs}
\usepackage{color}

\usepackage{xcolor,url}

\usepackage{cancel,dsfont} 

\pdfoutput=1

\newcommand{\dress}[1]{{\hat{#1}}}

\newcommand{\be}{\begin{equation}}
\newcommand{\ee}{\end{equation}}
\newcommand{\bea}{\begin{eqnarray}}
\newcommand{\eea}{\end{eqnarray}}
\newcommand{\barr}{\begin{array}}
\newcommand{\earr}{\end{array}}

\renewcommand{\Re}{{\rm Re}}

\newcommand{\SEFTofI}{${\cal S}$EFTofI}

\def \wt {\widetilde}

\def\fnleq{{f}_{\rm NL}^{\rm equil.}}
\def\fnlor{{f}_{\rm NL}^{\rm orthog.}}
\def\fnlrad{{f}_{\rm NL}^{\rm equil.,\,orthog.}}

\usepackage[colorlinks,bookmarks]{hyperref}
\definecolor{linkblue}{rgb}{0,0,0.8}
\definecolor{linkgreen}{rgb}{0,0.5,0}

\hypersetup{pdfpagemode=UseNone, pdfstartview=FitH, linkcolor=linkblue, %
            citecolor=linkgreen, urlcolor=linkblue}

\def\beq{\begin{equation}}
\def\eeq{\end{equation}}
\def\be{\begin{equation}}
\def\ee{\end{equation}}
\def\bea{\begin{eqnarray}}
\def\eea{\end{eqnarray}}
\def\d{{\partial}}

\def\mpl{M_{\rm Pl}}
\def\nn{\nonumber}

\begin{document}


\setcounter{page}{1} \baselineskip=15.5pt \thispagestyle{empty}

\begin{flushright}
\end{flushright}

\begin{center}

{\Large \bf The Supersymmetric
Effective Field Theory of Inflation}
\\[0.7cm]
{\large {Luca V. Delacr\'etaz}${}^{1}$, Victor Gorbenko${}^{1}$ and Leonardo Senatore${}^{1,2}$}
\\[0.7cm]
{\normalsize { \sl $^{1}$ Stanford Institute for Theoretical Physics,\\ Stanford University, Stanford, CA 94306}}\\
\vspace{.3cm}

{\normalsize { \sl $^{2}$ Kavli Institute for Particle Astrophysics and Cosmology, \\
Stanford University and SLAC, Menlo Park, CA 94025}}\\
\vspace{.3cm}

\end{center}

\vspace{.8cm}

\hrule \vspace{0.3cm}
{\small  \noindent \textbf{Abstract} \\[0.3cm]
\noindent  We construct the Supersymmetric Effective Field Theory of Inflation, that is the most general theory of inflationary fluctuations when time-translations and supersymmetry are spontaneously broken. The non-linear realization of these invariances allows us to define a complete SUGRA multiplet containing the graviton, the gravitino, the Goldstone of time translations and the Goldstino, with no auxiliary fields. Going to a unitary gauge where only the graviton and the gravitino are present, we write the most general Lagrangian built out of the fluctuations of these fields, invariant under time-dependent spatial diffeomorphisms, but softly-breaking time diffeomorphisms and gauged SUSY. With a suitable St\"uckelberg transformation, we introduce the Goldstone boson of time translation and the Goldstino of SUSY. No additional dynamical light field is needed. In the high energy limit, larger than the inflationary Hubble scale for the Goldstino, these fields decouple from the graviton and the gravitino, greatly simplifying the analysis in this regime. We study the phenomenology of this Lagrangian. The Goldstino can have a non-relativistic dispersion relation. Gravitino and Goldstino affect the primordial curvature perturbations at loop level. The UV modes running in the loops generate three-point functions which are degenerate with the ones coming from operators already present in the absence of supersymmetry. Their size is potentially as large as corresponding to $f_{\rm NL}^{\rm equil.,\,orthog.}\sim 1$ or, for particular operators, even $\gg 1$. The non-degenerate contribution from modes of order $H$ is estimated to be very small.

 \vspace{0.3cm}
\hrule

\newpage
\tableofcontents

\newpage

\section{Introduction and main ideas}

This paper is rather long and technical. For this reason, we provide here quite a long introduction that presents all the main ideas and results in a as least technical way as possible. For an executive summary, see the Conclusions in~Sec.~\ref{sec:conclusions}.

{\bf The Effective Field Theory of Inflation:} Most of the information that we have about inflation comes from the observation of the cosmological perturbations. It is therefore very important to study the most general dynamics for the fluctuations. Doing this amounts to studying the so-called Effective Field Theory of Inflation (EFTofI)~\cite{Cheung:2007st}. In the EFTofI, inflation is thought of as a period of time in the early universe where there is a physical clock whose evolution defines a preferred time-slicing. In this case, time diffeomorphisms (diffs.) are spontaneously broken, while time-dependent spatial diffs.~are not. This means that in the spectrum of the theory there must be a Goldstone boson, usually referred to as $\pi$, that non-linearly realizes the spontaneously broken time-diffs.. As typical with the case of non-linearly realized symmetries, the Lagrangian for the fluctuations around the vacuum is highly constrained by the symmetries, and can be constructed without any knowledge of the mechanism spontaneously breaking the symmetry. 

This is particularly important from both an in-principle and a practical point of view. In principle, we could imagine that inflation is realized by a theory that is fundamentally Lorentz invariant, by which we mean that the theory admits a formulation where all matter fields transform linearly under diffs.. Usually there is a vacuum solution which gives a maximally symmetric spacetime such as Minkowski or (A)dS. The description of the theory around this maximally symmetric vacuum might be very inappropriate to describe the theory around the inflationary solution. Over the years, we have become familiar to the phenomenon that theories are very different around different vacua by studying dualities in supersymmetric field theory, where the power of unbroken supersymmetry has allowed us to confidently study the regime of theories at strong and at weak coupling. Historically, this is not a novel effect: since coupling constants depend on energy, we have already experimentally verified that at high energy quarks dynamics is efficiently described by QCD, but at low energies this description is inappropriate, while a much more appropriate description is given by the chiral Lagrangian. It is indeed extremely complicated to describe the dynamics of pions using QCD. Not always the description of a given physics changes radically within some interval of energies or coupling constants: for example the standard model Electroweak Theory offers an appropriate descriptions practically for all phenomena of experimental relevance. But this does not necessarily need to be the case, as QCD and supersymmetric field theories have indeed taught us. Coming back to the theory of inflation, when we describe it in terms of a slow rolling inflationary scalar field, we are indeed assuming that we can describe the inflationary background in terms of a perturbatively small deviation of the theory around the Minkowski, or the maximally symmetric, solution. But this does not necessarily need to be the case, and the EFTofI allows us to cover all in one Lagrangian even the cases where the theory around the inflationary background is very different than the one around the maximally symmetric vacuum (see Fig.~\ref{fig:moduli_space} for a pictorial, somewhat historical, representation).  Because of this, the EFTofI is important also from a practical point of view: it allows us to study the observational signatures of inflationary models without the unnecessary and limiting constraint of being able to describe the theory around the maximally symmetric vacuum.

\begin{figure}[h!]
\begin{center}
\includegraphics[width=12cm]{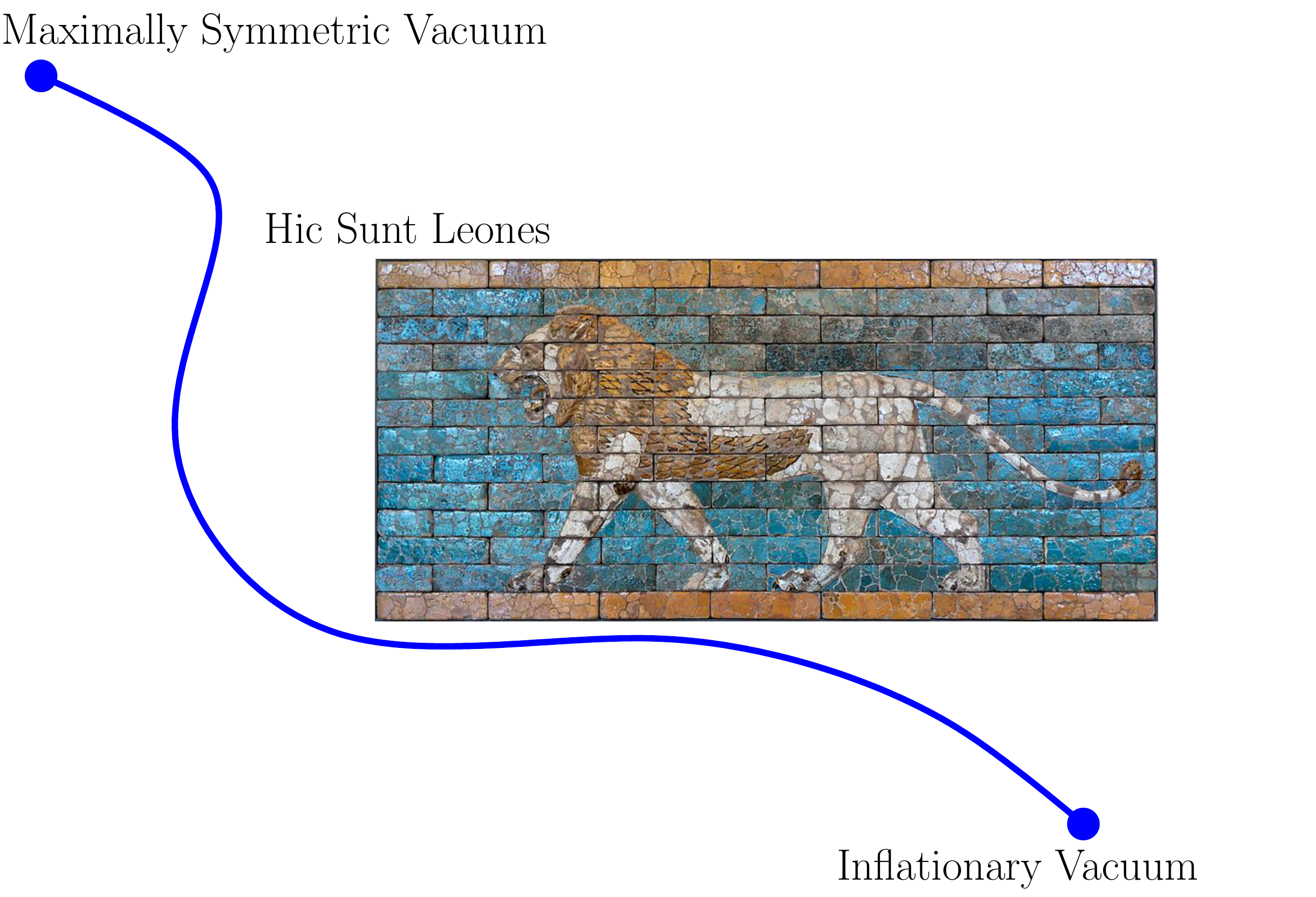}
\caption{\label{fig:moduli_space} \small The description of the theory around the maximally symmetric vacuum can be very different than the one around the inflationary vacuum. It might not be possible to perturbatively describe the transition from one vacuum to the other.}
\end{center}
\end{figure}

As for any EFT, in constructing the EFTofI it is essential to declare a priori the symmetries that are non-linearly realized. Of particular interest, for several different reasons that go from solving the hierarchy problem of the electroweak interactions to formulating a consistent UV completion of gravity, is the case in which we include supersymmetry among the symmetries of Nature. This is highly motivated from an additional aesthetic point of view, as supersymmetry is the only possible extension of the spacetime symmetries given by the Poincar\'e group (see for example~\cite{Wess:1992cp}). However, supersymmetry must be at best spontaneously broken at energies accessible at current particle colliders, and therefore it is conceivable that it could be better probed by higher energy observables, such as inflation in case its energy scale is sufficiently high.   The implications of supersymmetry as a symmetry of Nature  for the theory of the fluctuations during inflation will be the subject of this paper.

{\bf Gauge redundancies:} In our construction, gravity is self-evidently very important. In the presence of gravity, the Poincar\'e group is gauged~\footnote{Sometimes, gauge redundancies are erroneously assumed to be symmetries, but they are not symmetries because the action of the gauge group on a state leads to the same state.}, the gauge group being spacetime diffs., if one uses the metric as the dynamical field, or spacetime diffs.~times local Lorentz, if one uses the vierbein field. Since supersymmetry is a non-trivial extension of the Poincar\'e group, in the sense that the supersymmetry currents and the Poincar\'e currents do not commute, in the presence of gravity also supersymmetry is gauged to become what is known as supergravity~(SUGRA). So, in this paper we will be mostly dealing with SUGRA. 

This leads to what we consider a deeper point of view, that will be very important in our construction. A gauge  redundancy is a nuisance that is usually included in a theory to allow us to construct a local Lorentz-invariant Lagrangian that describes the dynamics of massless particles. To describe massless helicity-1 particles, we have the usual gauge redundancy of Yang-Mills theory, where the gauge group commutes with Poincar\'e. In the case of gravity, the graviton is a massless helicity-2 particle, and to construct its Lagrangian we introduce the gauge redundancy associated to spacetime diffs.. Finally, if we want to describe a massless helicity-3/2 particle, we need to introduce another gauge redundancy, which turns out to be the gauging of the supersymmetry transformations. Quite interestingly, the SUGRA algebra requires the gauging of Poincar\'e, that, as we mentioned, is associated to the presence of a helicity-2 massless particle. So, a massless helicity-3/2 particle requires the presence of a massless helicity-2 particle. It is for this reason that the helicity-3/2 massless particle is usually called the gravitino. One can make this statement somewhat more intuitive by realizing that the SUGRA algebra can be thought of as diffs.~acting on a spacetime manifold with additional Grassmann directions, {\it i.e.} a supermanifold: as the graviton is the gauge boson associated to the gauging of ordinary diffs.~the gravitino is one associated to the diffs.~in the Grassmann directions (see for example~\cite{Wess:1992cp}).

{\bf Softly Broken Gauge Invariances:} Let us consider now a massive particle either of spin 1, 2 or 3/2. Clearly, there is no need to introduce a gauge redundancy to construct a local Lorentz-invariant Lagrangian that describes the dynamics of such a particle. This Lagrangian can in general be split into terms that, if we were to reintroduce the gauge redundancy, will be gauge invariant, and terms that instead are not gauge invariant. If it happens that all the operators that break gauge invariance are relevant and the mass scale identified by these operators is larger than the mass of the states, then we say that (linearly-realized) gauge invariance is softly broken. It is then useful to think of this theory as actually the theory where the gauge redundancy is non-linearly realized. In fact one can introduce the gauge redundancy that would normally be introduced to describe massless particles, and find that, at energies above the mass of the states, the spectrum of the theory effectively splits into helicity 1, 3/2 or 2 particles plus an additional particle, of lower helicity, the latter being effectively decoupled. The presence of this additional particle can be understood to be required to preserve the overall number of degrees of freedom of the theory: a massive particle of spin 1, 2 or 3/2 has more than two degrees of freedom. Furthermore, as we will see, the Lagrangian of the smaller helicity particle typically coincides with the one of the Goldstone boson associated to the spontaneous breaking of the global group that is obtained by taking the global limit of the gauge group, up to small mixing terms~\footnote{Subtleties associated with this statement will be discussed in Sec.~\ref{FRW}.}.   Evidently, the introduction of the gauge redundancy is consistent only if one imposes the additional massive particle to transform non-linearly under the gauge transformation. For this reason, we say that in this case the gauge invariance is non-linearly realized, or, with a slight misuse of language, spontaneously broken~\footnote{We say spontaneously broken because if a field transforms linearly under the gauge transformations and takes a vev that is not invariant under the gauge transformations, the fluctuations of this field around the new vev will transform non-linearly  under the gauge transformation. So, everytime  gauge invariance is spontaneously broken, it is non-linearly realized and therefore (linearly-realized) gauge invariance is softly broken. Of course, physical states that are related by a gauge transformations are still identified.}. In this case, the introduction of the gauge redundancy significantly simplifies the analysis. In fact, for example, it is the Goldstone boson that first becomes strongly interacting as we move to higher energies.
Notice that the existence of a regime of energies where the Goldstone boson is decoupled is nontrivial. However, this happens when the gauge coupling is weak and the operators that break the gauge symmetry are all relevant operators (while the kinetic term is a marginal operator). This means that the effects of these operators on the quadratic Lagrangian becomes smaller and smaller as we push the energy scale above some scale, allowing to think of the spectrum as a split one.

{\bf Unitary Gauge Lagrangians:} Such softy broken gauge theories have played a crucial role in the history of Physics. Most notable, the Lagrangian that describes the massive $W$-bosons of the Standard Model of Particle Interactions is usefully thought of as the Lagrangian of a softly broken gauge symmetry at energies below the Higgs mass. A simplified Lagrangian that contains the relevant physical information is  given by
\be\label{eq:GaugeBosonUnitary}
S[A]=\int d^4 x\;\left\{ {\rm Tr}\left[F_{\mu\nu}^2\right]+ m^2 {\rm Tr}\left[A_\mu^2\right]+\ldots\right\}\ , \qquad{\rm with}\qquad A_\mu=A_\mu^a \,\tau^a\ ,
\ee
where $\tau^a$ are the generators of the gauge group and $\ldots$ represents higher order terms in the number of fields and derivatives. The mass term clearly breaks Gauge invariance. In the Standard Model, the mass $m$ arises from the Higgs field taking a vev and softly breaking the gauge symmetry.

The same idea is what leads to the construction of the EFTofI (in the absence of supersymmetry). Here, as anticipated, we assume that during the epoch of inflation the gauge theory of spacetime diffs.~is spontaneously broken by the presence of some physical clock that controls the duration of inflation to a theory where only time-dependent spacial diffs.~are unbroken. Analogously to the case of massive gauge bosons, the EFTofI is written by writing the most general Lagrangian invariant under the residual gauge invariance and using the fields at our disposal, in this case just the metric, and being careful in ensuring that the time diffs.~are broken only by relevant operators. Schematically, neglecting, in this introduction, irrelevant numerical factors and indices, this leads to~\cite{Cheung:2007st}
\bea\label{eq:EFTofIUnitary}
&&S_{\rm EFTofI}[g]=\int d^4x\;\sqrt{-g} \;{\cal L}\left[R_{\mu\nu\rho\sigma},g^{\mu\nu}, K_{\mu\nu},\nabla_\mu, t \right]=\\ \nonumber
&&\quad=\int d^4x\;\sqrt{-g}\left\{\mpl^2 R\right.\\ \nonumber
&&\quad \left.+ \mpl^2\dot H(t)  g^{00}+\mpl^2(H(t)^2+\dot H(t)) + M(t)^4 \left(g^{00}+1\right)^2+\ldots+\bar M(t)^3 \left(g^{00}+1\right)\delta K+\ldots  \ \right\}\ ,
\eea
where $\delta K_{\mu\nu}$ represents the fluctuations of the extrinsic curvature of the equal time slices, and $H(t)$ is the Hubble rate in Friedmann Robertson Walker (FRW) spacetime, and where all uncontracted tensor indices  are upper 0's. $(g^{00}+1)$ is first order in the fluctuations as the background value of $g^{00}$ is equal to $-1$. $\ldots$ represent higher order terms in the fluctuations or the derivatives.

The same logic applies unaltered to the case in which we declare that the ultimate spacetime gauge invariance is not just the gauging of Poincar\'e, but the gauging of supersymmetry (SUSY). The SUSY algebra, schematically of the form
\be
\{Q,\bar Q\}=P^\mu \sigma_\mu\ ,
\ee
where $Q$ are the SUSY charges, tells us that if time diffs.~are not a gauge redundancy of the theory, SUSY cannot be a redundancy either. Therefore, in an inflationary background where we assume that time diffs.~are are spontaneously broken, so is SUSY. According to this logic, we do not need to force the theory to be invariant under SUSY, but simply we write the most general Lagrangian invariant under the residual gauge invariance (which is time-dependent spacial diffs.), using the fields at our disposal, which in this case are the metric and the gravitino, $\psi_\mu$, and being careful in ensuring that the time diffs.~and SUSY are broken only by relevant operators. As usual when dealing with spinors in a curved spacetime, it is useful to introduce local Lorentz invariance and the vierbein $e_\mu^a$ as
\be
g_{\mu\nu}=e^a_\mu e^b_\nu \eta_{ab}\ ,
\ee 
where $\mu,\nu,\ldots$ represent spacetime indices , and $a,b,\ldots$ represent local Lorentz indices. Using the notation of Wess and Bagger~\cite{Wess:1992cp}, we are schematically led to 
\begin{equation}\label{eq:SEFTofIUnitary}
\begin{split}
& S_{\rm {\cal S}EFTofI}= \int d^4x\;\sqrt{-g}\; {\cal L}\left[R_{\mu\nu\rho\sigma},e_{\mu}^a, K_{\mu\nu},\psi^\mu, \nabla_\mu, t \right]=\\    
&	=  S_{\rm EFTofI}[g]\, +\\ 
	& \mpl^2\int d^4 x \, e \left[ \frac{1}{2}\varepsilon^{\mu\nu\rho\sigma}\bar{ \psi}_\mu \bar\sigma_\nu D_\rho  \psi_\sigma +  m_{3/2} (\psi_\mu\sigma^{\mu\nu}\psi_\nu)+  m_0 (\psi_\mu\sigma^{\mu 0}\psi^0)+ m_\star (\psi_\mu\psi^\mu + \psi^0\psi^0)+ \hbox{c.c.} \right.\\
&\qquad \qquad \qquad \left.\vphantom{\frac{1}{2}}+ i\wt m_1 (\bar\psi^0\bar\sigma^\mu \psi_\mu-\bar\psi_\mu\bar\sigma^\mu \psi^0) + i\wt m_2 \varepsilon^{\mu\nu\lambda 0}(\bar\psi_\mu\bar\sigma_\nu\psi_\lambda)+  \delta g^{00} (m_{(3)} \psi_\mu\psi^\mu+\hbox{c.c.}) + \ldots \right]\, ,
\end{split}
\end{equation}
where $\ldots$ represent higher order terms in the fluctuations or the derivatives, of which we wrote just one representative, the term proportional to $m_{(3)}$.
Setting $m_\star=0$ for the rest of the introduction, it will be later useful to notice that the terms in the action (\ref{eq:SEFTofIUnitary}) that are quadratic in the gravitinos can be recast as 
\bea\label{eq:SEFTofIUnitarySimpl}\nn
&& S_{\rm {\cal S}EFTofI}[g,\psi] =  S_{\rm EFTofI}[g]\,
 +\frac{1}{2}\mpl^2\int d^4 x \, e \, \left[ \varepsilon^{\mu\nu\rho\sigma}\bar{ \psi}_\mu \bar\sigma_\nu \mathcal D_\rho  \psi_\sigma + \delta g^{00} (m_{(3)} \psi_\mu\psi^\mu+  \hbox{c.c.} )+ \ldots \right] \, , \\
\eea
where $\mathcal D_\mu$ is a covariant derivative that acts on spinors $\chi$ as  
\begin{equation}\label{cov_D_intro}
\begin{split}
\mathcal D_\mu \chi
	&= D_\mu \chi - \frac{i}{2} \left[\left(m_{3/2}+\frac{1}{2}m_0 g^{00}\right)\sigma_\mu - m_0  t_\mu (t_\nu\sigma^\nu) \right]\bar\chi - \frac{i}{2} \left[\vphantom{\frac{1}{2}}\wt m_2t_\mu - i\wt m_1 (t_\nu \sigma^\nu) \bar\sigma_\mu\right]\chi\, ,
\end{split}
\end{equation}
where $t_\mu = \delta_\mu^0$ and $D_\mu$ is the ordinary covariant derivative in curved spacetimes.

 The action in (\ref{eq:SEFTofIUnitary}) is one of the main results of this paper. It is a Lagrangian for the metric and the gravitino, invariant only under time-dependent spatial diffs.. It is the most general Lagrangian for the fluctuations in an FRW background where time diffs.~{\it and} SUSY are spontaneously broken. The first term in the third line represents the standard SUSY invariant kinetic term for the gravitino. The remaining terms in the third line and the terms in the fourth line represent mass terms for the gravitino, interaction terms between the metric and the gravitino, and higher order terms in an expansion in the number of the fluctuations and of the derivatives. It is an action directly for the fluctuations around this symmetry breaking background. No fields need to take a vev in this action in order to obtain the physical action. 

Notice, contrary to the normal treatment of supersymmetric field theories, that there are no auxiliary fields in this action. Indeed, as we will explain in detail later, when SUGRA is non-linearly realized, the auxiliary fields, that for linearly realized SUSY are introduced to enforce the equal number of bosonic and fermionic off-shell degrees of freedom and closure of the algebra, are not needed. This is an additional simplification of our formalism. Additionally, notice that the presence of the gravitino has enforced no constraints on the terms we could write in the bosonic case: all of the terms that are allowed in absence of supersymmetry are allowed now, no restriction is imposed on them by SUSY.

{\bf Reintroducing Gauge Invariance:} Since all the Lagrangians in~(\ref{eq:GaugeBosonUnitary}), (\ref{eq:EFTofIUnitary}) and (\ref{eq:SEFTofIUnitary}) represent Lagrangians where a gauge invariance is softly broken, as we discussed earlier it is useful to make the theory gauge invariant by reintroducing a matter field that transforms non-linearly under the gauge transformation.

In general, for any gauge invariance, this is achieved by performing what is normally called the St\"uckelberg trick: one performs a gauge transformation under which the Lagrangian is not invariant. Since the Lagrangian is not gauge invariant to start with, one then promotes the parameters of the gauge transformation that are present in the Lagrangian after the gauge transformation to fields that transform non-linearly under the gauge transformation. These fields are slightly improperly called Goldstone fields. This procedure makes the Lagrangian manifestly gauge invariant, at the cost of reintroducing fields that transform non-linearly under the gauge group. In practice, this amounts to replacing the fields that are present in the action with so-called St\"uckelbergized fields, that are combinations of the original fields and the Goldstone fields.

For the Gauge Boson example of eq.~(\ref{eq:GaugeBosonUnitary}), we can restore the gauge invariance by introducing Goldstone bosons $\alpha\equiv \pi^a\tau_a$, packaged in $U=e^{-\alpha}$ for simplicity, and replacing the fields $A_\mu$ with the St\"uckelbergized fields $\dress A_\mu$ given by
\begin{equation}\label{eq:stu}
A_\mu \quad \xrightarrow{\hbox{\scriptsize St\"u}} \quad  \dress A_\mu[A_\mu,\pi^a] \equiv U(A_\mu +\frac{i}{g} \d_\mu)U^{-1}\, .
\end{equation}
Notice $\dress A_\mu^a$ is then gauge invariant upon assigning the non-linear transformation of the Goldstone fields~\footnote{Indeed,
\begin{equation}
\begin{split}
\dress A_\mu^a \tau_a 
	&= UA_\mu U^{-1} + U\d_\mu U^{-1} \\  \nn
	&\to U \left[(A_\mu+\d_\mu)w^{-1}\right]wU + Uw^{-1}\d_\mu(w U^{-1})=UA_\mu U^{-1} + U\d_\mu U^{-1} + U (\d_\mu w^{-1}w + w^{-1}\d_\mu w)U^{-1}\\
	&= \dress A_\mu^a \tau_a\, .
\end{split}
\end{equation}
}
\begin{equation}\label{NLR}
U\quad\to\quad U {{w}}^{-1}\, .
\end{equation}
At this point, 
\bea\label{eq:GaugeBosonStuck}
S[A,\pi]=\int d^4 x\;\left\{ {\rm Tr}\left[F_{\mu\nu}^2\right]+ m^2 {\rm Tr}\left[\dress A_\mu^2\right]+\ldots\right\}\ ,
\eea
is gauge invariant, with gauge invariance non-linearly acting on the Goldstone bosons. Notice that since  ${\rm Tr}\left[F_{\mu\nu}^2\right]$ is gauge invariant, one has that $ {\rm Tr}\left[\dress F_{\mu\nu}^2\right]= {\rm Tr}\left[F_{\mu\nu}^2\right]$. We can now understand why the action in~(\ref{eq:GaugeBosonUnitary}) is sometimes called unitary gauge action: since gauge invariance is non-linearly realized, one gauge, called unitary gauge, is the one obtained by setting the Goldstone bosons to zero. In this gauge, the action in~(\ref{eq:GaugeBosonStuck}) takes the form~(\ref{eq:GaugeBosonUnitary}). 

The construction in terms of St\"uckelbergized fields has led us to realize of an additional (though equivalent) way to look at the building of gauge invariant Lagrangians where gauge invariance is non-linearly realized. One builds St\"uckelbergized fields that, under a non-linearly realized transformation, transform as under a linearly realized one (if none is present, than the St\"uckelbergized fields are invariant), and one builds an action using the  St\"uckelbergized fields, $\dress A_\mu$, that is invariant only under the linearly realized transformations. Once expressed in terms of the original fields, $A_\mu$ and $\pi$ for example, the resulting action will be automatically invariant under the additional non-linearly realized gauge transformation. This is the renown CCWZ construction~\cite{Coleman:1969sm,Callan:1969sn}, applied to the case of gauge transformations. In the case of the gauge bosons in (\ref{eq:GaugeBosonUnitary}), there was no linearly realized gauge transformation, and the St\"uckelbergized fields $\dress A_\mu$ transformed as scalars under the non-linearly realized gauge symmetry.

Similarly, for the case of the bosonic EFTofI, we can replace $g_{\mu\nu}$ with the diffs. St\"uckelbergized fields $\dress g_{\mu\nu}$ given by~\cite{ArkaniHamed:2002sp}
\begin{equation}\label{eq:stu2}
g^{\mu\nu}\quad \xrightarrow{\hbox{\scriptsize St\"u}}\quad \dress g^{\mu\nu}[g,\pi] \equiv \d_\alpha(x^\mu+\pi^\mu)\d_\beta(x^\nu+\pi^\nu) g^{\alpha\beta}\ , 
\end{equation}
where the Goldstone fields associated to the four diffs., $\pi^\mu$, are defined as $\pi^{\mu}=\{\pi,\vec\pi \}$. Under a diff.~of parameter $\xi^\mu$, $\tilde x^\mu=x^\mu-\xi^\mu(x)$, they transform as
\be
\pi^\mu(x)\quad\to\quad \tilde\pi^\mu(\tilde x) = \pi^\mu( x(\tilde x))+\xi^\mu(x(\tilde x)) \ .
\ee
Since our Lagrangian does not break spatial diffs., the Goldstone bosons associated to spatial translations will disappear from the action and only $\pi$ will appear, however the introduction of the Goldstones of spatial diffs.~$\vec\pi$ is useful to write (\ref{eq:stu2}) in a covariant form. In practice, only the fields that appear in operators that are not invariant under time diffs.~need to be St\"uckelbergized, in fact, St\"uckelbergization in diff.~invariant operators will cancel.
The action resulting from this procedure is invariant under non-linearly realized time diffs.~and schematically reads~\cite{Cheung:2007st}:
\bea\label{eq:EFTofIStuck}\nonumber
&&S_{\rm EFTofI}[g,\pi]=\int d^4x\;\sqrt{-g}\left\{\mpl^2 R+ \mpl^2\dot H(\dress t)  \dress g^{00}[g_{\mu\nu},\pi]+\mpl^2(H(\dress t)^2+\dot H(\dress t)) \right.\\ 
&&\quad \left.+ M(\dress t)^4 \left(\dress g^{00}[g_{\mu\nu},\pi]+1\right)^2+\ldots+\bar M(\dress t)^3 \left(\dress g^{00}[g_{\mu\nu},\pi]+1\right)\delta \dress K[g_{\mu\nu},\pi]+\ldots  \ \right\}\ ,
\eea
where $\dress t=t+\pi$ and a similar definition holds for $\delta  {\dress K}$ (see~\cite{Cheung:2007st}).

Similarly to the gauge boson case, an additional, equivalent, way to look at the construction of this Lagrangian is to write a Lagrangian in terms of $\dress g$ and $\dress t$, that is just invariant under time-dependent spatial diffs.. The resulting Lagrangian, once expressed in terms of $g$ and $\pi$, will be invariant under non-linearly realized time diffs..

Again, a similar construction can be done to make the theory invariant under non-linearly realized SUGRA transformations. Again, we are going to perform a St\"uckelberg trick: we introduce St\"uckelbergized fields that are obtained after acting on the fields with a SUGRA transformation, and promoting the parameter of the transformation to be a dynamical field. Contrary to the case of gauge transformations where the parameter is a boson, in the case of local supersymmetry the parameter is a Grassmann variable, so that the dynamical field that is introduced is a spin-1/2 spinor, it takes the name of Goldstino, and we denote it with the symbol $\lambda$. We also introduce a field $\pi$, the Goldstone boson of time translations, when we perform a time diff.~as part of a general SUGRA transformation. The action in (\ref{eq:SEFTofIUnitarySimpl}) takes the form
\bea\label{eq:SEFTofIStuck}
&&S_{\rm {\cal S}EFTofI}[e,\psi,\pi,\lambda] =   \\ \nn
&&\qquad\qquad S_{\rm EFTofI}[e,\pi]+ \frac{1}{2}\mpl^2\int d^4 x \, e \,\left\{ \varepsilon^{\mu\nu\rho\sigma}\bar{\dress{ \Psi}}_\mu[\psi,\lambda,e,\pi] \bar\sigma_\nu \hat{\mathcal D}_\rho \dress \Psi_\sigma[\psi,\lambda,e,\pi] \right. \\ \nn 
&&\qquad\qquad\qquad\qquad\qquad\qquad\qquad \left.+m_{(3)} \hat{\delta g^{00}}[e,\pi] \hat\Psi_\mu[\psi,\lambda,e,\pi]\hat\Psi^\mu[\psi,\lambda,e,\pi]+ \ldots+\hbox{c.c.} \right\} \ ,
\eea
where
\bea\label{eq:gravitino_stuck}
&& \dress \Psi^\mu 	\equiv \psi^\mu -2\dress{\mathcal  D}^\mu\lambda + (\hbox{3 fermion})\,  ,
\eea
where ``(\hbox{3 fermion})'' means terms that are cubic in fermionic fields $\psi$ and $\lambda$ (and that will be negligible for our phenomenological applications). 
$\hat \Psi_\mu$ is the full (including time diffs.) SUGRA-st\"uckelbergized gravitino $\psi_\mu$ (where we did not st\"uckelbergize the $\mu$ index because it is contracted in a diff.~invariant way). 
Notice, as we will discuss in the text, that there is no need to SUGRA-st\"uckelbergize the veilbein.

This is one of the major results of this paper. It is the most general action where local supersymmetry and time-diffs.~are non-linearly realized. In particular, it is the action that describes the model independent signatures of spontaneously broken supersymmetry in inflation. It is an action that contains the following physical degrees of freedom: two helicity-2 states in the graviton $g$, two helicity-3/2 states in the gravitino $\psi$, one helicity-0 state $\pi$ non-linearly realizing time diffs., two helicity-1/2 state $\lambda$ non-linearly realizing local SUSY. No additional scalar partner of the Goldstone boson $\pi$ (and in particular of the inflaton), is required to be light by supersymmetry. No auxiliary fields are present. No fields take a vev: this is the action directly for the fluctuations. Because of these features, this action is remarkably simple.

{\bf No Auxiliary Fields:} So far, we have neglected to include any auxiliary field and we have constructed a Lagrangian realizing SUGRA as a symmetry. Since St\"uckelbergizartion is an off-shell procedure, the reader familiar with the Supersymmetric literature might wonder how it is possible to realize SUGRA without the inclusion of auxiliary fields. When SUGRA is non-linearly realized, it is possible to construct operators out of the Goldstino that, under SUGRA, transform identically to the auxiliary fields. This means that one can substitute the auxiliary fields with these operators in any expression where they appear, without changing the algebra and the transformation properties of the remaining fields. At this point, the auxiliary fields have disappeared yet the algebra still closes off shell~\footnote{It is quite common in the SUGRA literature to dispense of the auxiliary fields by integrating them out from a specific Lagrangian and substituting the solutions to their equations of motion in the transformations of the physical fields. This procedure leads to transformation laws for these fields that are Lagrangian dependent and the algebra does not close off shell. This is different from our construction, that instead leads to Lagrangian-independent transformations, which close off shell.}. In fact, in (\ref{eq:gravitino_stuck}), the St\"uckelberg transformation should be meant as the SUGRA transformation obtained after this substitution.

{\bf Decoupling Limit:} As anticipated, the reason why it is useful to introduce a non-linearly realized gauge redundancy is that, when it is softly broken, there is a limit of energies where  the Goldstone bosons (including the fermionic Goldstino) decouple from the gauge bosons (generally meant to include the vector bosons, gravitons and gravitinos). Let us see how this happens first in the case of internal gauge invariance.

If we expand the action in~(\ref{eq:GaugeBosonStuck}), we find that the mass term of the gauge bosons induces many terms, among which a kinetic term for the $\pi$'s, a mixing term with the gauge bosons, and self interactions for the Goldstone. We have, dropping group indices  for simplicity:
\bea\label{eq:GaugeBosonStuckExtract}
S[A,\pi]\supset \int d^4 x\;\frac{m^2}{g^2}\left[   (\d\pi)^2+\pi^2(\d\pi)^2+ g A_\mu\d^\mu \pi+\ldots \right]\ .
\eea
The canonically normalized Goldstone field is given by $\pi_c=m/g \cdot \pi$. The quartic interaction tells us that the theory becomes strongly coupled at $\Lambda\sim 4\pi\, m/g$. The mixing term tells us that the Goldstone boson mixes with a helicity-0 component of $A_\mu$, $\d_\mu A^\mu$. Since the theory is gauge invariant, $A_\mu$ carries only two helicity-1 dynamical degrees of freedom, and therefore $\d_\mu A^\mu$ is a constrained variable. In fact, in the gauge where $A^0=0$, the equation of motion for $\d_\mu A^\mu=\d_i A^i$ is not dynamical, and it can be solved in terms of $\pi$ to give
\be
\d_j^2 A_i\sim \frac{m^2}{g}\d_i\pi\ ,\quad\Rightarrow\quad \d_\mu A^\mu\sim \frac{m^2}{g} \pi\ .
\ee
Plugging this in the action, we see that the contribution of the mixing term becomes arbitrarily irrelevant for energies $E^2\gg m^2$. It is also easy to check that interactions between the Goldstones and the gauge bosons are also parametrically suppressed with respect to the Goldstone self-interactions in the same limit.  As advertised, if $g\ll1$, there is a parametrically large window in which the Goldstone bosons are weakly coupled and decoupled from the gauge bosons. This is the essence of why it was useful to think of this theory as the theory of a spontaneously broken gauge invariance.

As the reader should by now expect, a similar story applies to the (bosonic) EFTofI. After St\"uckelbergization, the action~(\ref{eq:EFTofIStuck}) contains kinetic, mixing, and interaction terms. The only complication is that here there are more relevant operators, so the parameter space is larger. For example, if we focus on the tadpole term $\dress g^{00}$, we have
\bea\label{eq:EFTofIStuckExtract}
S_{\rm EFTofI, tad}\sim \int d^4x\,\sqrt{-g}\;\dot H \mpl^2 \left[(\d\pi)^2+g^{0i}\d_i\pi+\ldots\right]\ .
\eea
Because the theory has a gauge redundancy, $g^{0i}$ is a constrained variable whose solution to the equation of motion reads schematically of the form
\be
\mpl^2  \d_j^2 g^{0i}\sim \dot H\mpl^2 \d_i\pi\ ,\quad\Rightarrow\quad \d_ig^{0i}\sim \dot H \pi\ .
\ee
Notice that the LHS of the first equation above comes from the Einstein-Hilbert term of the action (we just need to notice that it has two (spatial) derivatives acting on $g^{0i}$), while the source on the RHS comes from the matter action, and is therefore proportional to $\dot H$. Plugging back in the action, we see that the mixing term between $\pi$ and $g^{0i}$ leads to a mass term for $\pi$ of size $\dot H$. This mass becomes more and more irrelevant as we move to higher energies starting from $\dot H$: $E^2\gg \dot H$. Since in inflation $\dot H\ll H^2$ and we typically compute correlation functions at energies of order $H$, this tells us that the Goldstone is decoupled from the gravitons~\cite{Cheung:2007st}.

Let us observe the same phenomenon for the case of the supersymmetric EFTofI of~(\ref{eq:SEFTofIStuck}).  Expanding the action in terms of gravitino $\psi$ and Goldstino $\lambda$ fields, the  part of the action that is quadratic in the fermionic fields reads
\bea\label{eq:SEFTofIStuckExtract}
&& S_{\rm {\cal S}EFTofI}\nn
	= \int d^4x\, e\;\mpl^2 \, \varepsilon^{\mu\nu\rho\sigma}\left(\frac{1}{2}\bar\psi_\mu \bar\sigma_\nu \mathcal D_\rho \psi_\sigma -  \bar\psi_\mu \bar\sigma_\nu [\mathcal D_\rho, \mathcal D_\sigma]\lambda
	+  \mathcal D_\mu\bar\lambda \bar\sigma_\nu [\mathcal D_\rho, \mathcal D_\sigma]\lambda\right) + {\rm c.c.}  \\ 
&&\qquad\qquad\qquad\qquad\qquad\qquad	\qquad+ m_{(3)} \delta g^{00} \left(\psi^\mu \psi_\mu -2\mathcal D_\mu \lambda\psi^\mu+4\mathcal D_\mu \lambda\mathcal D^\mu \lambda +{\rm c.c.}\right)+\ldots\ .
\eea
This expression is remarkably simple, and is again one of the main results of our paper. The Goldstino receives a kinetic term and a mixing term with the gravitino, whose prefactors are proportional to $[\mathcal D_\rho, \mathcal D_\sigma]$. Since this is a commutator of two covariant derivatives defined in (\ref{cov_D_intro}), it contains no ordinary derivatives, but contains several contributions, one of which is clearly the Riemann tensor that is indeed defined by the commutator of the standard curved-spacetime covariant derivatives on helicity-1/2 fields: $D_\mu$. In an inflationary background, this is of size at least $H^2$. This is an important fact to which we will return later. The Goldstino mixes with a helicity-1/2 component of the gravitino that, because of gauge invariance, must be a constrained variable. In the SUGRA gauge where $\bar \sigma^i\psi_i=0$, which kills one helicity-1/2 component, the equation of motion for the remaining helicity-1/2 component, $\psi^0$, is a constrain equation, which leads to the following solution:
\be
\mpl^2\bar\sigma^i \d_i \psi^0\sim \mpl^2[\mathcal D,\mathcal D] \lambda\ ,\quad\Rightarrow\quad 
\bar\sigma^i \d_i \psi^0\sim H^2 \lambda\ ,
\ee
where we replaced $[\mathcal D,\mathcal D]\sim H^2$. Plugging back in \eqref{eq:SEFTofIStuckExtract}, we see that gravitino-Goldstino mixing is subleading when $E^2\gg H^2$. Unless there are (at-least-apparently) tuned cancellations in $[\mathcal D_\rho, \mathcal D_\sigma]$, the decoupling between the Goldstino and the gravitino occurs only at energies at least above the Hubble scale. This originates from the fact that de Sitter space already breaks SUGRA, which in turn means that the Goldstino kinetic and mixing terms are born with a large wavefunction (of order $H^2\mpl^2$). This means that generically in inflation there is no sense in computing observable consequences of the Goldstino without including the gravitino, limiting the usefulness of the decoupling limit in the context of inflation. This is to be contrasted with the case of the Goldstone of time translation $\pi$: the fact that time translations are minimally broken by $\dot H$, rather than by $H$, often allows to use the decoupling limit for $\pi$ all the way to the Hubble scale.

{\bf Subtleties in the decoupling limit:} Let us focus on the very important decoupling at the level of the quadratic Lagrangian. We notice that in all of the gauge redundancies that we have considered here, the decoupling becomes better and better as we move to higher energies in a way which is quadratic with the energy. However, the decoupling limit occurs in a much more non-trivial way in the supersymmetric theory than in the bosonic cases. In fact, in the bosonic cases, the Goldstones are introduced in the action with one more derivative than the gauge boson (e.g. $\dress A_\mu\sim A_\mu+\d_\mu\pi$), so that, modulo cancellations that indeed are not present, the kinetic term of the Goldstones have more derivative than the mixing term (ex. $A^\mu\d_\mu\pi$). This induces a quadratic suppression of the mixing term at high energies. In fact, the sourcing of the constrained variable has one less derivative than the gauge-invariant two-derivative kinetic term (e.g. $\d^2_j A_i\sim\d_i \pi$), so that the sourcing of the gauge boson by the Goldstones is suppressed by one derivative at high energies. This leads to a solution for the constrained variable that, once plugged back in the mixing term, leads to a term doubly suppressed to the kinetic term of the Goldstone (e.g.~$A^\mu\d_\mu\pi\sim\pi^2$).

The situation is much more subtle for the supersymmetric case. Let us concentrate to the case where only $m_{3/2}$ is non-vanishing for simplicity. Let us try to run the same logic and focus on the leading derivative terms introduced by the St\"uckelbergization $\Psi_\mu\sim \psi_\mu+\d_\mu\lambda$.   The structure of the gravitino mass term makes the Goldstino kinetic term with the highest number of derivatives vanish (${\cal L}_{m_{3/2}}\sim m_{3/2}\d_\nu\lambda\sigma^{\mu\nu}\d_\nu\lambda=0$ ). This is indeed a fortunate event as otherwise the Goldstino would contain a ghost. But this has the consequence that there is no kinetic term at all for the Goldstino, while there is just a mixing term ({\it e.g.} $m_{3/2}\psi_\mu \sigma^{\mu\nu}\d_\nu\lambda$). If this were the end of the story, the Goldstino would inherit its kinetic term just from mixing, and therefore at high energies we would never recover a regime where the Goldstino is decoupled, no matter how softly we were to break the gauge invariance. Luckily, there is a solution. In the supergravity transformation that needs to be non-linearly realized, it is useful to keep track of the zero derivative terms, and in particular to perform a field redefinition of the gravitino field, so that its St\"uckelberg transformation reads schematically as $\Psi_\mu\sim \psi_\mu+{\cal D}_\mu\lambda\sim \psi_\mu+\d_\mu\lambda+ m_{3/2} \lambda$, where the ${\cal D}_\mu$ can be easily read off by writing the gravitino kinetic term as $\epsilon^{\mu\nu\rho\sigma}\psi_\mu \sigma_\nu {\cal D}_\rho\psi_\sigma$. Once this is done, one obtains the action in (\ref{eq:SEFTofIStuckExtract}) which has manifest decoupling. We learn that decoupling is not a simple power counting argument, but it can be important to keep the subleading terms in the gauge or St\"uckelberg transformation. This is what led to our expression (\ref{eq:SEFTofIStuck})~\footnote{The same phenomenon can happen also for bosonic gauge invariances, we will describe an example later in Sec.~\ref{sec:bosonic_stuck}.}.

{\bf Multifield:} Our formalism allows for a very straightforward generalization to the case of multifield inflation. Here, by multifield, we mean the fact that there are additional light degrees of freedom on top of the ones that are strictly required by the symmetry breaking pattern~\cite{Senatore:2010wk}. Once the additional fields are specified, the procedure to construct the general effective Lagrangian with non-linearly realized SUGRA is very simple. In unitary gauge, we write the most general Lagrangian function of the graviton, gravitino and the additional fields, that is invariant just under time-dependent spatial diffs., softly breaking time-diffs.~and SUGRA.  Then, we reintroduce the Goldstone and Goldstino by the same St\"uckelberg trick we have performed so far. This procedure can be implemented without any knowledge of the transformations of the matter fields under SUSY, as it amounts to first dressing them with the Goldstino so that they transform under SUSY as under a linearly-realized transformation, and then working directly with the dressed fields. It is immediately clear from this procedure that, as in the single-field bosonic case, all the terms that were allowed in the absence of SUSY are allowed even in this case. Furthermore, in this basis for the fields, any coupling between the Goldstino and the new fields will only be inherited from those unitary-gauge operators which describe interactions between the new fields and the gravitino.

{\bf Reheating:}  In constructing general predictions for inflation when there are multiple degrees of freedom that are light, we need to take into account of possible dependence of $\zeta$ from these additional fields induced after horizon crossing or at the time of reheating~\cite{Senatore:2010wk}. In general, at $t=t_{\rm rh}$ being the reheating time, we should write the most general local relation between $\zeta$ and the fluctuating fields invariant under time-dependent spatial diffs.:
\bea\label{eq:reheating_intro}
&&\zeta(\vec x,t_{\rm rh})=\zeta_{\rm bosonic}\left( \dress g_{\alpha\beta}(e(\vec x,t_{\rm rh}),\pi(\vec x,t_{\rm rh}))\right)\\ \nn
&&\ \qquad\qquad +a_1\; \dress\psi_\mu\left(e(\vec x,t_{\rm rh}),\pi(\vec x,t_{\rm rh})\right)\,\dress\psi^\mu\left(e(\vec x,t_{\rm rh}),\pi(\vec x,t_{\rm rh})\right)\\ \nn
&&\  \qquad\qquad +a_2\;\dress\psi^0\left(e(\vec x,t_{\rm rh}),\pi(\vec x,t_{\rm rh}))\right)\,\dress\psi^0\left(e(\vec x,t_{\rm rh}),\pi(\vec x,t_{\rm rh})\right) +\ldots \ ,
\eea
where $\ldots$ represents higher order terms in the derivatives and the number of fields, and $a_i$ represent unknown parameters that are determined by the specific way in which the $\psi$ fluctuations are converted into metric fluctuations in the sixty or so $e$-foldings from horizon crossing to reheating. This formula is useful because it needs to be evaluated at reheating time, when the gradients are negligible, so that the functional form of the contributions is highly constrained and cannot completely alter the predictions from horizon crossing time.  By substituting $\hat \psi^\mu$ with $\dress \Psi^\mu$ in (\ref{eq:reheating_intro}), one can evaluate this formula in every SUGRA-gauge.

\vspace{1cm}

{\bf Phenomenology:} After having developed the formalism to construct the effective Lagrangian associated to the spontaneous breaking of SUGRA during inflation, we are ready to study its observational consequences. We first study the dispersion relations of the Goldstino, finding that it can either have a linear dispersion relation, $\omega\sim c_\lambda k$, with $c_\lambda-1$ as large as ${\cal O}(1)$, or, in the presence of a non-zero $m_\star$, a quadratic dispersion relation, $\omega\propto m_\star\, k^2$. Requiring subluminal propagation of the sound waves imposes some bounds on the coefficients of~(\ref{eq:SEFTofIUnitarySimpl}), and in particular that the quadratic dispersion relation can occur only for an order one range of wavenumbers, at scales shorter than~$H$. 

The model-independent signature of SUGRA is associated to the presence of the gravitino (and of the Goldstino). Since these are fermions, they can only affect $\zeta$ correlation functions at loop level where, at leading level, they appear in internal lines, while $\zeta$ (or $\pi$) appear as external lines. We consider two different class of couplings. The so-called `minimal couplings' between the fermions and the metric or $\pi$, which arise from covariantization of the mass terms in the unitary gauge Lagrangian in~(\ref{eq:SEFTofIUnitarySimpl}), and the `non-minimal' ones that come from additional couplings that we write in unitary gauge. For all of these coupling, the nature of the loop-induced signal splits into two distinct contributions. When in the internal lines we consider momenta that are much greater than $H$, the resulting contribution to the correlation function has the same functional form of the one produced by the operators in the bosonic single-field case, parametrized by $f_{\rm NL}^{\rm equil.}$~\cite{Creminelli:2005hu} and $f_{\rm NL}^{\rm orthog.}$~\cite{Senatore:2009gt,Behbahani:2014upa} for the case where we assume an approximate shift symmetry for $\pi$. The fact that the contribution has this shape must be so by renormalizability of the theory. How large this contribution is cannot be reliably estimated within the EFT, because it depends on the UV completion. However, it is possible to estimate an upper bound to its size, which is obtained by cutting off the loops at the unitarity bound of the theory. Barring the presence of unexpected cancellations, the induced signal can be as high as to produce $\fnlrad\sim1$ for the minimal couplings, while it can be $\fnlrad\gg 1$ for the non-minimal couplings. 

When instead in the loop we run momenta of order $H$, the shape of the induced non-Gaussianity is not expected to be exactly degenerate with the one induced by the  operators of the bosonic EFT. Detecting their signal would probably be a `smoking gun' of the presence of supersymmetry as a non-linearly-realized symmetry of nature. We do not compute explicitly these shapes because we find that the induce signal is very small in practically all cases. We can summarize our findings by saying that the most natural signal of SUGRA being spontaneously broken during inflation seems to be the natural induction of sizable non-Gaussianities of the shape produced by the bosonic sector, up to the level of $\fnlrad\sim 1$, or, for non-minimal couplings, even $\gg 1$. Detecting such a shape would not be a direct indication of additional light fields present during inflation, but it would still be a spectacular signal that would teach us about the interacting structure of the theory of Inflation, and will motivate additional theoretical and observational studies.

The prospects of detecting $\fnlrad\sim 1$ seems rather hard at the moment. The current best limits come from the Planck analysis~\cite{Ade:2015ava}, that constrain $\fnlrad\lesssim 10^2$. Reaching $\fnlrad\sim 1$ requires us to be able to access many additional modes. At the moment, the best reasonably-short term opportunity seems to come from Large Scale Structure surveys. Recently, much progress in understanding their dynamics from an analytical point of view has been made with the introduction of the so-called Effective Field Theory of Large Scale Structures~\cite{Baumann:2010tm,Carrasco:2012cv,Porto:2013qua,Senatore:2014via}. If the new understanding conveyed by this novel theoretical approach, or by improving on former numerical approaches, will be powerful enough for us to reach such observational bounds is yet to be established, even though it will probably be challenging~\cite{Baldauf:2016sjb}. However, since Cosmology and cosmologists have faced many challenges in their past, we hope that they will successfully address this novel trial.

\vspace{1.1cm}

{\bf Structure of the paper:} After this ample introduction, the rest of the paper will be devoted to present explicit formulas and construction of the ideas and the logic presented in this section. Since many of the subtle points needed to construct the Supersymmetric EFTofI are already present in the simpler case where the spacetime is maximally symmetric and only SUSY is spontaneously broken, we will start by describing this case in Sec.~\ref{sec:max_sym}. We will then move to the FRW case in Sec.~\ref{FRW}. In~Sec.~\ref{sec:pheno} we will describe the observational signatures of the models we described. We conclude in Sec.~\ref{sec:conclusions}.

\vspace{1cm}

{\bf Relation to former literature:} The Inflationary supersymmetric and supergravity literature is, needless to say, extremely large. Our paper starts with investigating the non-linear realization of SUGRA and the supersymmetric Higgs mechanism in the context of maximally symmetric spacetimes. This of course traces back to the remarkable work of~\cite{Deser:1977uq}, which was then completed to recover the full group structure of the non-linear realization~\cite{Kapustnikov:1981de} and all the literature that followed. A short sample of the very most recent literature on SUGRA in inflation is~\cite{Bergshoeff:2015tra,Kallosh:2015sea,Kallosh:2015tea,Ferrara:2015tyn,Dudas:2015eha,Ferrara:2015gta,Carrasco:2015iij,Bergshoeff:2016psz,Kallosh:2016ndd,Kallosh:2016gqp,Ferrara:2016ajl,Antoniadis:2016aal}.

In the context of inflation and EFT, after the advent of the EFTofI~\cite{Cheung:2007st}, the first attempt to apply the same techniques to study the observable consequences of SUSY in inflation was made in~\cite{Senatore:2010wk}, where a supersymmetric multifield sector that was interacting with the inflaton only gravitationally  was considered. Since the inflaton breaks SUSY at a scale $\dot H\mpl^2$, stronger-than-gravity interactions with it are extremely dangerous. Restricting to gravitational interactions offered instead a sufficient sequestering from the SUSY breaking sectors. Ref.~\cite{Senatore:2010wk} found that some relations between observables were protected by unbroken supersymmetry. Our results show that these findings were model dependent. Indeed, if one asks for just a nob-linear realization of SUSY, every multifield Lagrangian can be made in this sense supersymmetric.

Subsequently, Ref.~\cite{Baumann:2011nk} attempted to describe in an EFT way also the inflaton sector. They found that a second scalar field, a partner of the Goldstone $\pi$, was generically present with mass of order~$H$. Its interactions with the Goldstone could lead to potentially large non-Gaussianities with the typical shapes of quasi-single-field inflation~\cite{Chen:2009zp}. We find that these results are very model dependent. We find there is no reason for a second scalar to be light during inflation: we find that only the Goldstino (or equivalently the gravitino) are sufficient to non-linearly realize SUGRA. From this, needless to say, many additional differences follow, as for example in the structure of the operators. Concerning the presence of an additional second scalar, direct interaction with the Goldstone will generically make an additional light field that is unprotected by any additional symmetry very massive, unless it is just gravitationally interacting, which tends to imply the lack of observable signatures.

The spectrum that we find here is in fact coincident with the one already pointed out in~\cite{Kahn:2015mla} as sufficient for non-linearly realizing SUGRA during inflation: a Goldstone and a Goldstino (see also~\cite{Antoniadis:2014oya, Kallosh:2014via,Dall'Agata:2014oka} for having already showed through particular models that this minimal spectrum can non-linearly realize SUGRA). However, our findings are quite different with respect to the structure of the Lagrangian and the resulting phenomenology. In fact, our construction is remarkably simple: there is no superspace needed, nor auxiliary fields, nor do we make any use of constrained superfields. This is not just a formal achievement, because the resulting simplified formalism allows us to explore in more generality the phenomenology of inflation. In fact, the main difference between our construction and the one in~\cite{Kahn:2015mla} resides in the organization of the perturbative expansion. In~\cite{Kahn:2015mla}, the theory is written around the maximally symmetric spacetime. The inflationary background is then obtained giving small vevs to the operators, so that the inflationary solution is perturbatively close to the maximally symmetric one~\footnote{In particular, higher dimension operators that contribute to the theory of the fluctuations by substituting some fields with their vevs are counted as a small perturbation.}. We call these models of inflation as slow-roll inflation, and, in this sense, Ref.~\cite{Kahn:2015mla} constructs the EFT of supersymmetric Slow-Roll Inflation. On the contrary, as we stress also in the introduction, the connection between the theory around the maximally symmetric spacetime and the inflationary one can be very complicated, and in general not even describable with perturbative techniques. We build instead directly the theory of the fluctuations, with no need of any knowledge of the background solution. This makes the assessment of the importance of an operator straightforward: every operator starts with a given number of fluctuations and derivatives.  This allows us to explore in full generality the phenomenology of the consequences of SUSY in Inflation. This is indeed made possible by the fact that our construction makes use of very analogous techniques that are used to construct the bosonic EFTofI, such as writing the theory of the fluctuations by going to unitary gauge, performing the St\"uckelberg trick and identifying the decoupling limit, generalizing from diff.~breaking to SUGRA breaking. In practice, the difference with~\cite{Kahn:2015mla} can be also shown with a series of examples. At a given order, even at the leading ones, we have many more operators and therefore physical effects. Furthermore, we find that the Goldstino can have a linear dispersion relation $\omega=c_\lambda k$ with $c_\lambda-1$ that can be up to order one; or it can even have a quadratic dispersion relation $\omega\propto k^2$. On the contrary,~\cite{Kahn:2015mla} finds only a linear dispersion with $c_\lambda-1\sim {\cal O}(\epsilon)$. We find that the Goldstino is always mixed with the gravitino at energies of order $H$, so that it does not make sense to study the gravitino without the Goldstino in making inflationary prediction. Instead Ref.~\cite{Kahn:2015mla} does not seem to discuss the relevance of the mixing between Goldstino and gravitino in inflationary phenomenology. Additionally, we find that our EFT can predict non-Gaussianites with $\fnlrad\sim 1$ and even $\gg 1$, while Ref.~\cite{Kahn:2015mla} finds only very small corrections. This makes it clear that our approach is very different to the one of~\cite{Kahn:2015mla}, which is correct if one is trying to describe what we define to be slow-roll inflation, but not a general inflationary solution.

\section{Maximally symmetric spacetimes\label{sec:max_sym}}

Some of the ideas and formalisms that we need to construct the Supersymmetric EFTofI~(\SEFTofI)~are already relevant in the simpler case in which diffeormorphisms are unbroken, but supergravity is softly broken. This situation is interesting on its own, and we present it here. The logic will be exactly as spelled out in the Introduction, but now the details will be given.

\subsection{Non-linear representation of the Supergravity algebra}\label{sec_redef}

We start by considering the pure supergravity multiplet $\{e_\mu^a,\, \psi_\mu,\,  m,\, b_\mu\}$, realizing local supersymmetry as~\cite{Wess:1992cp}~\footnote{\label{footnote:Ddef}We follow the notation of~\cite{Wess:1992cp} throughout this paper. The covariant derivative $D_\mu$ in the $(\frac{1}{2},0)$ and $(0,\frac{1}{2})$ representations is given by
\begin{equation*}
D_\mu^{(\frac12,0)} = \d_\mu + \frac12{\omega}_\mu^{ab}\sigma_{ab} \, , \qquad
D_\mu^{(0,\frac12)} = \d_\mu + \frac12{\omega}_\mu^{ab}\bar\sigma_{ab} \, , 
\end{equation*}
where 
\begin{equation*}
\sigma_{ab}=\frac{1}{4} \left(\sigma_a\bar\sigma_b-\sigma_b\bar\sigma_a\right)\, , \qquad 
\bar\sigma_{ab}=\frac{1}{4} \left(\bar\sigma_a\sigma_b-\bar\sigma_b\sigma_a\right)\, ,
\end{equation*}
and $\sigma_a$ are the ordinary Pauli matrices. The spin connection can be taken to be
\begin{equation*}
\omega_\mu^{ab}  = e^b{}_\nu \nabla_\mu e^{a\nu} \ .
\end{equation*}
This form of the spin connection does not include torsion terms, and, contrary to what would result from using the second order formalism, it does not depend on the Lagrangian. Different forms of the spin connection are degenerate with adding different higher dimension operators to the Lagrangian, which are included anyway.}
}
\begin{subequations}\label{delta_susy_aux}
\begin{align}
\delta_\epsilon e_\mu^a \label{delta_susy_aux_e}
	&= i \psi_\mu \sigma^a \bar \epsilon + \hbox{c.c.}  \, ,\\
\delta_\epsilon \psi_\mu \label{delta_susy_aux_psi}
	&= -2D_\mu \epsilon - K_\mu^{ab}\sigma_{ab}\epsilon + i\left[ m  \sigma_\mu\bar \epsilon +  b_\mu \epsilon + \frac{1}{3} b^\nu ( \sigma_\mu\bar \sigma_\nu\epsilon)\right] \, , \\
\delta_\epsilon  m \label{delta_susy_aux_M}
	&=  - \frac{1}{3}\epsilon \left(\sigma^\mu\bar\sigma^\nu  \psi_{\mu\nu} + i b^\mu \psi_\mu - 3i\sigma^\mu \bar \psi_\mu  m\right) \, , \\
\delta_\epsilon  b^a \label{delta_susy_aux_b}
	&= \frac{3}{8} (\bar \psi_{\mu\nu} \bar \sigma^a\sigma^\mu \bar\sigma^\nu \epsilon)
	- \frac{1}{8} (\bar \psi_{\mu\nu}\bar\sigma^\mu \sigma^\nu \bar\sigma^a \epsilon)
	- \frac{3i}{2} {m}^* (\epsilon \psi^a) \\ \nonumber
	& \ - \frac{i}{8}  b_c (\epsilon \sigma^c \bar \sigma^a \sigma^\mu \bar\psi_\mu)
	+ \frac{i}{4}   b^a(\epsilon \sigma^\mu \bar\psi_\mu) + \frac{i}{8} b^c (\bar \psi_\mu \bar \sigma^a \sigma_c \bar \sigma^\mu \epsilon) + {\rm c.c.}\, ,
\end{align}
\end{subequations}
Here $\epsilon$ is a Weyl fermion parametrizing the SUSY transformation, 
\begin{equation}\label{wouldbe_spin_connection}
K_\mu^{ab} =  \left[\left(-\frac{i}{4}\psi^a\sigma^b\bar\psi_\mu - \frac{i}{8}\psi^a\sigma_\mu\bar\psi^b + \hbox{c.c.}\right) - (a \leftrightarrow b)\right] \ .
\end{equation}
and $\psi_{\mu\nu} = D_{\mu}\psi_\nu + \frac{1}{2}K_\mu^{ab}\sigma_{ab}\psi_\nu - (\mu \leftrightarrow  \nu)$. Notice that $D_\mu$ contains only the (torsion-free) connection $\omega_\mu^{ab}(e)=e^b{}_\nu \nabla_\mu e^{a\nu}$~(\footnote{In~\cite{Wess:1992cp}, the spin connection is chosen to be
\begin{equation}\nn
\omega_\mu^{ab}{}_\text{Wess-Bagger}  = e^b{}_\nu \nabla_\mu e^{a\nu} + \left[\left(-\frac{i}{4}\psi^a\sigma^b\bar\psi_\mu - \frac{i}{8}\psi^a\sigma_\mu\bar\psi^b + \hbox{c.c.}\right) - (a \leftrightarrow b)\right] = \omega_\mu^{ab}(e) + K_\mu^{ab}\, .
\end{equation}
so that $D_\mu{}_\text{Wess-Bagger}\epsilon=D_\mu \epsilon + \frac{1}{2}K_\mu^{ab}\sigma_{ab}\epsilon$.
 As we explain in footnote~\ref{footnote:Ddef}, here we take the spin connection $\omega_\mu^{ab}$ to be $\omega_\mu^{ab}  = e^b{}_\nu \nabla_\mu e^{a\nu}$, and we consider~(\ref{delta_susy_aux}) with~(\ref{wouldbe_spin_connection}) simply as a definition of the transformation of the gravitino field.  }).
 
 The SUGRA algebra is given by 
\begin{align}\label{algebra}
[\delta_{\epsilon'},\delta_{\epsilon}]
	& =  \left(\delta^{\rm diff}_{y} + \delta^{L}_\Lambda + \delta_{\hat \epsilon}\right)\, ,\\ \nn
 [\delta_\epsilon,\delta^{\rm diff}_\xi] & =  \delta_{\xi^\mu\d_\mu\epsilon}\, ,\\ \nn
 [\delta_\epsilon,\delta^L_{\Lambda}]  & =   \delta_{\frac{1}{2}\Lambda^{ab}\sigma_{ab}\epsilon}\ ,
\end{align}%
with
\begin{eqnarray}\label{algebra2}\nn
&&y^\mu=-2i(\epsilon'\sigma^\mu\bar\epsilon - \epsilon\sigma^\mu\bar\epsilon')\, , \\ \nn
&& 
	\Lambda^{ab}= y^\mu \left(\omega_\mu^{ab}(e) + K_\mu^{ab}\right) -\left[4m\bar\epsilon \bar\sigma^{ab}\bar\epsilon' + \frac{2}{3}\epsilon\sigma^{[a}\cancel b \sigma^{b]}\bar\epsilon'+\hbox{c.c.}\right]\ ,  \\ 
	&&
\hat \epsilon = \psi_\mu y^\mu/2\, ,
\end{eqnarray}
where $\cancel b=b^\mu \bar\sigma_\mu$. Notice that the algebra closes on field dependent transformations: this peculiar fact is quite standard when we add internal gauge symmetries to spacetime diffs..

In order for the supermultiplet in (\ref{delta_susy_aux}) to linearly realize the SUGRA algebra, two auxiliary fields $m$ and $b_\mu$ needed to be added on top of the ones that we are interested in, $e^a_\mu$ and $\psi_\mu$. This requirement can be easily seen to be necessary to enforce the equality of fermionic and bosonic off-shell degrees of freedom or equivalently off-shell closure of the algebra.  However, here we are interested in the case in which SUGRA is softly broken. Therefore, we add to this multiplet a Goldstino $\lambda$. To derive the transformation of the Goldstino, one can implement the CWZ construction~\cite{Coleman:1969sm}, which gives
\begin{equation}
\delta_{\epsilon}\lambda 
	= -\epsilon + i(\lambda\sigma^\mu \bar\epsilon - \epsilon \sigma^\mu \bar\lambda)\left(-\frac{1}{2}\left(\psi_\mu + \delta_\lambda \psi_\mu\right) + \frac{i}{3}m \sigma_\mu \bar\lambda \right) + (\lambda\epsilon) \left(2m^* \lambda - \frac{4}{3}\cancel b \bar\lambda\right) +(\hbox{4 ferm.})\epsilon\ ,
\end{equation}%
where here and henceforth ``($n$ fermion)'' refers to terms containing $n$ physical fermions ($\psi_\mu$ or $\lambda$) which will not play a role in our analysis~\footnote{\label{footnote:CCWZconstruction}If the structure constants depend on the fields, it is not obvious that the proof of CWZ~\cite{Coleman:1969sm} still works, but it does. Let us repeat here the relevant steps. If the group $G$ is broken to a subgroup $H$, let us parametrize the coset $G/H$ with $e^{-\lambda \cdot Q}\in G/H$ and a group element as $e^{-\lambda \cdot Q} h \in G$, with $h\in H$, and $Q$ representing the broken generators. Then for $g\in G$, $g\,e^{-\lambda\cdot Q}$ is a group element and as such can be uniquely parametrized (close to the identity) as
\begin{equation}\label{CWZ}
g\, e^{-\lambda\cdot Q} = e^{-f_{(a,g)}\left(\lambda\right)\cdot Q} h (\lambda,a,g)\, ,
\end{equation}
where $a$ represents some set of fields on which the structure constants of the algebra depend, transforming under some representation of the group $G$: $a\to a'=D_G(g)\circ a$. In our case the fields $a$ are given by the $\{e,\psi,m,b\}$ supermultiplet. Eq.~(\ref{CWZ}) defines a non-linear realization of $G$ on the multiplet $\{ \lambda,a \}$, $\lambda(x)\to \lambda'(x')=f_{(a,g)}\left(\lambda(x)\right)$, $a(x)\to a'(x')=D_G\left(g\right)\circ a(x)$.  The difference with the usual CWZ construction here is that the group element $h (\lambda,a,g)$ and the tranformation $f_{(a,g)}$ depend explicitly on the matter fields $a$. This in particular means that the fields $a$ participate to the multiplet of the Goldstone bosons. To show that this construction is a faithful representation of $G$, we act with an additional transformation on~(\ref{CWZ}):
\begin{equation}\nn
\tilde g\, g\, e^{-\lambda \cdot Q} = \tilde g\, e^{-f_{(a,g)}\left(\lambda\right)\cdot Q}  h (\lambda,a,g)= e^{-f_{(a',\tilde g)}\left(f_{(a,g)}\left(\lambda\right)\right) \cdot Q} \, h \left(f_{(a,g)}\left(\lambda\right),a',\tilde g\right)\;h (\lambda,a,g)\, ,
\end{equation}
where in the second step we acted on $e^{-f_{(a,g)}\left(\lambda\right) \cdot Q}$ and where we remind that $a'=D_G(g)\circ a$.
On the other hand, this same object may also be written as
\begin{equation}\nn
(\tilde g g)\, e^{-\lambda \cdot Q} = e^{-f_{(a,(\tilde g g))}\left(\lambda\right) \cdot Q} h (\lambda,a,\tilde g g) \, .
\end{equation}
Comparing these two expressions, uniqueness of the parametrization shows that
\begin{equation}\nn
f_{(a,(\tilde g g))}\left(\lambda\right) = f_{(a',\tilde g)}\left( f_{(a,g)} (\lambda)\right) \quad \hbox{and} \quad 
h (\lambda,a,\tilde g g) = h \left(f_{(a,g)}(\lambda),a',\tilde g\right)\;h (\lambda,a,g)\ ,
\end{equation}
which tells us that this is a reparemetrization. In particular, $h (\lambda,a,g)$ is an homomorphism from $G$ to $H$. Uniqueness of the parametrization holds in a basis of generators of the group such that $[Q,T]\sim Q$, where $T$ are the unbroken generators. The SUGRA algebra of (\ref{algebra}) satisfies this requirement.

If now we introduce matter fields $B$ that transform under a (non-necessarily linear) representation of $H$: $B\to B'=D_H(h)\circ B$, then the same fields inherit a representation of the full group~G through their transformations under $h\in H$: $B\to B'=D_H\left(h (\lambda,a,g)\right)\circ B$. In particular, the fields $B$ can be obtained by decomposing the multiplet $A = D_G\left[e^{\lambda(x)\cdot Q}\right]\circ a$ into submultiplets that transform irreducibly under $H$. 
Just for clarity, we check that $B\to B'=D_H\left(h (\lambda,a,g)\right)\circ B$ is a faithful non-linear representation of $G$ by simply noticing that
\be\nn
D_H\left(h (\lambda,a,\tilde g g)\right)=D_H\left(h \left(f_{(a,g)}(\lambda),a',\tilde g\right)\circ h (\lambda,a,g)\right)=D_H\left(h \left(f_{(a,g)}(\lambda),a',\tilde g\right)\right)\, D_H\left(h (\lambda,a,g)\right) \ .
\ee
We will call the fields $B$ as `dressed' or `St\"uckelbergized' fields.
}.  The first explicit realization of this construction was obtained long ago by Kapustnikov in~\cite{Kapustnikov:1981de}.

The presence of the Goldstino allows us to construct dressed fields from the matter fields in $a$ as
\begin{equation}\label{eq:dresssedfields}
 A = D_G\left[e^{\lambda(x)\cdot Q}\right]\circ a\, .
\end{equation}
Under a general symmetry, these will transform as under a linearly realized symmetry $h$ (this is the renown CCWZ construction~\cite{Callan:1969sn}): 
\begin{equation}
A \quad \to \quad A' =  D_G\left[e^{\lambda'(x')\cdot Q}\right] \circ a' =  D_G\left[ h(\lambda,g,\psi)\,e^{\lambda(x)\cdot Q}\,g^{-1}g\right]\circ a = D_G\left[h(\lambda,g,\psi)\right]\circ  A\, ,
\end{equation}
where $h$ is a diff.~and a local Lorentz transformation. In the second passage, we have used the transformation property of the Goldstino (see footnote~\ref{footnote:CCWZconstruction}): $e^{\lambda'(x)Q}= h(\lambda,g,\psi)e^{\lambda(x)Q}\,g^{-1}$. This is particularly useful for us because  it allows us to impose the following constraints, invariant under all symmetries:
\begin{equation}
 M = 0 \, , \qquad
 B_\mu = 0 \, .
\end{equation}
These can be solved for the original auxiliary fields $m$ and $b_\mu$ in terms of the remaining fields
\begin{equation}\label{kill_auxiliary}
m = \frac{2}{3}(\lambda\sigma^{\mu\nu}\psi_{\mu\nu}) + \ldots\, , \qquad 
b^a = -\frac{3}{8}(\bar\psi_{\mu\nu}\bar\sigma^a\sigma^{\mu}\bar\sigma^\nu \lambda) +  \frac{1}{8} (\bar \psi_{\mu\nu}\bar\sigma^\mu \sigma^\nu \bar\sigma^a \lambda)+\hbox{c.c.}+\ldots\, ,
\end{equation}
where $\ldots$ stands for higher fermion terms, which can be obtained in a perturbative manner.
Upon substituting these values for the auxiliary fields everywhere they appear in the transformation of the fields and in the algebra, we construct a representation of SUGRA that does not contain the auxiliary fields, which we can forget from now on~\footnote{One might wonder if this replacement of the auxiliary fields might change the SUGRA algebra. The answer is `no', once we express the auxiliary fields in terms of the dynamical fields, as in (\ref{kill_auxiliary}). Let us give more details on how this happens. Consider a gauge invariant constraint $F(\chi,\phi_i)=0$ through which we express $\chi$ in terms of the other fields: $\chi=\tilde{\chi}(\phi_i)$. We would like to show that after substituting $\chi$ with $\tilde{\chi}$ the algebra does not change. In fact, an  even a stronger statement holds: the variation of an arbitrary functional of fields $f(\tilde{\chi}(\phi_i),\phi_i)$ under gauge transformations remains unchanged:
\begin{equation}
\frac{\partial f(\chi,\phi_i)}{\partial \chi}\delta \chi+\frac{\partial f(\chi,\phi_i)}{\partial \phi_i}\delta \phi_i= \frac{\partial f(\chi,\phi_i)}{\partial \chi}\frac{\partial \tilde{\chi}(\phi_i)}{\partial \phi_i}\delta \phi_i+\frac{\partial f(\chi,\phi_i)}{\partial \phi_i}\delta \phi_i\ .
\label{eq:footnote}
\end{equation}
Since the transformations are unchanged, so it will be the algebra. Eq.~(\ref{eq:footnote}) follows from showing that $\delta\chi=\tfrac{\partial \tilde{\chi}(\phi_i)}{\partial \phi_i}\delta \phi_i$. In turn, this follows from the implicit function theorem which we review for the case in hand. Indeed, for an arbitrary variation of fields $\phi_i$
\begin{equation} \nn
\left.\frac{\partial F(\chi,\phi_i)}{\partial \chi}\right|_{\chi=\tilde\chi(\phi)}d \tilde{\chi}(\phi_i)=-\left.\frac{\partial F(\chi,\phi_i)}{\partial \phi_i}\right|_{\chi=\tilde\chi(\phi)}d \phi_i\ ,
\end{equation}
Also, since $F(\tilde\chi(\phi_i),\phi_i)=0$ is gauge invariant, for an infinitesimal gauge transformation $\delta$ we have
\begin{equation}\nn
\left.\frac{\partial F(\chi,\phi_i)}{\partial \chi}\right|_{\chi=\tilde\chi(\phi)}\delta \chi=-\left.\frac{\partial F(\chi,\phi_i)}{\partial \phi_i}\right|_{\chi=\tilde\chi(\phi)}\delta \phi_i.
\end{equation}
Now setting $d \phi=\delta\phi$ and dividing the two last equations by $\left.\partial F(\chi,\phi_i)/\partial \chi\right|_{\chi=\tilde\chi(\phi)}$, we get:
\begin{equation}\nn
d\tilde{\chi}(\phi_i) \Big|_{d \phi_i=\delta \phi_i}\equiv\frac{\partial \tilde{\chi}(\phi_i)}{\partial \phi_i}\delta \phi_i=\delta\chi.
\end{equation}
Hence (\ref{eq:footnote}). 
}. Notice that imposing this constraint is similar, but not equivalent, to integrating out the auxiliary fields. In fact, integrating out the auxiliary fields would set them to the value that satisfies their equations of motion, which are Lagrangian dependent. At this point, the transformations of the remaining fields would become Lagrangian dependent, but this would imply that the algebra would close only on-shell, or equivalently only on a residual Lagrangian-dependent transformation (see for example~\cite{Freedman:2012zz}). The constraint we impose here instead is Lagrangian independent, and, since it is invariant under all symmetries, the algebra obtained by substituting this constraint into the original algebra still closes off shell. The fact that our construction had to be possible could have been expected from noticing that, in the context of global SUSY, the Volkov-Akulov Lagrangian~\cite{Volkov:1973ix} does not contain auxiliary fields, and in fact, our construction is related to the analogous one done for global SUSY long time ago by Rocek~\cite{Rocek:1978nb}.

The construction we just implemented can also be thought of in the following way. Once the Goldstino is in the spectrum, we can construct operators out of the Goldstino, the vierbein and the gravitino that transform under a general SUGRA transformation as the auxiliary fields $m$ and $b_\mu$. At this point, nothing forbids us to substitute $m$ and $b_\mu$ with these operators, and obtaine a closed algebra and multiplet.

It will be useful when studying the action to perform the following field redefinition
\begin{equation}\label{eq:field_red1}
\psi_\mu' = \psi_\mu - i\,m_{3/2}^*\sigma_\mu\bar\lambda\, .
\end{equation}
The gravitino now transforms as
\begin{equation}
\begin{split}
\delta_{\epsilon}\psi_\mu'
	&= -2\mathcal D_\mu \epsilon - K_\mu^{ab}\sigma_{ab}\epsilon + i\left[ m  \sigma_\mu\bar \epsilon +  b_\mu \epsilon + \frac{1}{3} b^\nu ( \sigma_\mu\bar \sigma_\nu\epsilon)\right] + (\hbox{4 ferm.})\epsilon \\
	&= -2\mathcal D_\mu \epsilon  + (\hbox{2 ferm.})\epsilon\ ,
\end{split} 
\end{equation}
where $K_\mu^{ab}$ is the $\psi\psi$ bilinear defined in \eqref{wouldbe_spin_connection}, $m$ and $b_\mu$ are the $\psi\lambda$ bilinears (to leading order in fermions) defined in \eqref{kill_auxiliary}, and where we have defined a generalized covariant derivative as
\begin{subequations}\label{cov_D_dS}
\begin{align}
\mathcal D_\mu \lambda &= D_\mu \lambda - \frac{i}{2} m_{3/2}^*\sigma_\mu \bar\lambda\, , \\
\mathcal D_\mu \bar \lambda &\equiv \overline{{\mathcal D}_\mu \lambda}= D_\mu \bar\lambda - \frac{i}{2} m_{3/2}\bar\sigma_\mu\lambda\, .
\end{align}
\end{subequations}
The convenience of introducing the covariant derivative (\ref{cov_D_dS}) has been explained in the introduction and we will comment on it again later.

The auxiliary fields $m,\,  b_\mu$ appearing in the local SUSY transformation of the gravitino $\psi_\mu$ are necessary for the off-shell closure of the algebra \eqref{delta_susy_aux} in the pure supergravity multiplet, but do not otherwise participate in the dynamics. When SUSY is non-linearly realized, there is no need of fermion-boson degeneracy, so it is expected that these fields should not be needed. This is what we have achieved in this construction. This is particularly convenient for our purposes, as it allows us to get rid of the auxiliary fields from the outset without spoiling the algebra.  Specifically, the multiplet ${e{}_\mu^a,\psi',\lambda}$ transforms as 
\begin{subequations}\label{delta_susy}
\begin{align}
\delta_\epsilon e{}_\mu^a
	&= i \left(\psi'_\mu  -i m_{3/2}^* \bar\lambda \bar \sigma_\mu\right)\sigma^a\bar \epsilon + \hbox{c.c.} + (\hbox{3 fermion})\cdot \epsilon  \, ,\\
\delta_\epsilon \psi'_\mu 
	&= -2\mathcal D_\mu \epsilon + (\hbox{2 fermion})\cdot \epsilon\, , \\
	\delta_{\epsilon}\lambda 
	& = -\epsilon + i(\lambda\sigma^\mu \bar\epsilon - \epsilon \sigma^\mu \bar\lambda)(\mathcal D_\mu \lambda - \frac{1}{2}\psi'_\mu ) + (\hbox{4 ferm.})\epsilon
\end{align}
\end{subequations}

We conclude this section with some important comments. Field redefinitions do not change the algebra~\footnote{Here is the proof. Imagine we have a set of fields $\varphi^a$ satisfying the algebra 
\be
\left[\delta_{\epsilon},\delta_{\epsilon'}\right]\,\varphi^a=\delta_{\epsilon''(\epsilon,\epsilon',\{\varphi^l\})}\varphi^a\ ,
\ee
where the form of $\epsilon''(\epsilon,\epsilon',\{\varphi^l\})$ is dictated by the group structure. Notice that, as it happens for SUGRA, we have allowed for a dependence on the fields themselves in the closure of the algebra. Consider a field redefinition of the fields $\wt\varphi^a=\wt\varphi^a(\{\varphi^b\})$. The new fields $\wt \varphi^a$ satisfy the algebra
\bea
&&\left[\delta_{\epsilon},\delta_{\epsilon'}\right]\wt\varphi^a=\delta_{\epsilon}\left(\frac{\d \wt\varphi^a}{\d \varphi^b}\,\delta_{\epsilon'}\varphi^b\right)- \delta_{\epsilon'}\left(\frac{\d \wt\varphi^a}{\d \varphi^b}\,\delta_{\epsilon}\varphi^b\right) \\ \nn
&&\qquad\qquad=\frac{\d^2 \wt\varphi^a}{\d \varphi^c\d\varphi^b}\left(\delta_{\epsilon}\varphi^c\delta_{\epsilon'}\varphi^b- \delta_{\epsilon'}\varphi^c\delta_{\epsilon}\varphi^b\right)+\frac{\d \wt\varphi^a}{\d \varphi^b}\left(\delta_\epsilon\delta_{\epsilon'}\varphi^b- \delta_{\epsilon'}\delta_{\epsilon}\varphi^b\right)\\ \nn
&&\qquad\qquad=\frac{\d^2 \wt\varphi^a}{\d \varphi^c\d\varphi^b}\left(\delta_{\epsilon}\varphi^c\delta_{\epsilon'}\varphi^b - (\pm)  \delta_{\epsilon'}\varphi^b\delta_{\epsilon}\varphi^c\right)+\frac{\d \wt\varphi^a}{\d \varphi^b}\left(\delta_{\epsilon''(\epsilon,\epsilon',\{\varphi^l\})}\varphi^b\right)\\ \nn
&&\qquad\qquad=\frac{\d^2 \wt\varphi^a}{\d \varphi^c\d\varphi^b}\left(\delta_{\epsilon}\varphi^c\delta_{\epsilon'}\varphi^b- \delta_{\epsilon}\varphi^c \delta_{\epsilon'}\varphi^b\right)+\delta_{\epsilon''(\epsilon,\epsilon',\{\varphi^l(\{\wt\varphi^m\})\})}\wt\varphi^a\\ \nn
&&\qquad\qquad=\delta_{\epsilon''(\epsilon,\epsilon',\{\varphi^l(\{\wt\varphi^m\})\})}\wt\varphi^a\ , \\ \nn 
&&\Rightarrow \quad \left[\delta_{\epsilon},\delta_{\epsilon'}\right]\,\wt\varphi^a=\delta_{\epsilon''(\epsilon,\epsilon',\{\varphi^l(\{\wt\varphi^m\})\})}\,\wt\varphi^a \ .
\eea
In the third line we have used that $\d^2 \wt\varphi^a/(\d \varphi^c\d\varphi^b)=\pm\d^2 \wt\varphi^a/(\d \varphi^b\d\varphi^c)$ to relabel the indices  in the second term, with the minus sign holding when both fields are Grassmann. In the fourth line, we have commuted $\delta_{\epsilon}\varphi^c \delta_{\epsilon'}\varphi^b$ in the second term, keeping track of the relevant sign that cancels the former $\pm$.  We conclude that the field-redefined fields still satisfy the identical algebra,  namely with the same field-dependence in the structure constant once expressed in terms of the original fields $\varphi^l$.
}, so that the transformations of the new multiplet $\{e{}_\mu^a,\psi_\mu',\lambda\}$ still satisfy the commutation relation \eqref{algebra}, with $m$ and $b$ replaced by (\ref{kill_auxiliary}), which means they offer a non-linear representation of the SUGRA algebra without any auxiliary fields. This means that, if we work with this multiplet, auxiliary fields can be omitted in the construction of non-linearly realized SUSY invariant actions. This is in contrast to linear realizations of SUSY, where the appearance of auxiliary fields in the transformation of physical fields is inevitable.  Dropping from the start the auxiliary fields will be the approach we will take in this paper.

\subsection{St\"uckelberg trick}\label{sec_stuck}

The St\"uckelberg trick consists in using the fields which non-linearly realize the spontaneously broken symmetries to construct objects that, under any symmetry, transform as under a (field dependent) {\em linearly realized} symmetry. If these St\"uckelbergized objects are then assembled into an action invariant under the linearly realized symmetries, the action will automatically be invariant under the full symmetry group -- this is the strategy we will use in Sec.~\ref{sec_action} to construct SUSY invariant actions.

In the case at hand, the St\"uckelberg trick is implemented by using the Goldstino $\lambda$, which non-linearly realizes SUSY. As given in Eq.~(\ref{eq:dresssedfields}), we define dressed fields as
\begin{subequations}\label{stuckelberg}
\begin{align}
E_\mu^a 
	&\equiv e_\mu^a + \delta_\lambda e_\mu^a+\ldots = e_\mu^a + i \left(\psi_\mu  -D_\mu \lambda\right)\sigma^a\bar \lambda + \hbox{c.c.} + (\hbox{4 fermion})\, , \\
\Psi_\mu 
	&\equiv \psi'_\mu + \delta_\lambda \psi'_\mu+\ldots = \psi'_\mu -2\mathcal D_\mu\lambda + (\hbox{3 fermion})\, ,
\end{align}
\end{subequations}
where $\delta_\lambda\equiv \lambda Q$ represents a SUGRA transformation with parameter $\lambda$.
%
%
One can verify that symmetries now act on the St\"uckelbergized objects $E_\mu^a$, $\Psi_\mu$ as field-dependent {\em linearly realized} symmetries, i.e. diffeomorphisms or local Lorentz, as expected on general grounds for non-linearly realized symmetries~\cite{Coleman:1969sm,Callan:1969sn}. For example, the action of an infinitesimal local SUSY transformation on $E_\mu^a$ is given by 
\begin{equation}\label{stuck_trans}
\begin{split}
\delta_\epsilon E_\mu^a
	&= -i\mathcal D_\mu\epsilon \sigma^a \bar\lambda + i\mathcal D_\mu \lambda\sigma^a \bar\epsilon + \hbox{c.c.} + \hbox{(3 fermion)}\cdot \epsilon\\
	&= \frac{1}{2}\left(\delta_\epsilon\delta_\lambda - \delta_\lambda\delta_\epsilon\right)e_\mu^a + \hbox{(3 fermion)}\cdot \epsilon\\
	&= \frac{1}{2}\left(\delta^{\rm diff}_{\xi} + \delta^L_{\Lambda_\xi}\right)E_\mu^a  + \hbox{(3 fermion)}\cdot \epsilon\, ,
\end{split}
\end{equation}
where we used the supergravity algebra, and where the diff.~and local Lorentz transformation parameters are given by
\begin{equation}
\xi^\mu = -2i( \epsilon\sigma^\mu\bar\lambda - \lambda\sigma^\mu\bar\epsilon )\, , \qquad
\Lambda_\xi^{ab} = \xi^\mu\omega_\mu^{ab}\, .
\end{equation}
%
%
It is also easy to see that to this order in fermion fields
\begin{equation}
\delta_\epsilon \Psi_\mu = 0 + (\hbox{2 fermion})\cdot \epsilon\, ,
\end{equation}
which is compatible with $\delta_\epsilon \Psi_\mu = \frac{1}{2}\left(\delta^{\rm diff}_{\xi} + \delta^L_{\Lambda_\xi}\right)\Psi_\mu$. Consequently, any action $S[E, \Psi]$ written in terms of the St\"uckelbergized fields that is invariant under the linearly realized symmetries -- diffeomorphisms and local Lorentz transformations -- will automatically be invariant under local supersymmetry transformations as well.

\subsection{Construction of the action}\label{sec_action}

The St\"uckelberg trick (or CCWZ construction) guarantees that any action $S[E(e,\psi',\lambda),\Psi(e,\psi',\lambda)]$ invariant under  diffs.~is automatically invariant under local supersymmetry as well. We remind the reader that the reason auxiliary fields can be ignored in this construction is that since supersymmetry is non-linearly realized, the physical fields can be rearranged in a reduced multiplet that is closed under the action of SUSY, as shown in Sec.~\ref{sec_redef}~\footnote{We further remind the reader of what we discussed in the Introduction. Since SUSY is a gauge invariance that is non-linearly realized, one could circumvent the St\"uckelberg trick entirely by working in SUSY unitary gauge where $\lambda=0$ and where the action is only invariant under the linearly realized symmetries. Both approaches are equivalent. The advantage of the St\"uckelberg trick, as we will see, is that it makes it explicit the presence of the dynamical helicity-1/2 component of the gravitino (the Goldstino), which plays a particular role in the dynamics. }.

As a further simplification, we perform an additional field redefinition, and define the fundamental vierbein to be $E_\mu^a$. This is useful because the relevance of the St\"uckelberg trick  is to extract the helicity-1/2 component out of the gravitino, which is left with only the helicity-3/2 components; it does not appear that a particular simplification incurs from extracting a Goldstino out of the vierbein. We are then left with the simple task of constructing actions $S[E,\Psi(e(E,\psi',\lambda),\psi',\lambda)]$ that are invariant under diffeomorphisms and local Lorentz transformations. This can be further simplified if one neglects 4 fermion terms in the action. Since $E_\mu^a$ and $e_\mu^a$ differ by a 2 fermion term \eqref{stuckelberg}, we have
\begin{equation}
S[E,\Psi(e(E,\psi',\lambda),\psi',\lambda)] = 
	S[E,\Psi(E,\psi',\lambda)] + \hbox{(4 fermion)}\, .
\end{equation}
For ease of presentation, in the following the ``dressed'' vierbein $E_\mu^a$ will be denoted by a lower-case $e_\mu^a$ and we will drop the prime to identify the redefined gravitino $\psi_\mu'$. 

As anticipated in the introduction, we remind the reader that, contrary to the $S$ matrix, field redefinitions do not keep correlation functions invariant. Therefore, if one were to be interested in correlation functions of the original fields, one would have to keep track of the field redefinitions. We will come back to this point in the context of inflation.

We illustrate our method by constructing the low energy effective theory of supergravity in maximally symmetric spacetimes. This can be seen as a novel way to construct the Volkov-Akulov Lagrangian~\cite{Volkov:1973ix, Komargodski:2009rz} in these spacetimes, and to couple it to a dynamical metric (see for e.g.~\cite{Bergshoeff:2016psz})~\footnote{In this paper we will stop at quadratic order in the fermionic fields, but one could carry on our procedure at higher orders.}.  In our formalism, this is simply achieved with the unitary gauge action
\begin{equation}\label{dS_action}
S_{\rm dS}[e,\psi] = S_{\bar{\rm SG}} + S_{m_{3/2}} - \int d^4 x \, e\,  \Lambda\, ,
\end{equation}
where the pure supergravity action without auxiliary fields is given by~\footnote{We define the Levi-Civita tensor as
\begin{equation*}
\varepsilon^{\mu\nu\rho\sigma} = e_a^\mu e_b^\nu e_c^\rho e_d^\sigma \epsilon^{abcd} = \epsilon^{\mu\nu\rho\sigma} /e\, ,
\end{equation*}
with $e=\det (e_\mu^a)$ and $\epsilon^{0123}=1$. For example, in this notation
\begin{equation*}
F\wedge F \propto \epsilon^{\mu\nu\rho\sigma}F_{\mu\nu}F_{\rho\sigma} d^4 x = e\,  \varepsilon^{\mu\nu\rho\sigma}F_{\mu\nu}F_{\rho\sigma} d^4 x \, .
\end{equation*}
}
\begin{equation}\label{SG_action}
S_{\bar{\rm SG}}[e, \psi]
	= \frac{1}{2}\mpl^2\int d^4 x \, e \left[   -R + \varepsilon^{\mu\nu\rho\sigma}(\bar{ \psi_\mu} \bar\sigma_\nu D_\rho  \psi_\sigma + \hbox{c.c.}) \right] \, ,
\end{equation}
and
\begin{equation}
S_{m_{3/2}} = \mpl^2\int d^4 x \, e \, m_{3/2} (\psi_\mu\sigma^{\mu\nu}\psi_\nu) + \hbox{c.c.}\, .
\end{equation}
$S_{m_{3/2}}$  can be neatly repackaged into the gravitino kinetic term by making use of the generalized covariant derivative \eqref{cov_D_dS}.
In this case we simply have
\begin{equation}
S_{\bar{\rm SG}} + S_{m_{3/2}} = \frac{1}{2}\mpl^2\int d^4 x \, e \left[   -R + \varepsilon^{\mu\nu\rho\sigma}(\bar{ \psi_\mu} \bar\sigma_\nu \mathcal D_\rho  \psi_\sigma + \hbox{c.c.}) \right] \, .
\end{equation}
The unitary gauge action \eqref{dS_action} is manifestly invariant under the linearly realized symmetries; invariance under all symmetries can be restored with the St\"uckelberg trick described in Sec.~\ref{sec_stuck}. As a result, the action
\begin{equation}\label{dS_action_stu}
S_{\rm dS}[e,\Psi(e,\psi,\lambda)]\, ,
\end{equation}
with $\Psi_\mu= \psi - 2 \mathcal D_\mu \lambda +(3\,\text{fermions}$), is invariant under local SUSY transformations in addition to diffeomorphisms and local Lorentz transformations. A nice feature of this construction is that the operators that in the standard supersymmetric formalism are generated or receive contributions by spontaneous SUSY breaking, here can be written directly in the Lagrangian. An example is a positive cosmological constant, that we can simply and directly write in~(\ref{dS_action}), while in the normal formalism it cannot be written, and instead it emerges after plugging the solutions of the equations for the auxiliary variables and matter in the action. More generally, there need not to be any relation between $m_{3/2}$ and $\Lambda$.

The action~(\ref{dS_action_stu}) can be spelled out more explicitly: after integration by parts the fermionic part of the action is quite elegantly found to be~\footnote{In obtaining this expression, it is useful to realize that the generalized covariant derivatives appearing in this expression can be integrated by parts. More explicitly, we have:
\be\nn
\int d^4x \,e\; \mpl^2 \,  \varepsilon^{\mu\nu\rho\sigma} \mathcal D_\mu\bar\lambda \bar \sigma_\nu \mathcal D_\rho\mathcal D_\sigma \lambda+ \hbox{c.c.}=- \int d^4x \,e\; \mpl^2 \,  \varepsilon^{\mu\nu\rho\sigma} \bar\lambda \bar \sigma_\nu \mathcal D_\mu \mathcal D_\rho\mathcal D_\sigma \lambda+ \hbox{c.c.}  \ .
\ee}
\begin{equation}\label{fermbil_dS}
\frac{1}{e}(
\mathcal L_{\psi\psi} + \mathcal L_{\psi\lambda} + \mathcal L_{\lambda\lambda})
	= \mpl^2 \,  \varepsilon^{\mu\nu\rho\sigma}\left(\frac{1}{2}\bar\psi_\mu \bar\sigma_\nu \mathcal D_\rho \psi_\sigma -  \bar\psi_\mu \bar\sigma_\nu [\mathcal D_\rho, \mathcal D_\sigma]\lambda
	+  \mathcal D_\mu\bar\lambda \bar\sigma_\nu [\mathcal D_\rho, \mathcal D_\sigma]\lambda\right) + {\rm c.c.}\, .
\end{equation}
Notice the appearance of the commutator
\begin{equation}\label{com_dS}
[\mathcal D_\mu, \mathcal D_\nu]\lambda = \left(\frac{1}{2}R_{\mu\nu\alpha\beta} - |m_{3/2}|^2 g_{\mu\alpha}g_{\nu\beta}\right)\sigma^{\alpha\beta}\lambda\, .
\end{equation}
that we inserted thanks to the antisymmetry of the $\varepsilon$ symbol.

In particular this commutator vanishes if 
\begin{equation}
R_{\mu\nu\alpha\beta} = |m_{3/2}|^2 (g_{\mu\alpha}g_{\nu\beta} - g_{\mu\beta}g_{\nu\alpha})\, ;
\end{equation}
The Riemann tensor takes this form in Anti-de Sitter (AdS) space with the $m_{3/2}$-related cosmological constant $\Lambda=-3\mpl^2 |m_{3/2}|^2$ -- the fact that no kinetic term for the Goldstino is introduced in the St\"uckelberg procedure if the cosmological constant takes this value (since $\left.[\mathcal D_\mu ,\mathcal D_\nu]\right|_{\rm AdS\ background}=0$ in \eqref{fermbil_dS}) reflects the well known result that~\eqref{dS_action} is already invariant under local SUSY transformations, and describes pure supergravity around AdS \cite{PhysRevD.15.2802}. In this case, the Goldstino that we obtain is, at best, strongly coupled at all energies, and, since we know that the Lagrangian in~(\ref{dS_action}) linearly realizes SUGRA in this specific AdS spacetime, every appearance of the Goldstino in this Lagrangian will be cancelled by a field redefinition of the vierbein~\footnote{Notice that since we are not introducing any auxiliary field, the linearly realized SUGRA under which our Lagrangian is invariant in this limit closes only on shell, and not off shell. Still, this is enough to guarantee that the propagating degrees of freedom have just helicity 2 and 3/2 (see~\cite{Freedman:2012zz} for details).}. For a generic cosmological constant the background solution to the action~\eqref{dS_action} is instead de Sitter or Anti de-Sitter, with
\begin{equation}
R_{\mu\nu\alpha\beta} = -\frac{\Lambda}{3\mpl^2} (g_{\mu\alpha}g_{\nu\beta} - g_{\mu\beta}g_{\nu\alpha})\, ,
\end{equation}
so that the commutator does not generically vanish, and the quadratic fermionic action is
\bea\label{spelledout_fermbil_dS}
&& \frac{1}{e}\mathcal L_{\psi\psi}
	= \mpl^2  \left(\frac{1}{2}\varepsilon^{\mu\nu\rho\sigma}\bar\psi_\mu \bar\sigma_\nu D_\rho \psi_\sigma + m_{3/2}\psi_\mu\sigma^{\mu\nu}\psi_\nu\right) + \hbox{c.c.}\, ,\\ \nn
&& \frac{1}{e}\mathcal L_{\psi\lambda}
	= -i \left(\Lambda + 3\mpl^2|m_{3/2}|^2\right) \bar\psi_\mu\bar\sigma^\mu\lambda   + \hbox{c.c.}\, , \\ \nn
&& \frac{1}{e}\mathcal L_{\lambda\lambda}
	= -i \left(\Lambda + 3\mpl^2|m_{3/2}|^2\right) \bar\lambda\cancel D\lambda    + 2 \left(\Lambda + 3\mpl^2|m_{3/2}|^2\right) m_{3/2}\lambda \lambda+ \hbox{c.c.}\, ,
\eea
with $\cancel D = \bar\sigma^\mu D_\mu$. In particular, in de Sitter space $\Lambda>0$, none of the terms above vanish and a Goldstino is required to realize supergravity in this spacetime.

  Notice that the action in (\ref{spelledout_fermbil_dS}) manifestly shows that the Goldstino and the gravitino decouple at high energies: the mixing term has the form $\bar\psi_\mu\bar\sigma^\mu\lambda$, which has one less derivative than the kinetic terms, and becomes more and more irrelevant at energies
  \be
  E\gg E_{\rm mix}\sim \left|\frac{\Lambda}{\mpl^2} + 3 |m_{3/2}|^2\right|^{1/2}\ .
  \ee 
  Though expected on general grounds, obtaining this was not obvious, as already emphasized in~\cite{ArkaniHamed:2004yi}. In fact, this was achieved only thanks to a careful redefinition the gravitino, so that its SUGRA transformation is proportional to the covariant derivative in~(\ref{cov_D_dS}) of the SUGRA parameter. This implies that in the mixing term the ordinary derivatives only appear in the form of a commutator of covariant derivatives, which has no ordinary derivative left acting on the fermions. In a sense, this construction achieves the fact that SUGRA is spontaneously broken not directly by the mass term operators, but by the commutator of the covariant derivatives, which gives rise to a lower dimension operator, proportional to the spacetime curvature or to the square of the gravitino mass (instead of the gravitino mass to the single power). Indeed, after the field redefinition, the gravitino transformation is the one such that the Lagrangian in AdS with $\Lambda=-3\mpl^2 |m_{3/2}|^2$ would be invariant. The breaking is proportional to $\Lambda/\mpl^2-m_{3/2}^2$. In this way, SUGRA can be thought of as being softly broken by an operator which is more relevant that the naive effect generated by the mass terms: an operator with the same dimensions as the one associated to the spacetime curvature. This fact makes decoupling manifest.  
  

It is interesting to make the following observation. Naively, in the decoupling limit, we are expected to obtain the renowned Volkov-Akulov Lagrangian~\cite{Volkov:1973ix}. However, this is achieved only in the limit $E^2\gg {\rm Max}(m^2_{3/2},\Lambda/\mpl^2)$ where spacetime is flat. There is an intermediate parametric regime where we still have a decoupling Lagrangian, but it is different from the Volkov-Akulov one. This is obtained when we tune $E_{\rm mix}\sim \left|\frac{\Lambda}{\mpl^2} + 3 |m_{3/2}|^2\right|^{1/2}$ to be much smaller than the curvature of the spacetime and the gravitino mass ($3|m_{3/2}|^2\simeq -\Lambda/\mpl^2$ in this limit). In this case, the Goldstino decouples from the gravitino, but  the action is the one of a massless Goldstino in rigid AdS space. 

As an application of the action \eqref{dS_action_stu} in the rigid dS limit, we will derive in App.~\ref{app:Goldstino_dispersion} the dispersion relation of the helicity-1/2 mode in de Sitter space.

 \section{Breaking time diffeomorphisms\label{FRW}}
    
In the former section we constructed the most general Lagrangian where SUGRA is softly broken while keeping diffs.~unbroken. We started by writing an action in unitary gauge that respected only the unbroken symmetries. Then a Goldstino was introduced by performing a St\"uckelberg transformation to reintroduce the non-linearly realized SUGRA gauge invariance. From this, we were able to identify a decoupling limit where the Goldstino decouples from the gravitino. Several subtleties were addressed in this construction -- for example care had to be taken to make the algebra and the field transformations independent of the auxiliary fields, and to chose a particular basis of fields which made decoupling manifest. 

We  now turn to the main case of interest for this paper, which is the one in which we decide to break time diffs.~in addition to SUGRA. This will be relevant for FRW spacetimes and to describe the theory of inflation, which we postulate to be a period of FRW expansion where time diffs.~are spontaneously broken. As most of this section will mirror the construction in the previous section, only the subtleties specific to breaking both SUGRA and time diffs.~will be explained at length here.

\subsection{Unitary gauge action}

The construction follows very similarly the logic of the former section, with the differences that were explained in the introduction. We start by writing a unitary gauge action where the only dynamical fields are the graviton and the gravitino. In this case, the action will be invariant only under time-dependent spatial diffs.. The bosonic unitary gauge action was studied in Ref.~\cite{Cheung:2007st}, to which we add the gravitino kinetic term and all the relevant ({\it i.e.} non-marginal) operators involving the gravitinos that are allowed under the linearly realized symmetries. The full unitary gauge action is  
\begin{equation}\label{S_unitary}
\begin{split}
& S_{\rm {\cal S}EFTofI}\,  
	=  S_{\rm EFTofI}[g]\, +\\ 
	& \mpl^2\int d^4 x \, e \left[ \frac{1}{2}\varepsilon^{\mu\nu\rho\sigma}\bar{ \psi}_\mu \bar\sigma_\nu D_\rho  \psi_\sigma +  m_{3/2} (\psi_\mu\sigma^{\mu\nu}\psi_\nu)+  m_0 (\psi_\mu\sigma^{\mu 0}\psi^0)+ m_\star (\psi_\mu\psi^\mu + \psi^0\psi^0)+ \hbox{c.c.} \right.\\
&\qquad \qquad \qquad \left.\vphantom{\frac{1}{2}}+ i\wt m_1 (\bar\psi^0\bar\sigma^\mu \psi_\mu-\bar\psi_\mu\bar\sigma^\mu \psi^0) + i\wt m_2 \varepsilon^{\mu\nu\lambda 0}(\bar\psi_\mu\bar\sigma_\nu\psi_\lambda)+  \delta g^{00} (m_{(3)} \psi_\mu\psi^\mu+\hbox{c.c.})+ \ldots\right]  \, ,
\end{split}
\end{equation}
where $\ldots$ denotes higher derivative terms and terms with a number of fluctuations greater or equal to three (the term in $m_{(3)}$   being an example of these).
The absence of ghosts imposes the specific form of the $m_\star$ term as well as of the $\wt m_1$ term. There is no loss of generality in taking $\wt m_1,\;ñ\wt m_2  \in \mathbb R$. The other coefficients are complex $m_{3/2},\, m_0,\, m_\star \in \mathbb C$. Notice, quite interestingly, that $\wt m_{1,2}$ preserve the $U(1)$-chiral symmetry $\psi_\mu\to e^{i\alpha}\psi_\mu$. The term in $m_{(3)}$ starts cubic in the fluctuations, and represents just one of the many that we could write, and which are included in the $\ldots$. Notice furthermore that since no linear term in the gravitino is allowed, the tadpoles of the Lagrangian are the same as in the case of the standard EFTofI.

For the moment, we will ignore the $m_\star$ term. If we focus first only on the part of the action that contributes to the quadratic Lagrangian of the gravitino, $S_{\psi\psi}^{(k)}$, the action can again be packaged into a modified kinetic term
\begin{equation}
S_{\psi\psi}^{(k)} = 
	\frac{1}{2}\mpl^2\int d^4 x \, e \, \varepsilon^{\mu\nu\rho\sigma}\bar{ \psi}_\mu \bar\sigma_\nu \mathcal D_\rho  \psi_\sigma +  \hbox{c.c.}\, ,
\end{equation}
where now
\begin{equation}\label{cov_D}
\begin{split}
\mathcal D_\mu \lambda
	&= D_\mu \lambda - \frac{i}{2} \left[\left(m_{3/2}^*+\frac{1}{2}m_0^* g^{00}\right)\sigma_\mu - m_0^*  t_\mu \sigma^0 \right]\bar\lambda - \frac{i}{2} \left[\vphantom{\frac{1}{2}}2\wt m_2t_\mu - i\wt m_1 \sigma^0 \bar\sigma_\mu\right]\lambda\, ,\\
\mathcal D_\mu \bar\lambda
	&= D_\mu \bar\lambda - \frac{i}{2} \left[\left(m_{3/2}+\frac{1}{2}m_0 g^{00}\right)\bar \sigma_\mu - m_0  t_\mu \bar \sigma^0 \right]\lambda + \frac{i}{2} \left[\vphantom{\frac{1}{2}}2\wt m_2t_\mu + i\wt m_1  \bar \sigma^0 \sigma_\mu\right]\bar \lambda\, ,
\end{split}
\end{equation}
with $t_\mu = \delta_\mu^0$. Some of the terms appearing in~(\ref{cov_D}) already appeared in Ref.~\cite{Dumitrescu:2012ha}, though in a very different context. Mirroring the maximally symmetric case, we work with the redefined field $\psi_\mu'$,
\be
\psi_\mu'= \psi_\mu - i \left[\left(m_{3/2}^*+\frac{1}{2}m_0^* g^{00}\right)\sigma_\mu - m_0^*  t_\mu \sigma^0 \right]\bar\lambda - i \left[\vphantom{\frac{1}{2}}2\wt m_2t_\mu - i\wt m_1 \sigma^0 \bar\sigma_\mu\right]\lambda\, ,
\ee
 whose transformations are still given by \eqref{delta_susy}, with now $\mathcal D_\mu$ given by the more general form \eqref{cov_D} above. We will drop the primes from now on unless when they are needed for clarity.

\subsection{St\"uckelberg trick: time diffeomorphims and local SUSY\label{sec:FRWstuck}}
Following Ref.~\cite{Cheung:2007st}, invariance under time diffeomorphisms can be restored by performing a broken transformation and promoting the transformation parameter to a field $\pi$. This amounts to replacing any upper ``0'' index in \eqref{S_unitary} as $A^0 = A^\mu t_\mu \to \hat A^0=A^\mu \d_\mu (t+\pi)$. Similarly, local supersymmetry can then be introduced by performing a SUGRA transformation and promoting the transformation parameter to a field $\lambda$, as described in the former section. Since Poincar\'e is a subgroup of SUGRA, this  procedure can be performed in two steps: first reintroduce Poincar\'e and then reintroduce SUGRA (the opposite order is not allowed as SUSY is not a subgroup of SUGRA). For example, on the field $g^{00}$, this procedure amounts to the following transformation:
\bea\label{eq:two_step_stuck}
&&g^{00} = \eta^{ab}e_a^0 e_b^0\quad \xrightarrow{\hbox{\footnotesize diff. St\"uck.}}  \quad \dress g^{00}=\eta^{ab}e_a^\mu e_b^{\nu}\d_\mu(t+\pi)\d_\nu(t+\pi) \\ \nn
&&\quad\quad\quad\quad\quad\quad\quad \xrightarrow{\hbox{\footnotesize SUSY St\"uck.}} 
  \quad \dress G^{00}= \eta^{ab}E_a^\mu E_b^{\nu}\d_\mu(t+\pi)\d_\nu(t+\pi)\, ,
\eea%
where $\hat G^{00}$ indicates the St\"uckelbergized field $g^{00}$ (not to be confused with the Einstein tensor). The similar expression for the gravitino is given in (\ref{eq:gravitino_stuck}) that we repeat here:
\bea\label{eq:gravitino_stuck_2}
&& \dress \Psi^\mu 	\equiv  \psi'{}^\mu -2\dress{\mathcal  D}^\mu\lambda + (\hbox{3 fermion})\, .
\eea
Here we neglected to St\"uckelbergize the overall ${}^\mu$ index, whose St\"uckelbergization we keep explicit for convenience. 
Notice the appearance of $\psi'$ and of the particular covariant derivative. 
The undressed $\mathcal D_\mu$'s are given by \eqref{cov_D} and the dressed $\dress{\mathcal D}_\mu$'s are obtained by the standard procedure, but neglecting to St\"uckelbergize the overall ${}_\mu$ index, whose St\"uckelbergization we keep explicit for convenience. 
 More explicitly, we have
\begin{equation}\label{hatcov_D}
\begin{split}
\dress {\mathcal D}_\mu \lambda
	&= D_\mu \lambda - \frac{i}{2} \left[\left(m_{3/2}^*+\frac{1}{2}m_0^* \dress g^{00}\right)\sigma_\mu - m_0^*  \dress \delta^0_\mu \dress \sigma^0 \right]\bar\lambda - \frac{i}{2} \left[\vphantom{\frac{1}{2}}2\wt m_2\dress \delta^0_\mu - i\wt m_1 \dress \sigma^0 \bar\sigma_\mu\right]\lambda\, ,\\
\end{split}
\end{equation}
where $\dress \delta_\mu^0=\d_\mu (t+\pi)$.
 As we will comment later, the appearance of the covariant derivative will be essential to obtain manifest decoupling of the Goldstino from the gravitino. In summary, we work with a SUGRA multiplet that contains the metric, the gravitino, the Goldstone of time diffs.~and the Goldstino.

We observe that the SUSY St\"uckelberg transformation in (\ref{eq:two_step_stuck}) did not affect the Goldstone field~$\pi$. In other words, we assumed that we could assign a transformation law to $\pi$ under SUGRA such that $\pi$ transform as under a time-diff., without needing to field-redefine it with the Goldstino. This is indeed expected to be possible from what we discussed in footnote~\ref{footnote:CCWZconstruction} where we generalized the CWZ, construction. In fact, we explained there that if we have matter fields that transform under $H$ in a non-linear representation $D_H$, they inherit automatically a non-linear representation of $G$ where they transform under $g\in G$ through the homomorphism $h(\lambda,a,g)$ that goes from $G$ to $H$. In our case, at the step of introducing SUGRA,  time-diffs.~are part of $H$, and therefore it is possible to assign to $\pi$ a transformation law under SUGRA such it transforms as under a time diff.. In particular, the explicit parameter of the time-diffs.~induced by SUGRA  can be read off from the transformation of the other St\"uckelbergized fields, see for e.g.~\eqref{stuck_trans}. We have that the parameter of the diff.~is $\xi^\mu = -2i(\epsilon \sigma^\mu \bar\lambda - \lambda\sigma^\mu\bar\epsilon)$ (the parameter of the Lorentz transformation is irrelevant because $\pi$ is a Lorentz scalar). Therefore, we have the following transformation property for $\pi$:~\footnote{It is straightforward to show that this transformation satisfies the algebra (\ref{algebra}):
\begin{equation*}
\begin{split}
[\delta_{\epsilon'},\delta_\epsilon] \pi
	&= -2i(\epsilon' \sigma^0 \bar\epsilon - \epsilon\sigma^0\bar\epsilon') - 2i ( \epsilon' \sigma^\mu\bar\epsilon- \epsilon\sigma^\mu \bar\epsilon')\d_\mu\pi  + (\hbox{4 fermion})\\
	&= \delta^{\rm diff}_{y^\mu}\pi   + (\hbox{4 fermion}) = (\delta^{\rm diff}_{y^\mu}\pi+\delta_\Lambda^L + \delta_{\hat\epsilon})\pi   + (\hbox{4 fermion})\, ,
\end{split}
\end{equation*}
with parameters defined in equation (\ref{algebra2}).
}
\be\label{pi_transf}
\delta_\epsilon \pi
   =  \frac{1}{2}(\delta^{\rm diff}_{\xi} + \delta^L_{\Lambda_\xi}) \pi + \ldots
   = \frac{1}{2}(\xi^0+\xi^\mu\d_\mu\pi) + \ldots
   =-i(\epsilon \sigma^0 \bar\lambda - \lambda\sigma^0\bar\epsilon) - i ( \epsilon \sigma^\mu\bar\lambda- \lambda\sigma^\mu \bar\epsilon)\d_\mu\pi + \ldots  \, ,
\ee
%
where $\ldots$ denote four fermion terms. Notice furthermore that the $\pi$ transformation must be proportional to $\lambda$, because one cannot break time-diffs.~without at the same time breaking SUSY.

As in the maximally symmetric case we will define $E_\mu^a$ to be the physical vierbein, and denote it by $e_\mu^a$ in the following. After both St\"uckelberg steps, the action
\begin{equation}\label{S_stuck}
S_{\rm {\cal S}EFTofI} = S_{\rm EFTofI}[g,\pi] + \frac{1}{2}\mpl^2\int d^4 x \, e \, {\varepsilon}^{\mu\nu\rho\sigma}\bar{\dress \Psi}_\mu \bar{\sigma}_\nu \dress{{\mathcal D}}_\rho {\dress \Psi}_\sigma +m_{(3)} \hat{\delta G^{00}} \hat\Psi_\mu\hat\Psi^\mu+  \hbox{c.c.} + \ldots\ ,
\end{equation}
is fully invariant under diffeomorphisms, local Lorentz transformations and local supersymmetry. Note that $S_{\rm EFTofI}[g,\pi]$ is purely bosonic, and all interactions between $\pi$ and $\psi_\mu$ or $\lambda$ are contained in the terms involving the gravitino in \eqref{S_stuck}. Action (\ref{S_stuck}) represents the most general low energy action when SUSY and time-diffs.~are spontaneously broken.

\subsection{Decoupling limit\label{sec:dec}}
As we have already described, the St\"uckelberg trick is particularly useful because it extracts from the gravitino the spin-1/2 degree of freedom, which we will show decouples at high energies. To show that this is the case, we focus for the moment on the kinetic and mass terms for the gravitinos $S_{\psi\psi}^{(k)}$, which are the relevant terms. Diff.~invariance of these terms induces couplings to gravity and $\pi$ that we also discuss. The part of the kinetic action of the gravitinos which is quadratic in the fermions has the same form as eq.~\eqref{fermbil_dS}: 
\begin{equation}\label{fermbil}
\frac{1}{e}
(\mathcal L_{\psi\psi}^{(k)} + \mathcal L_{\psi\lambda}^{(k)} + \mathcal L_{\lambda\lambda}^{(k)} )
	= \mpl^2 \,  {\varepsilon}^{\mu\nu\rho\sigma}\left(\frac{1}{2}\bar{ \psi}_\mu \bar{ \sigma}_\nu \dress{\mathcal D}_\rho { \psi}_\sigma -  \bar{ \psi}_\mu \bar{ \sigma}_\nu [\dress{\mathcal D}_\rho, \dress{\mathcal D}_\sigma]\lambda
	+  \dress{\mathcal D}_\mu\bar\lambda \bar{ \sigma}_\nu [\dress{\mathcal D}_\rho, \dress{\mathcal D}_\sigma]\lambda\right) + {\rm c.c.}\, ,
\end{equation}
%
%
 The commutator of derivatives is
\begin{equation}\label{com_gen}
\begin{split}
&[\dress{\mathcal D}_\mu,\dress{\mathcal D}_\nu]\lambda=\\
&\left\{\frac{1}{2}R_{\mu\nu\alpha\beta} - \left[|\dress M|^2-\wt m_1^2\dress{g}^{00}\right]g_{\mu\alpha}g_{\nu\beta}  + 2\left[\wt m_1^2 + \dot{\wt m_1} - \Re(\dress M m_0^*)\right] \dress{\delta}_{[\mu}^0 g_{\nu]\alpha}\dress{\delta}_\beta^0 - 2\wt m_1 \,\left[{\nabla_{[\mu} \dress\delta_{\alpha}^0}\right] g_{\nu]\beta} \right\}\sigma^{\alpha\beta}\lambda\\
&+ \left\{\vphantom{\frac{}{}}-im_0^* \dress{\delta}_{[\mu}^0 \left[\nabla_{\nu]}\dress \delta_\alpha^0\right]   - \left[2\dress M^* \wt m_2+i\wt m_1(\dress M^* -m_0^* \dress{g}^{00}) \right] \dress{\delta}_{[\mu}^0 g_{\nu]\alpha} - i \d_{[\mu}\hat M^*g_{\nu]\alpha}  \right\}\sigma^\alpha\bar\lambda
\end{split}
\end{equation}
with $\dress M= m_{3/2}+\frac{1}{2}m_0 \dress g^{00}$. We kept $\hat g$ instead $\hat G$ in (\ref{com_gen}) because at the order at which we work, where we neglect four-fermion terms, the difference is negligible.

In the case of maximally symmetric spacetimes, we found that in the limit in which the space time was AdS with $\Lambda=-3\mpl^2 |m_{3/2}|^2$, SUGRA was linearly realized. This revealed itself in our formalism by the fact that no healthy Goldstino was introduced by the St\"uckelberg procedure. In turn, this originated from the fact that $\left.[{\mathcal D}_\mu,{\mathcal D}_\nu]\right|_{\rm background}=0$. We are now going to ask the same question for FRW spacetimes where time diffs.~are broken. This amounts to finding metrics for which the commutator \eqref{com_gen}, with $\pi=0$ and $\delta g_{\mu\nu}=0$, vanishes. We are not able to find an FRW solution for the most generic case, but we find a few interesting examples when restricting ourselves to certain special parameters. If we set $m_{3/2}= c\, e^{i \widetilde m_2 t}$, $\widetilde m_2={\rm const}$, and all other parameters equal to zero, the resulting spacetime is the usual AdS space with $\Lambda=-3\mpl^2 |m_{3/2}|^2$. For $m_0\neq 0$ or $\, \wt m_2\neq 0$ and constant, and all other parameters equal zero, we find Minkowski space. Finally, if we set $\wt m_1\neq 0 ,\wt m_2\neq 0$ and constant, and all other parameters equal to zero, we find a closed FRW with vanishing Hubble rate. The condition that $\left.[{\mathcal D}_\mu,{\mathcal D}_\nu]\right|_{\rm background}=0$ implies that on these spacetimes there exists, barring topological obstructions, a killing Weyl spinor $\chi$ (${\mathcal D}_\mu\chi=0$).  In turn, this guarantees that in the rigid limit one can define a SUSY invariant theory living on this manifold~\cite{Dumitrescu:2012ha}. The case of $m_{3/2}= c\, e^{i \widetilde m_2 t}$, $\widetilde m_2={\rm const}$ in AdS with $\Lambda=-3\mpl^2 |m_{3/2}|^2$ has a stronger property. In fact, this theory still possesses linearly realized gauge invariance. This can be seen by performing the field redefinition $\psi\to  e^{i\widetilde m_2 t}\psi$. In this case, the Lagrangian reduces to the one of standard SUGRA in AdS. Therefore it is SUGRA invariant without the introduction of a Goldstino and a Goldstone~\footnote{One might wonder if on the other manifolds with non-vanishing killing spinors one can define a full SUGRA invariant theory. In order to do so, one needs to be able to add matter to support the relevant manifold as a solution, while at the same time preserving time diffs.~and boson-fermion degeneracy, thus preserving SUGRA. For the case of Minkowski, in principle the Lagrangian could just break boosts and not time translations. In this case, one would just need to introduce the three Goldstone bosons of boost breaking (see for example~\cite{Nicolis:2015sra, Delacretaz:2015edn}). The counting of on-shell degrees of freedom tells us that additional massless  matter needs to be added to make this theory SUGRA invariant. It seems hard to achieve this. Similar considerations apply to the case of $S^3\times R$. We leave further explorations on this topic to future work.}. Notice that the theory is not strictly equivalent to AdS SUGRA, because the original gravitino has a gap, giving an effect similar to a chemical potential (see related discussion in~\cite{Kahn:2015mla}).

Around generic geometries, the commutator \eqref{com_gen}, with $\pi=0$, does not vanish, and the theory contains a Goldstino. The Goldstino-gravitino mixing is given by
\begin{equation}\label{psi_lambda3}
\begin{split}
&\frac{1}{\mpl^2 e}\mathcal L_{\psi \lambda}^{(k)}= \\	
	& -i \left\{
	R_{\mu\nu} - \frac{1}{2}R g_{\mu\nu} 
	- 2 \left[\wt m_1^2 + \dot{\wt m_1} - \Re(\dress M m_0^*)\right]\d_\mu(t+\pi)\d_\nu(t+\pi)- 2\wt m_1 \,\nabla_\mu\d_\nu(t+\pi)\right.\\ 
	&\quad\left.\vphantom{\frac{1}{2}}+ \left[3|\dress M|^2 + g^{\rho\sigma}\d_\rho(t+\pi)\d_\sigma(t+\pi)\left(-\wt m_1^2 + 2\dot{\wt m_1} - 2\Re(\dress M m_0^*)\right) + 2\wt m_1\, \nabla^\rho \d_\rho(t+\pi)\right]g_{\mu\nu} \right\} \bar\psi^\mu \bar\sigma^\nu \lambda \\
	&\quad + \left\{4 \left(-2i\dress M^*\wt m_2 + \wt m_1(\dress M^*-m_0^* g^{\rho\sigma}\d_\rho(t+\pi)\d_\sigma(t+\pi)) + \frac{1}{2}m_0^* \,\nabla^\rho \d_\rho(t+\pi)\right)g_{\mu\alpha}\d_\beta(t+\pi) \right. \\ 
	&\qquad\quad\left. \vphantom{\frac{1}{2}}+4g_{\mu\alpha} \d_\beta \hat M^*-2m_0^* \left[(\nabla_\mu \d_\alpha(t+\pi))\d_\beta(t+\pi) + g_{\mu\alpha} \d^\gamma(t+\pi) \nabla_\beta \d_\gamma(t+\pi)\right]\right\}\bar\psi^\mu \bar\sigma^{\alpha\beta}\bar\lambda  \\ 
	& \quad + {\rm c.c.} + \hbox{4 fermion}\, ,
\end{split}
\end{equation}
where $\nabla_\mu$ is the usual covariant derivative, and $D_\mu$ was defined in footnote~\ref{footnote:Ddef}. Here and in the following equations~(\ref{kinkin3}),~(\ref{eq:mass}), and~(\ref{eq:mass2}), every mass coefficient should be evaluated at a time equal to $t+\pi$.
 The Goldstino kinetic term is similarly given by
\begin{equation}\label{kinkin3}
\begin{split}
&\frac{1}{\mpl^2 e}\mathcal L_{\lambda\lambda}^{(k),{\rm kin}} 
	= \frac{1}{e}\mathcal L_{\psi\lambda}[\lambda, \psi_\mu\to -D_\mu\lambda ]\\	
	& = i \left\{
	R_{\mu\nu} - \frac{1}{2}R g_{\mu\nu} 
	- 2 \left[\wt m_1^2 + \dot{\wt m_1} - \Re(\dress M m_0^*)\right]\d_\mu(t+\pi)\d_\nu(t+\pi)- 2\wt m_1 \,\nabla_\mu\d_\nu(t+\pi)\right.\\ 
	&\ \left.\vphantom{\frac{1}{2}}+ \left[3|\dress M|^2 + g^{\rho\sigma}\d_\rho(t+\pi)\d_\sigma(t+\pi)\left(-\wt m_1^2 + 2\dot{\wt m_1} - 2\Re(\dress M m_0^*)\right) + 2\wt m_1\, \nabla^\rho \d_\rho(t+\pi)\right]g_{\mu\nu} \right\} D^\mu\bar\lambda \bar\sigma^\nu \lambda \\
	&\  - \left\{4 \left(-2i\dress M^*\wt m_2 + \wt m_1(\dress M^*-m_0^* g^{\rho\sigma}\d_\rho(t+\pi)\d_\sigma(t+\pi)) + \frac{1}{2}m_0^* \,\nabla^\rho \d_\rho(t+\pi)\right)g_{\mu\alpha}\d_\beta(t+\pi) \right. \\ 
	&\qquad\quad\left. \vphantom{\frac{1}{2}}+4g_{\mu\alpha} \d_\beta \hat M^*-2m_0^* \left[(\nabla_\mu \d_\alpha(t+\pi))\d_\beta(t+\pi) + g_{\mu\alpha} \d^\gamma(t+\pi) \nabla_\beta \d_\gamma(t+\pi)\right]\right\}D^\mu\bar\lambda \bar\sigma^{\alpha\beta}\bar\lambda  \\ 
	& \quad + {\rm c.c.} + \hbox{4 fermion}\, ,
\end{split}
\end{equation}
where in (\ref{psi_lambda3}) and (\ref{kinkin3}), $\dress M= m_{3/2}+\frac{1}{2}m_0 g^{\mu\nu }\d_\mu(t+\pi)\d_\nu(t+\pi)$.

Finally, the Goldstino mass term is easily, though lengthly, obtainable from  (\ref{psi_lambda3}) with the following replacement
\bea\label{eq:mass}
&& \!\!\!\!\!\!\!\!\! \mathcal L_{\lambda\lambda}^{(k),{\rm mass}} = \mathcal L_{\psi\lambda}^{(k)}[\lambda, \psi_\mu\to -(\dress{\mathcal D}_\mu - D_\mu)\lambda ] \\ \nn
&&\!\!\!\!\!\!\!\!\!  =\mathcal L_{\psi\lambda}^{(k)}\left[\lambda, \psi_\mu\to  - \frac{i}{2} \left[\left(m_{3/2}^{{ *}}+\frac{1}{2}m_0^{{ *}} \hat g^{00}\right)\sigma_\mu - m_0^{{ *}}  \hat \delta^0_\mu (\hat \delta^0_\nu\sigma^\nu) \right]\bar\lambda - \frac{i}{2} \left[\vphantom{\frac{1}{2}}{{ 2}}\wt m_2\dress\delta^0_\mu - i\wt m_1 (\hat \delta^0_\nu \sigma^\nu) \bar\sigma_\mu\right]\lambda \right]\ .
\eea%
This term contains additional couplings between $\lambda$ and $\pi$, schematically of the form $\lambda\lambda\d\pi$ and $\lambda\lambda\d\pi\d\pi$. For example, in the case where only $\wt m_2\neq 0$, the mass term is given by
\begin{equation}\label{eq:mass2}
\frac{1}{\mpl^2e}\mathcal L_{\lambda\lambda}^{(k),{\rm mass}}
	= - { 2}\wt m_2\left(R^{\mu\nu} - \frac{1}{2}R g^{\mu\nu}\right) (\bar\lambda \bar\sigma_\mu \lambda)\d_\nu(t+\pi)\, .
\end{equation}
Expressions~(\ref{psi_lambda3}),~(\ref{kinkin3}) and~(\ref{eq:mass}) are exact up to quadratic orders in the fermionic fields, and to all orders in bosonic perturbations, i.e.~the metric was not assumed to be FRW. On the other hand, in the FRW metric, these expressions drastically simplify, as shown in App.~\ref{app:Goldstino_prop}.

Since the mixing terms between $\lambda$ and $\psi_\mu$ do not have any derivative acting on the fermions, decoupling is manifest at sufficiently high energies. Again, as in the case of maximally symmetric spacetimes, this was achieved thanks to a careful redefinition of the fields so that the St\"uckelberg transformation contained subleading mass terms with the covariant derivative ${\cal D}$~(\footnote{The fact that this covariant derivative exists can be understood in the following way. In the maximally symmetric case, the covaraint derivative was chosen so that it was equal to the transformation of the gravitino in the linearly-realized SUSY in AdS. This made evident that it was the curvature, and not the mass, to break SUSY.  In the FRW case, we discussed just above that, if we turn on one parameter at the time, we can find manifolds for which $\left.[{\mathcal D}_\mu,{\mathcal D}_\nu]\right|_{\rm background}=0$. This makes it clear that, the breaking of SUSY in the quadratic Lagrangian will be either proportional to the product of two masses, which has dimension two instead of one as it would be for the mass, or the curvature of the spacetime, which has dimension two. Effectively this makes manifest that the breaking of SUSY is due to a operator as relavent as the curvature.}).  

Even though the expressions are a bit lengthy, estimates are readily simplified if we realize the following. For the purpose of inflation, we are usually interested in energy scales of order Hubble. Exceptions to this case are present when the approximate continuous shift symmetry of $\pi$ is softly broken to a discrete one~\cite{Behbahani:2011it,Flauger:2016idt}. Apart for these exceptions, that we do not study here and that can be studied in a similar way in future work, if the mass parameters that we introduce in~(\ref{S_unitary}) are much larger than $H$, then the gravitino can be integrated out and we are left with the standard EFT of Inflation~\footnote{This scale might have to be modified in the case the Goldstino has a speed of sound much smaller than one. As we will discuss next, this is not expected to be the case unless there are unexpected cancellations among the different parameters of the EFT.}. Therefore, we can limit ourself to study the case where all the parameters 
\be\label{eq:parametersleqH}
m_{3/2}\sim m_0\sim \tilde m_1\sim \tilde m_2\sim m_\star \quad\lesssim \quad H \ .
\ee
This is a technically natural choice, as we discuss later.
In this regime, we have that the canonical gravitino and Goldstino fields are
\be\label{eq:canonical_leo}
\psi^\mu = \frac{\psi^\mu_c}{\mpl}\ , \qquad \lambda\sim \frac{\lambda_c}{H\mpl} \ ,
\ee
so that, schematically
\be
{\cal L}_{\psi\lambda}^{(k)}\sim \mpl^2 H^2 \psi^\mu \sigma_\mu\lambda \sim H \psi_{c}^{\mu}\sigma_\mu \lambda_c \ .
\ee
We find that, barring cancellations that we do not find to be justified by any symmetry reason if we are close to an inflationary background, decoupling between the Goldstino and the gravitino will occur above energies of order $E_{\rm mix}\sim H$. In particular, this tells us that for inflationary calculations where $\pi$ is protected by an approximate continuous shift symmetry, calculations of inflationary correlation functions cannot be reliably performed by considering only the Goldstino and not the gravitino. This is to be contrasted with the purely bosonic inflationary case, where the mixing scale between $\pi$ and the graviton is often much smaller than $H$, and so inflationary correlation functions can be computed in the decoupling limit.

It is interesting at this point to study the speed of propagation $c_\lambda$ of Goldstino fluctuations. Superluminal propagation would imply that the effective theory cannot be UV completed by a Lorentz-invariant local unitary theory~\cite{Adams:2006sv}. We study the properties of the Goldstino propagation in App.~\ref{app:Goldstino_prop}, and the requirement of subluminal propagation on the parameters introduced in~(\ref{S_unitary}). We find that subluminal propagation requires 
\be
c_\lambda\leq 1 \quad\Leftrightarrow \quad \beta^2 \geq \alpha^2 + |\gamma|^2\,
\ee
where $\beta,\alpha$ and $\gamma$ are combinations of the parameters in~(\ref{S_unitary}) that are defined in App.~\ref{app:Goldstino_prop}, eq.~(\ref{abc_coeffs}). We also find that, even if we move away from the parameter region identified by~(\ref{eq:parametersleqH}), the speed of sound of the Goldstino is order one unless there are unexpected cancellations among the different parameters of the EFT, justifying the choice made in (\ref{eq:parametersleqH}). Our finding is in contrast with former literature of the subject, which argued that $c_\lambda- 1\sim{\cal O}(\epsilon)$, while we argue that the technically-natural parameter range includes $c_\lambda-1\sim {\cal O}(1)$. Therefore, we focus on the regime $c_\lambda-1\sim {\cal O}(1)$ for the rest of the paper.

\subsubsection{$\delta K \delta g^{00}$ and subtle decoupling in the bosonic case\label{sec:bosonic_stuck}}
In order to obtain manifest decoupling, we had to perform a careful field redefinition of the gravitino. Such subtleties with decoupling are of course not unique to SUGRA. Similar situations, where the subleading terms in the St\"uckelberg transformation are important, also happen in the bosonic EFTofI. For example, let us consider the following term in the unitary gauge Lagrangian $S_{\rm EFTofI}[g]$ in \eqref{S_unitary}:
\be
\int\sqrt{-g}\bar M^3 \delta K \delta g^{00}\ .
\ee
Since the breaking of time diffs.~is due to a dimension three operator, while the standard terms such as $(\delta g^{00})^2$ are dimension two, we expect that manifest  decoupling will not happen after the usual replacement $g^{\mu\nu}\to \hat{g}^{\mu\nu}$ given in (\ref{eq:stu2}). Indeed, with the transformation in~(\ref{eq:stu2}), one finds terms of the form
\be\label{eq:mixing_K}
\sim \bar M^3\left(H (\d_i\pi)^2-\d_i\delta g^{00}\d^i\pi-2\d_i\delta g^{0i}\dot{\pi}-\delta \dot{g}^i_{\;i}\dot{\pi}\right)\ .
\ee
The mixing term above has the same number of derivatives as the kinetic terms for $\pi$ and $\delta g_{\mu\nu}$, which shows that decoupling will not happen manifestly. Indeed, the equations for the constrained variables will read $\mpl^2\d^2\delta g\sim \bar M^3\d^2\pi\ \ \Rightarrow\ \ \delta g\sim \frac{\bar M^3}{\mpl^2} \pi$. Upon plugging back this solution into~(\ref{eq:mixing_K}), we see that the mixing term contributes to a term in $(\d\pi)^2$ which has the same number of derivatives as the original kinetic term: decoupling has not manifestly happened. 

All of these harming mixing terms can be removed by means of a field redefinition for $g_{\mu\nu}$ that involves $\pi$ with no derivatives acting on it. If we redefine
\be
g'^{\mu\nu}={g}^{\mu\nu}+4\frac{\bar M^3}{\mpl^2} g^{0\mu}g^{0\nu}\pi+2\frac{\bar M^3}{\mpl^2} g^{\mu\nu}\pi\ ,
\ee
then the transformation of $g'^{\mu\nu}$ under time diffs.~gets modified accordingly, so that the St\"uckelberg transformation for it reads:
\be\label{eq:new_stuck_boson}
\hat{g}'^{\mu\nu}=\d_\alpha(x^\mu+\pi^\mu)\d_\beta(x^\nu+\pi^\nu) g'^{\alpha\beta}-4\frac{\bar M^3}{\mpl^2} g'^{0\mu}g'^{0\nu}\pi-2\frac{\bar M^3}{\mpl^2} g'^{\mu\nu}\pi+\ldots\ ,
\ee
where we neglected quadratic terms in $\pi$.
The action written directly in terms of $ \hat{g}'$ rather than $\hat g$ has manifest decoupling in terms of the fields $\delta g'$ and $\pi$. 
This is completely analogous to what we saw in Secs.~\ref{sec:max_sym} and~\ref{FRW} where in the presence of the $m$ terms it was important to perform suitable field redefinitions on $\psi$ and $g$ so that the action after a St\"uckelberg step that kept track of the  terms proportional to the $m$'s inside $\dress{\mathcal D}$ had manifest decoupling~\footnote{The only non-analogous aspect of this construction is the notation: in the main text, we dropped the prime`` ' " in $\hat g$ in~(\ref{eq:new_stuck_boson}) because in the case of SUSY St\"uckelberg there was no risk of confusion as we used capitalized letters for the St\"uckelbergized fields.}.


\subsection{$m_\star\neq 0$ and a very non-relativistic dispersion relation}
The remaining kinetic unitary gauge term that has not yet been discussed is the one proportional to $m_\star$ in (\ref{S_unitary})
\begin{equation}
S_{\mathcal S \rm EFTofI} \supset
	\mpl^2 \int d^4 x \, e \, m_\star (\psi_\mu\psi^\mu + \psi^0\psi^0) + \hbox{c.c.}\, .
\end{equation}
This particular linear combination of $\psi_\mu\psi^\mu$ and $\psi^0\psi^0$ is required to avoid a Goldstino kinetic term of the form $\dot\lambda\dot\lambda$ which would propagate a ghost.  As mentioned earlier, this term cannot be repackaged into the gravitino kinetic term with the help of a modified covariant derivative~$\mathcal D$. Consequently, decoupling in this theory is not automatic by construction, so we check it explicitly here. The St\"uckelberg transformation leads to
\begin{equation}
S_{\mathcal S \rm EFTofI} \supset
	\mpl^2 \int d^4 x \, e \, m_\star \dress\Psi^\mu\dress\Psi^\nu(g_{\mu\nu} +\dress\delta_\mu^0\dress\delta_\nu^0) + \hbox{c.c.}\, .
\end{equation}
For the purposes of showing decoupling, the action above can be simplified. First, we will focus on the action quadratic in fermions (i.e.~set the bosonic perturbations $\pi,\delta g$ to zero), so that $\dress \Psi_\mu \to \Psi_\mu$ and $\dress \delta_\mu^0 \to \delta_\mu^0$. Second, since the FRW background alone implies that the Goldstino and gravitino are mixed up to $E_{\rm mix} \sim H$ and $m_\star$ can at most increase the mixing energy, we will assume $E_{\rm mix}\gtrsim H$ from the start, which allows us to neglect the mass terms in the covariant derivative
\begin{equation}
\Psi_\mu = \psi_\mu - 2\mathcal D_\mu \lambda \sim \psi_\mu + \d_\mu \lambda\, .
\end{equation}
We are thus lead to the following form of quadratic action
\begin{equation}\label{mstar_stuck}
\frac{1}{e\mpl^2}\mathcal L^{\rm 2f} \sim 
	\psi\d\psi + H^2 \lambda\d\lambda 
	+ m_\star \left[\d_i\lambda \d^i\lambda + \psi^i\d_i\lambda\right]\, .
\end{equation}
The term in $H^2\lambda\d\lambda$ comes from the gravitino kinetic term. Since $m_\star$ alone does not generate any kinetic term $\lambda \dot \lambda$, it is important to include the curvature contribution above in order to have a healthy theory. The Goldstino dispersion has therefore the very non-relativistic form 
\begin{equation}
\omega\sim k + \frac{m_\star}{H^2}k^2 \, .
\end{equation}
Notice that despite the kinetic Goldstino-gravitino mixing $\psi^i\d_i\lambda$ in \eqref{mstar_stuck}, the Goldstino decouples at high energies.  Indeed, in the gauge $\bar\sigma^i\psi_i=0$ and for wavevectors $k\gg
H$, the gravitino equation of motion leads to
\begin{equation}
\d^i \psi_i \sim \sigma^{0i}\d_i \psi_0 \sim m_\star
\sigma^i\d_i\bar\lambda \qquad \Rightarrow  \qquad
\psi^i\sim m_\star\sigma^i \bar \lambda\   .
\end{equation}
So the mixing term is negligible at high wavevectors $k\gg m_\star$:
\begin{equation}
m_\star \psi^i\d_i \lambda \sim m_\star^2 \bar\lambda\bar\sigma^i\d_i\lambda \ll m_\star (\d_i\lambda)^2\, .
\end{equation}
The mixing energy in this case is
\begin{equation}
E_{\rm mix} \sim k_{\rm mix} + \frac{m_\star}{H^2}k_{\rm mix}^2 \sim m_\star \left(1+ \frac{m_\star^2}{H^2}\right)\, .
\end{equation}
For $m_\star\lesssim H$, which is the regime of interest, mixing due to the background curvature is comparable or more important, as it gives $E_{\rm mix}\sim H$. 

Notice finally that as we move to very high wavenumbers, the speed of propagation of Goldstino waves will become superluminal, forbidding the UV completion of this Lagrangian with a local, Lorentz invariant one. We will come back on this important point in Sec.~\ref{sec:mstarpheno}.


\subsection{Multifield inflation: including additional degrees of freedom}

So far, we have provided a formalism that allows us to construct the most general Lagrangian for the fluctuations in FRW spacetime where SUGRA is spontaneously broken and where the only relevant degrees of freedom are the graviton $e_\mu^a$, the gravitino $\psi_\mu$, the Goldstone $\pi$ and the Goldstino $\lambda$. In general, however, the theory could contain additional light degrees of freedom, whose inclusion is {\it not} required by the non-linear realization of SUGRA. Our formalism makes the inclusion of these additional degrees of freedom quite straightforward. We follow the same logic as used for constructing the EFT of Multifield Inflation~\cite{Senatore:2010wk} and our supersymmetric extension of the single field case. We start by writing an action in a gauge that is unitary both for the SUSY transformations and for time diffs.. On top of the graviton $e_\mu^a$ and the gravitino $\psi_\mu$, we now have additional light degrees of freedom. We therefore write the most general Lagrangian for the fluctuations with these fields, invariant just under time-dependent spatial diffs., softly breaking SUGRA and time diffs.. The result is a Lagrangian that is the sum of the single field supersymmetric EFTofI which we have developed so far, $S_{\rm {\cal S}EFTofI}$, of the non-SUSY multifield EFT of Inflation (removing of course the part that is common to the two), that we call $S_{\rm \widetilde{multiEFTofI}}$, the tilde signifying the removal of the single field part, and of a new part. The new part consists of terms that couple the additional fields to the gravitino. If for simplicity we focus on one single scalar additional fields, $\sigma$,  endowed by a shift symmetry, the leading terms will be bilinear in the gravitino field, taking schematically the form
\begin{equation}\label{S_unitary_multi}
\begin{split}
& S_{\rm {\cal S}multiEFTofI}\,  
	=  S_{\rm {\cal S}EFTofI}\, + S_{\rm \widetilde{multiEFTofI}}\\ 
	&+ \mpl^2\int d^4 x \, e \; \left[  c_{3/2,1} \dot\sigma (\psi_\mu\sigma^{\mu\nu}\psi_\nu)+  \left(c_{0,1} \dot\sigma (\psi_\mu\sigma^{\mu 0}\psi^0)+ \d_\rho\sigma \left(c_{0,2}\psi_\mu\sigma^{\mu \rho}\psi^0+c_{0,3}  \psi_\mu\sigma^{\mu 0}\psi^\rho\right)\right)\right.\\
	&\left.\qquad+\left( c_{\star,1}  \dot\sigma\psi_\mu\psi^\mu+c_{\star,2}\d_\mu\sigma \psi^\mu\psi^0+c_{\star,3}\dot\sigma  \psi^0\psi^0 \right)+\ldots+ \hbox{c.c.} \right]\ .
\end{split}
\end{equation}
In (\ref{S_unitary_multi}), we have not been careful in writing all the leading terms, as the systematic construction, including also additional fields, is at this point very simple and will be done elsewhere. 

It is now easy to reintroduce non-linearly realized SUGRA with exactly the same procedure as we performed in Sec.~\ref{sec:FRWstuck} for the single field case. In particular, the Goldstino is reintroduced by performing the replacement in (\ref{eq:gravitino_stuck})
\be
\psi^\mu\to  \dress \Psi^\mu= \psi^\mu -2\dress{\mathcal  D}^\mu\lambda + (\hbox{3 fermion})\,  ,
\ee 
For the matter field $\sigma$, we have
\be
\sigma\quad \to\quad\hat \Sigma \ .
\ee
Similarly to what we did for the vierbein field, we can chose the fundamental field $\sigma'\equiv \hat \Sigma$ and work directly with them. This implies that no knowledge of the transformations of the matter fields under SUSY is required for this construction.

While we do not give the full details of the action, as it is lengthy but straightforward, it is important to highlight a few important lessons. First, every term that we could have written in the non-SUSY EFT of multifield inflation is still allowed. This is reminiscent of what we found in the single field case: all terms that were allowed in the non-SUSY single field case were still allowed by the SUSY case. Second, due to the presence of a new particle, the gravitino, SUGRA indirectly allows for new operators coupling the light fields in multifield inflation and the gravitino. Third, it is only these operators that introduce couplings between the Goldstino and the  light fields in multifield inflation. 

We add a fourth important comment. Though so far we have implicitly assumed that the spacetime algrebra is SUGRA, as described in~(\ref{algebra}), this procedure of coupling additional matter to our \SEFTofI~applies also to the case in which the additional matter is required by an enlargement of the spacetime symmetry group we consider. For example, we could add to SUGRA an $R$-symmetry, which is also spontaneously broken. This is very similar to the enlargement of the spacetime symmetry group we had when we passed from ordinary diffs. to SUGRA. From the point of view of the EFTofI, adding a spontaneously broken $R$-symmetry amounts to including in the EFT the additional Gauge bosons of this gauge invariance. Let us call them $V_\mu$. To write the EFT Lagrangian, we can go to a gauge that is unitary with respect to time diffs., SUSY and $R$-symmetry, and write an action invariant only under time-dependent spatial diffs.~out of the fields $e^a_\mu,\psi_\mu$ and $V_\mu$. By introducing the Goldstone boson associated to the $R$-symmetry, that we can call $\phi$, on top of the usual $\pi$ and $\lambda$, we can then define St\"uckelbergized fields that transform under a general SUGRA+$R$-symmetry as under a linearly realized diff.. The parameter of the time-dependent diff., now possibly depending also on $\phi$ and $V_\mu$,~will be such that the fields realize the extended SUGRA algebra. To obtain the decoupling limit, we can perform exactly the same field redefinitions as we did in the case of normal SUGRA and that leads to the introduction of the Goldstone $\pi$ and $\lambda$, with the only difference that now the Goldstone of $R$-symmetry will appear in the non gauge invariant terms containing $V^\mu$, schematically $\dress V^\mu\sim V_\mu+\d_\mu\phi$. The resulting action is therefore made of the same action as we add for the \SEFTofI, plus a novel contribution containing the terms in $V^\mu$ and $\phi$. It follows that all what we discussed in this section applies unaltered to the case in which we extend the spacetime symmetries.

\subsection{Reheating}

We have so far constructed a Lagrangian (or Lagrangians including the case of multifields), that describe the consequences of the spontaneous breaking of SUGRA in the context of Inflation. However, we have not yet shown how to compute observables. We do this in this subsection. In cosmology, observables corresponds to in-in correlation functions. In the case of purely adiabatic fluctuations, we are interested in in-in correlation functions of a variable $\zeta$ which is constant outside the horizon (the generalization to include isocurvature fluctuations is straightforward and will not be done explicitly here, see for e.g.~\cite{Senatore:2010wk}). In a setup where more than one light degree of freedom is present, as in this case, the fluctuation $\zeta$ is not determined just by the physics at horizon crossing, as in single field inflation, but can be affected by the fluctuating fields even when a certain mode is outside the horizon. However, the fact that this influence occurs when all modes of interest are outside of the horizon allows us to simply parametrize this effect in the following way~\cite{Senatore:2010wk}. At the reheating time $t=t_{\rm rh}$, we write the most general local relation between $\zeta(\vec x,t_{\rm rh})$ and the fields that are present in the theory, that is invariant under time dependent spatial diffs.. Schematically, we have
\bea\label{eq:reheating}
&&\zeta(\vec x,t_{\rm rh})=\zeta_{\rm bosonic}\left( \dress g_{\alpha\beta}(e(\vec x,t_{\rm rh}),\pi(\vec x,t_{\rm rh}))\right)\\ \nn
&&\ \qquad\qquad +a_1\; \dress\psi_\mu\left(e(\vec x,t_{\rm rh}),\pi(\vec x,t_{\rm rh})\right)\,\dress\psi^\mu\left(e(\vec x,t_{\rm rh}),\pi(\vec x,t_{\rm rh})\right)\\ \nn
&&\  \qquad\qquad +a_2\;\dress\psi^0\left(e(\vec x,t_{\rm rh}),\pi(\vec x,t_{\rm rh}))\right)\,\dress\psi^0\left(e(\vec x,t_{\rm rh}),\pi(\vec x,t_{\rm rh})\right) +\ldots \ ,
\eea
where $\ldots$ represents higher order terms in the derivatives and the number of fields, and $a_i$ represent unknown parameters that are determined by the specific way in which the $\psi$ fluctuations are converted into metric fluctuations in the sixty or so $e$-foldings from horizon crossing to reheating. $\zeta_{\rm bosonic}=-H\pi+\ldots$, where $\ldots$ represent higher order terms in the fluctuations (and, incidentally, slow roll corrections), is the relation between $\zeta$ and the metric fluctuations in the case of single field inflation~\cite{Cheung:2007st}.

We can write expression (\ref{eq:reheating}) in a generic SUGRA gauge by noticing that any scalar gravitino bilinear in the SUGRA-unitary gauge where $\lambda=0$, such as $\hat \psi^0\hat \psi^0$, can be written as
\bea
&&\left.\dress  \psi^0\dress \psi^0\right|_{x,\; {\rm SUGRA \;unitary}} =\left.\dress \Psi^0\dress \Psi^0\right|_{x+\xi^\mu/2 +\dots}=\left.\dress \Psi^0\dress \Psi^0\right|_{x}+\text{(4 fermions)}\ ,
\eea
which follows from the fact that in unitary gauge $\psi_\mu=\Psi_\mu$ and that, under a generic SUGRA transformation, $\dress \Psi^0\dress \Psi^0$ is a scalar.  Making this substitution for all the fields appearing in (\ref{eq:reheating}), we obtain
\bea\label{eq:reheating_redef}\nn
&&\zeta(\vec x,t_{\rm rh})=\zeta_{\rm bosonic}\left(\hat g_{\alpha\beta}(E(\vec x,t_{\rm rh}),\pi(\vec x+\vec\xi(\lambda)/2,t_{\rm rh}+\xi^0(\lambda)/2)+\xi^0(\lambda)/2)\right)\\ \nn
&&  +a_1 \hat\Psi_\mu\left(E(\vec x,t_{\rm rh}),\psi'(\vec x,t_{\rm rh}),\pi(\vec x,t_{\rm rh}),\lambda(\vec x,t_{\rm rh})\right)\hat\Psi^\mu\left(E(\vec x,t_{\rm rh}),\psi'(\vec x,t_{\rm rh}),\pi(\vec x,t_{\rm rh}),\lambda(\vec x,t_{\rm rh})\right)\\ \nn
&&   +a_2\hat\Psi^0\left(E(\vec x,t_{\rm rh}),\psi'(\vec x,t_{\rm rh}),\pi(\vec x,t_{\rm rh}),\lambda(\vec x,t_{\rm rh})\right)\hat\Psi^0\left(E(\vec x,t_{\rm rh}),\psi'(\vec x,t_{\rm rh}),\pi(\vec x,t_{\rm rh}),\lambda(\vec x,t_{\rm rh})\right)\\ 
&& +\ldots \, .
\eea
To avoid confusion, we have reverted to denoting the redefined vierbein  and gravitino as $E$ and $\psi'$, as in-in correlation functions are not invariant under field redefinitions. $\ldots$ represents additional terms that are quadratic in the fields (and that we did not write for brevity), and higher order terms.


\section{Phenomenology\label{sec:pheno}}

Having developed the general formalism to construct an inflationary Lagrangian where SUGRA is spontaneously broken, we are now ready to study its observational signatures. First, we will explore what are the natural parameter regions of the theory~(\ref{S_stuck}), and determine the associated unitarity bounds. In this section we will drop all Lorentz and spacetime indices except in the very few cases when neglecting them can cause confusion.

\subsection{Unitarity bounds}

In this section we determine the unitarity bounds of the different parameter regions of the theory (\ref{S_stuck}) by estimating the energies at which radiative corrections become order one. As discussed before, we can safely assume that all the mass terms in the unitary gauge are at most of order Hubble
\begin{equation}
m_{3/2},\, m_0,\, \wt m_1,\, \wt m_2,\, m_\star \quad \lesssim\quad H\, .
\end{equation}
This parameter range is radiatively stable, as is evident from the unitary gauge action~(\ref{S_unitary}). In fact, for any interacting operator of the form $m\, \psi_c {\cal O}/\Lambda^{n}$ in~(\ref{psi_lambda3}), where ${}_c$ refers to canonically normalized fields~(see in (\ref{eq:canonical_leo})), there is one in~(\ref{kinkin3}) where the gravitino is replaced by the Goldstino giving $\frac{m}{H}\d \lambda_c{\cal O}/\Lambda^{n}$, in formulas 
\be
\frac{m }{\Lambda^{n}}\psi_c {\cal O}\quad\to\quad \frac{m }{H\Lambda^{n}} \d\lambda_c {\cal O}\ .
\ee
Modulo cancellations, that one can easily check are generically not present, for $m\sim H$, the unitarity bound of the theory is $\Lambda$. Since in unitary gauge all mass terms softly break SUSY, radiative corrections need to be proportional to the mass itself and must be cutoff at most at $\Lambda$. 
This implies that the radiatively generated gravitino masses will be at most of order $H$, with potentially just a logarithmic correction in $\Lambda$, and the theory is therefore technically natural~\footnote{It is interesting to notice that the radiatively generated gravitino masses are in fact further suppressed than naive power counting would suggest. This can be understood by various discrete transformations (such as $P$, $T$ or $\psi\to i \psi$) which only leave the unitary gauge action invariant if certain masses are flipped $m\to -m$. Therefore, any radiative correction to the mass terms must have the same ``parity'' under the relevant $\mathbb Z_2$ symmetry as the bare mass. These discrete symmetries, that can even allow to consistently keep some mass terms to zero, are not required for radiative stability of the theory, so we will not discuss them further.
}. One can similarly check that in this parameter range the bosonic operators do not receive large radiative corrections. 

For each choice of the parameters, we identify the leading operators that set the unitarity bound.
In this sense, the action (\ref{S_stuck}) has four qualitatively different regimes which we will treat separately. First, we consider the case $\{\wt m_1,\, m_0\} = 0$, where the leading operator is $\d\pi\lambda\d\lambda$. Next, we assume at least one of $\wt m_1$ or $m_0$ is non-vanishing, in which case there is a $\d^2 \pi \lambda\d \lambda$ term which reduces the unitarity bound and thus changes the analysis. Another interesting regime occurs when the `non-minimal' cubic terms in the unitary gauge action of the form $\delta g^{00} \psi\psi$ are important. These `non-minimal couplings' to the metric reduce the unitarity bound by generating a $\d\pi \d\lambda\d\lambda$ term and lead to enhanced non-Gaussianities. Finally, we turn on $m_\star \neq 0$, which leads to a non-relativistic Goldstino at high energies $\omega\sim k^2$. 

Let us point out two interesting simplifications. First, it was shown in Sec.~\ref{sec:dec} that in the absence of unexpected cancellations among the different parameters, the technically natural region for the Goldstino speed of sound covers the values $c_\lambda -1\sim \mathcal{O}(1)$ in a FRW universe. Therefore, non-Gaussianities and other observables cannot be significantly enhanced by powers of $1/c_{\lambda}$, so we will assume $c_\lambda-1 \sim {\cal O}(1)$ in the following. On the other hand, the Goldstone can have a technically natural low speed of sound, the phenomenological consequences of which can be found in Ref.~\cite{Cheung:2007st,Senatore:2009gt}. In the present work, we will focus on the physical effects of non-linearly realized SUSY, and therefore only consider $c_\pi\simeq 1$. 

Second, the fact that the mixing energy between the Goldstino and the gravitino is of order $H$ implies that they are decoupled at energy scales much greater than $H$, which is the regime of interest for radiative corrections. So, while in order to make inflationary predictions we cannot neglect the mixing of $\lambda$ with $\psi_\mu$, we can neglect it to study radiative corrections, which greatly simplifies the treatment~\footnote{Another interesting regime where the decoupling might be usefully used is the case in which we spontaneously break the continuous shift symmetry of $\pi$ to a discrete subgroup. In this case, a resonance occurs at energies parametrically larger than $H$ (see~\cite{Behbahani:2011it} for a discussion of this in the EFT context). In this situation, the Goldstino decoupled Lagrangian might be employed. We leave the study of the phenomenology of this case to future work.}.

\subsubsection{Effects of $m_{3/2}\neq0 \text{ or }\wt m_{2}\neq 0 , \quad \left[\{\wt m_1,\, m_0,\, m_{(3)},\, m_\star\} = 0\right]$}

In this parameter region, the Lagrangian quadratic in fermions (and including for clarity also the $\pi$ kinetic term) has the form
\begin{equation}
\frac{1}{e\mpl^2} \mathcal L^{\rm 2f} \sim \dot H (\d\pi)^2 +\psi \d \psi + H^2 \bigl[\lambda \d \lambda + \d \pi \lambda\d \lambda + \d\pi \psi \lambda\bigr]\, ,
\end{equation}
where we are assuming $m_{3/2}(t)\not \propto e^{i\wt m_2 t}$, otherwise there are no interactions between $\pi$ and fermion bilinears~\footnote{The Goldstino kinetic term does not vanish in this limit because it gets a contribution from the spacetime curvature of order $H\mpl$. This means that the theory with $m_{3/2}(t) \propto e^{i\wt m_2 t}$ appears to be a quite uninteresting limit, as the coupling between the Goldstino and the Goldstone is only mediated by gravity.}. The Lagrangian for the canonical fields (see~(\ref{eq:canonical_leo}))
\begin{equation}\label{eq:canonical}
\psi_c = \mpl\, \psi \, , \qquad
\lambda_c \sim \mpl H\, \lambda \, , \qquad
\pi_c \sim \sqrt{\epsilon} \mpl H\,  \pi \, ,
\end{equation}
is given by
\begin{equation}
\mathcal L^{\rm 2f} \sim (\d\pi_c)^2+\psi_c \d \psi_c + \lambda_c \d \lambda_c  +  \frac{1}{\sqrt{\epsilon}\mpl H}\d \pi_c \lambda_c \d \lambda_c + \frac{1}{\sqrt{\epsilon}\mpl}\d\pi_c \psi_c \lambda_c\, .
\end{equation}
The unitarity bound can be read off from the Goldstone-Goldstino interaction and corresponds to the energy scale at which Goldstino or Goldstone loops should be cut off in order to give $\lesssim{\cal O}(1)$ radiative corrections. Accounting for the $1/(4\pi)^2$ suppression of loops, it is approximately given by
\begin{equation}\label{cutoff_case1}
\Lambda^4 \sim (4\pi)^2 \epsilon (\mpl H)^2 \quad \Rightarrow \quad \Lambda \sim H \left(\frac{4\pi}{\zeta}\right)^{1/2}\, ,
\end{equation}
where here $\zeta\equiv\langle\zeta^2\rangle^{1/2}\sim H/(\sqrt{\epsilon}\mpl)\sim 3\cdot 10^{-5}$ is the magnitude of curvature fluctuations.

Note that in this section we have set to zero certain mass terms which lead to higher derivative terms such as $(\d\pi)(\d\lambda)^2$ and $\d^2\pi(\lambda\d\lambda)$, and are treated in the following sections. These terms are radiatively generated in the present regime as well, but are suppressed with higher powers of the unitarity bound (\ref{cutoff_case1}).

\subsubsection{Effects of $\wt m_1\neq 0 \text{ or } m_0\neq 0 , \quad \left[\{m_{(3)},\, m_\star\} = 0\right]$}
When either $\wt m_1$ or $m_0$ are non-vanishing, the Lagrangian quadratic in fermions takes the form
\begin{equation}\label{eq:unitlow}
\frac{1}{e\mpl^2} \mathcal L^{\rm 2f}
	\sim  \psi \d \psi + H^2 \Bigl[\lambda \d \lambda + \bigl(\d \pi + (\d \pi)^2\bigr) \lambda \d \lambda + \bigl(\d \pi + (\d \pi)^2\bigr) \psi \lambda\Bigr] + H \Bigl[\d^2 \pi \lambda \d \lambda + \d^2 \pi \psi \lambda\Bigr]\, .
\end{equation}
Here we have assumed that if $m_0\neq0$, $m_0\sim H$. However, as discussed in Appendix~\ref{app:Goldstino_prop}, if only $m_0\neq 0$, $m_{3/2}=m_0/2$, and all the other mass parameters are set to zero, subluminally of the fluctuations requires $|m_0|\lesssim \epsilon H$. However, this restriction does not seem to apply in the generic case where all the parameters considered in this subsection are set to be comparable, and we restrict to this case here~\footnote{\label{footnote:subluminality}{The presence of superluminal fluctuations implies that the theory cannot be UV completed in a local Lorentz invariant one~\cite{Adams:2006sv}. An alternative, but somewhat more radical, point of view with respect to the one we develop here  would be to nevertheless study the phenomenology of this consistent EFT, giving up on the locality and Lorentz invariance of the UV completion. }}. 
After canonical normalization, the action in~(\ref{eq:unitlow}) becomes
\begin{equation}\label{case2_L_can}
\begin{split}
\mathcal L^{\rm 2f} \sim \ 
	&\psi_c \d \psi_c + \left[1 + \frac{\d\pi_c}{\sqrt{\epsilon}\mpl H} + \left(\frac{\d\pi_c}{\sqrt{\epsilon}\mpl H}\right)^2\right]\lambda_c \d \lambda_c + \frac{1}{\sqrt{\epsilon}\mpl H^2}\d^2 \pi_c \lambda_c\d\lambda_c \\
	& + \left[\frac{\d\pi_c}{\sqrt{\epsilon}\mpl H} + \left(\frac{\d\pi_c}{\sqrt{\epsilon}\mpl H}\right)^2\right]\psi_c \lambda_c + \frac{1}{\sqrt{\epsilon}\mpl H}\d^2 \pi_c\psi_c\lambda_c \, .
\end{split}
\end{equation}
With respect to the former case, the unitarity bound is lowered by the vertex $\d^2 \pi \lambda \d \lambda$ to
\begin{equation}
\Lambda^6 \sim (4\pi)^2 (\sqrt{\epsilon}\mpl H^2)^2 
\quad \Rightarrow \quad
\Lambda \sim H \left(\frac{4\pi}{\zeta}\right)^{1/3}\, .
\end{equation}
%

\subsubsection{Effects of $m_{(3)}\neq 0$, $\quad \left[ \{m_\star\}=0\right]$}

Let us now add the `non-minimal coupling' terms of the form $m_{(3)}\delta g^{00} \psi\psi$ in the unitary gauge action~\footnote{Here we are generically treating all unitary gauge terms that couple $\delta g^{00}$ to gravitino bilinears. In reality, however, the most interesting ones are those where the structure of the gravitino bilinear is such that, upon reintroduction of the Goldstino, the maximal derivative term does not cancel. In particular this implies that these terms could no be obtained by promoting  the mass terms $\wt m_1,\, \wt m_2,\, \ldots$ to functions of $g^{00}$. One example of such a term is $\delta g^{00}\psi^\mu\psi_\mu$ which cannot be absorbed into a mass term $\psi_\mu\psi^\mu$, since this mass term would lead to a ghost. }. These give contributions to the quadratic fermion action of the form
\begin{equation}\label{case3_L_can}
\mathcal L^{\rm 2f} \supset \frac{m_{(3)}}{\sqrt{\epsilon} \mpl H^3}(\d\pi_c)(\d\lambda_c)^2 
	+ \frac{m_{(3)}}{\epsilon \mpl^2 H^4}(\d\pi_c)^2(\d\lambda_c)^2\, .
\end{equation}
These terms, just like the $\wt m_1$ and $m_0$ terms of the previous section, lower the unitarity bound to
\begin{equation}\label{eq:cutoff}
\Lambda^3 \sim 4\pi\sqrt{\epsilon} \mpl H^2\cdot \frac{H}{m_{(3)}} \quad \Rightarrow \quad
\Lambda \sim H \left(\frac{4\pi}{\zeta}\right)^{1/3} \left(\frac{H}{m_{(3)}}\right)^{1/3}\, ,
\end{equation}
if $m_{(3)}\gtrsim H$. If $m_{(3)}\lesssim H$, the cutoff is still given by the one of the former section.
We need to check the radiatively generated gravitino mass to see if there is an upper bound on $m_3$. The operator $\delta g^{00}\psi^\mu\psi_\mu$ contains a term of the form 
\be
m_{(3)}\delta g^{00}\psi^\mu\psi_\mu\quad\supset\quad m_{(3)}\frac{\dot\pi_c^2}{\left(\sqrt{\epsilon}H\mpl\right)^2}\psi\psi\ ,
\ee
which induces a mass term of order
\be
m\sim \frac{m_{(3)}}{16\pi^2}\frac{\Lambda^4}{\left(\sqrt{\epsilon}H\mpl\right)^2}\ .
\ee
Once we plug (\ref{eq:cutoff}) in the above equation, we see that the correction is very small. Similarly, it is easy to check that the additional diagram obtainable by using two insertions of the $\dot\pi\psi\psi$ vertex induces a small gravitino mass as well. Therefore $m_{(3)}$ is not restricted to be $\lesssim H$.
It is nevertheless bounded, since we require the theory to be weakly coupled at the Hubble scale: 
\begin{equation}
H\ll \Lambda \quad \Rightarrow \quad m_{(3)}\ll H \,\frac{4\pi}{\zeta}\, .
\end{equation}
%

\subsubsection{Effects of $m_\star \neq 0$\label{sec:mstarpheno}}

Since the $m^\star$ term in the action does not give a kinetic term $\lambda\d_0\lambda$ for the Goldstino, the theory is sick around flat space. This pathology is regulated by the FRW (or even de Sitter) background. Around such a background, the action quadratic in fermions at energies $\omega \gg H$ has the form
\begin{equation}
\begin{split}
\frac{1}{e\mpl^2}\mathcal L^{\rm 2f} \sim \ &
	\psi\d\psi + H^2 \left[\lambda\d\lambda + \d\pi \lambda\d\lambda + \d\pi \psi\lambda\right]\\
	&+ m_\star \left[\d_i\lambda \d^i\lambda + \psi^i\d_i\lambda + \d \lambda\psi\d\pi  +  \d\pi (\d\lambda)^2+\psi(\d\pi)^2 \d\lambda + (\d\lambda\d\pi)^2\right]\, .
\end{split}
\end{equation}
The $m_\star$ term leads to the following Goldstino dispersion at high energies:
\begin{equation}
\omega\sim k + \frac{m_\star}{H^2}k^2   \qquad \hbox{for } \omega\gg H\, . 
\end{equation}
There is a cross-over between a linear and a quadratic dispersion at $E_{\rm cr} \sim H^2/m_\star\gtrsim H$:
\begin{equation}
\omega \sim 
\left\{
\begin{array}{cl}
k \quad &\hbox{for } H\lesssim \omega \lesssim E_{\rm cr}\, , \\
\dfrac{m_\star}{H^2} k^2 \quad &\hbox{for } E_{\rm cr} \lesssim \omega \lesssim \Lambda\, .
\end{array}
\right.
\end{equation}
Above $E_{\rm cr}$, the group velocity of Goldstino waves is approximately $c_{\lambda,{\rm group}}\sim \d\omega/\d k\sim 1+k/E_{\rm cr}$, where we neglected order one numbers. Imposing subluminal propagation tells us that the dispersion relation can be quadratic only for an order one range of wavenumbers, making the study of the phenomenology of this situation not much different than what we found with the other dispersion relations. In particular, it better be that superluminal propagation occurs only for energies above the unitarity bound, where we do not trust the theory. This requires $m_\star\lesssim H (\zeta/4\pi)^{1/2}$, which it can be checked is indeed the maximum size  that we expect it to be radiatively generated by loops of the interactions introduced earlier on (potentially with higher time derivative operators as well). Therefore, unless we tune $c_\lambda$ in $\omega=c_\lambda k$ to be very small, we expect the quadratic dispersion to be only marginally important, and we neglect it from now on (see footnote (\ref{footnote:subluminality})).


\subsection{Observational signatures}

We are now ready to study the observational signatures predicted by the supersymmetric EFTofI.  We will focus on non-Gaussian signatures. Since the Goldstino and the gravitino affect density perturbations only through their coupling to the Goldstone $\pi$, non trivial $\pi$-correlation functions are introduced only at loop level. These interactions in the decoupling limit are expected by the non-Abelian nature of the group being spontaneously broken, SUSY, in contrast for example with the case in which we break only time diffs., where $\pi$ can be free in the decoupling limit. 

 Given a certain loop diagram, there are two regimes of interest. One consists of the contributions from internal modes much greater than~$H$, and the opposite from modes of order $H$. Modes much longer than $H$ are not expected to contribute relevantly because of the nature of the couplings and because the fluctuations of fermions, even more so for massive one like ours, are expected to decay outside of the horizon; of course these modes will contribute to generate the (mildly) squeezed limit of correlation functions, which however is expected to be a subleading signal (See for e.g.~\cite{Senatore:2009cf,Senatore:2012nq,Pimentel:2012tw,Senatore:2012ya} for studies of radiative contribution in inflationary correlation functions with emphasis on an EFT interpretation.). Therefore we concentrate on external legs with momenta of order $H$. Apart for an exponentially small contribution, the contribution from modes running in the loops that are much shorter than~$H$ has the same functional form as the one from operators in the bosonic EFT. Indeed, this has to be so for the theory to renormalizable.  
The UV contribution of these loops cannot be reliably estimated within the EFT, and we know only that they have to be cutoff at scale $\Lambda_{\rm cut} $ below the unitarity bound~$\Lambda$. The origin of this fallacy is easy to realize by the simple fact that the UV contribution of the diagrams is UV dominated, and so it depends on the UV completion, and so is not model independent, which is what the EFT is able to parametrize. Assuming there is no tuned cancellation between the radiative corrections and the bosonic EFT parameters, an upper bound to the UV contributions is estimated by cutting off the loops at the unitarity bound. As mentioned, these contributions are degenerate with the ones produced by the bosonic EFTofI. At leading derivative level, for the three-point function, they are well described by a linear combination of the equilateral~\cite{Creminelli:2005hu} and orthogonal~\cite{Senatore:2009gt} templates, whose sizes are parametrized by the coefficients $\fnleq$ and $\fnlor$. Similar templates exist for the four-point function (see for example~\cite{Smith:2015uia}), though we leave the study of the four-point function to future work (See the results of the Planck analysis for the current best limits~\cite{Ade:2015ava}.).

The contribution from modes of order $H$ running in the loops leads instead to a functional form of the correlation function that is different, at least in principle, to the one induced by the bosonic EFTofI operators. This contribution, if detected, will be probably associated quite uniquely to Goldstinos and gravitinos running in the loops, and would therefore represent a `smoking gun' signature of supersymmetry as a non-linearly realized symmetry in our universe. We will not evaluate explicitly the loops, so we will not study possible accidental similarity in shape with bosonic EFTofI operators. We leave this to future work. Since we will assume no cancellation between the UV contribution to the shape and the terms in the bosonic EFTofI, the signal associated with the UV contribution will be generically much larger than the `smoking gun' one coming from internal modes of order $H$.

In brief, we will find the following. For the `minimal couplings' to the metric that are induced by diff.~invariance of the mass terms, we find that in general the UV contribution from loops leads to three-point functions of the shape produced by the bosonic EFTofI with $\fnleq$ or $\fnlor$ that can be as large as order one. The `smoking gun' signatures from low-energy momenta running in the loops are irremediably small. This leaves a reasonable detection prospect. The situation is even better for the `non-minimal' couplings of the form $\delta g\, \psi\psi$. In this case, the UV modes lead to a possibly very large $\fnleq$ or $\fnlor$, while the Hubble scale modes lead to a still very small, though much larger than in the other cases, $f_{\rm NL}\ll 1$. 
All of this is quite good news: it gives us a strong motivation to reach $\fnleq$ or $\fnlor$ of order one. 
Furthermore, at least for the non-minimal couplings, it gives a (still challenging) hope to detect the smoking gun signals of SUSY through cosmological observations.

\subsubsection{Effects of $m_{3/2}\neq0 \text{ or }\wt m_{2}\neq 0 , \quad \left[\{\wt m_1,\, m_0,\, m_{(3)},\, m_\star\} = 0\right]$\label{sec:first_pheno}}
The strength of non-Gaussianities can be estimated by finding the radiatively generated Goldstone three-point function. For example, the EFTofI operator $\dot \pi_c^3$ is generated by a Goldstone loop with three insertions of the $\d\pi_c \lambda_c \d \lambda_c/(\sqrt{\epsilon}\mpl H)$ vertex with the following coefficient:
\begin{equation}\label{case1_3piloop}
\sim \frac{1}{(\sqrt{\epsilon}\mpl H)^3} \frac{1}{(4\pi)^2}\int d^4p\, \frac{p^3}{p^3} \sim \frac{\Lambda^4_{\rm cut} }{(\sqrt{\epsilon}\mpl H)^3 (4\pi)^2} \lesssim \frac{1}{\sqrt{\epsilon}\mpl H}\, ,
\end{equation}
where in the last step we have cutoff the loop at the unitarity bound (\ref{cutoff_case1}).
The strength of non-Gaussianities is then estimated as 
\begin{equation}
\fnlrad \zeta \sim \left.\frac{\mathcal L_3}{\mathcal L_2}\right|_{E\sim H} 
	\lesssim \left.\frac{(\d \pi_c)^3}{(\sqrt{\epsilon}\mpl H) (\d \pi_c)^2}\right|_{E\sim H} 
	\sim \frac{H^2}{\sqrt{\epsilon}\mpl H}
	\sim \zeta \quad \Rightarrow \quad \fnlrad \lesssim 1\, ,
\end{equation}
with equality holding when the loop is cutoff at the unitarity bound.
This contribution is of course degenerate with the contribution from the bosonic theory, though it motivates a somewhat large signal from it. The ``smoking-gun'' signature of SUSY comes from the Goldstone three-point function that is generated by internal Goldstinos at energies of order~$H$, as this contribution is not degenerate with a local operator in the bosonic EFT. Its size can be estimated by taking the loop on the left hand side of~\eqref{case1_3piloop} at energies of order $H$, giving
\begin{equation}
\sim \frac{H^4}{(\sqrt{\epsilon}\mpl H)^3 } \sim \frac{\zeta^2}{\sqrt{\epsilon}\mpl H}\, ,
\end{equation}
which leads to the following very small estimate for non-Gaussianities
\begin{equation}
f_{\rm NL}^{\rm smok.\; gun} \zeta \sim \frac{H^6}{(\sqrt{\epsilon}\mpl H)^3 } \sim {\zeta^3}
	\qquad \Rightarrow \qquad
	f_{\rm NL}^{\rm smok.\; gun} \sim {\zeta^2} \sim {10^{-9}}\, .
\end{equation}
Here and in the rest of this section, for these contributions from modes of order~$H$, we will not be careful with factor of $(4\pi)^2$ or with the ambiguity of the scale at which to compute the loops (e.g. $\sim H$ or $\sim 2H$), as keeping track of these differences does not change relevantly the conclusion.
%


\subsubsection{Effects of $\wt m_1\neq 0 \text{ or } m_0\neq 0 , \quad \left[\{m_{(3)},\, m_\star\} = 0\right]$}

The Lagrangian (\ref{case2_L_can}) generates the following cubic operators at one loop
\begin{equation}
\mathcal L_3 \quad \supset \quad
(\d\pi)^3\, , \quad (\d^2\pi)(\d\pi)^2\, , \quad (\d^2\pi)^2 (\d\pi)\, , \quad (\d^2\pi)^3\, .
\end{equation}
Following very similar steps as in (\ref{sec:first_pheno}), the radiatively generated non-Gaussianities are in all cases small and are given by
\begin{equation}
\fnlrad \zeta \sim \left.\frac{\mathcal L_3}{\mathcal L_2}\right|_{E\sim H} 
	\lesssim \zeta \left(\frac{\zeta}{4\pi}\right)^{2/3} \qquad \Rightarrow \qquad \fnlrad \lesssim \left(\frac{\zeta}{4\pi}\right)^{2/3}\sim 10^{-4}\, .
\end{equation}
Since all fermion loops above have a quadratic divergence, the non-Gaussianities produced by fermions at energy $E\sim H$ are suppressed by $H^4/\Lambda^4\sim \zeta^{4/3}$, and are thus again given by
\begin{equation}
f_{\rm NL}^{\rm smok.\; gun} \sim \zeta^2\sim 10^{-9}\, .
\end{equation}
Again, this is very small.
%


\subsubsection{Effects of $m_{(3)}\neq 0$, $\quad \left[ \{m_\star\}=0\right]$}
The two extra terms in \eqref{case3_L_can} coming from the non-minimal couplings lead to more appreciable non-Gaussianities. The $(\d\pi)^3$ operator can be generated either with three $\d\pi (\d\lambda)^2$ vertices, or one $\d\pi (\d\lambda)^2$ vertex and one $(\d\pi)^2 (\d\lambda)^2$. The radiatively generated non-Gaussianities can be large, respectively  
\begin{equation}
\fnlrad{}^{(a)} \sim  \frac{m_{(3)}}{H}\left(\frac{\Lambda_{\rm cut}}{\Lambda}\right)^6\lesssim  \frac{m_{(3)}}{H} \, , \qquad \hbox{and} \qquad
\fnlrad{}^{(b)} \sim \left(\frac{\Lambda_{\rm cut}}{\Lambda}\right)^6\lesssim 1 \, ,
\end{equation}
Note that there is a slight subtlety in computing the first diagram, where the leading divergence cancels because the integrand is odd under inversion of the loop momentum. The contribution of the first term can be very large, and indeed is basically just constrained by current limits on $\fnlrad$ from the CMB~\cite{Ade:2015ava}: $m_{(3)}\lesssim 10^2 H\left(\Lambda/\Lambda_{\rm cut}\right)^6$. The signature SUSY non-Gaussianities is in this case given by 
\bea
&&f_{\rm NL}^{\rm smok.\; gun,(a)} \sim f_{\rm NL}^{\rm equil.,orthog.,(a)}\left(\frac{H}{\Lambda_{\rm cut}}\right)^6 \lesssim 10^{2}\left(\frac{H}{\Lambda_{\rm cut}}\right)^6\, , \\  \nn
&&f_{\rm NL}^{\rm smok.\; gun,(b)} \sim  \left(\frac{\Lambda_{\rm cut}}{\Lambda}\right)^6\left(\frac{H}{\Lambda_{\rm cut}}\right)^6\ll 1.
\eea
The signature SUSY non-Gaussianities are much larger than in former cases, but still, unless we keep $\Lambda_{\rm cut}$ close to $H$, or we evaluate the loops at a somewhat larger scale than $H$ (where however the difference with respect to the $\fnlrad$ terms is decreased), they appear hardly detectable in the not so distant future.

\subsubsection{Slow-roll power counting}

As we mentioned in the introduction, we have constructed the EFT based only on the symmetries that are present in the most general inflationary solution. However, one might consider the case in which the inflationary solution is perturbatively close to the maximally symmetric one. By perturbatively close, we mean that, in constructing the theory of the fluctuations from a complete theory, every insertion of a vev of the time derivatives of the fields is taken as a suppression. We call these models of inflation as slow-roll inflation models, and we call the parameter representing this suppression as $\alpha_{\rm sl}\ll1$. In the case of a slowly rolling scalar field, we expect $\alpha_{\rm sl} \sim \dot\phi/\Lambda^2$, with $\Lambda$ representing some high energy scale. 

In this case, we expect an hierarchy in size among the different operators. Basically, every upper zero index that is associated to the breaking of Lorentz invariance carries a factor of $\alpha_{\rm sl}$. In this counting, we naturally expect, roughly, $m_0\sim \wt m_1\sim \wt m_2\sim \alpha_{\rm sl}\, H$, and $m_\star\sim \wt m_{(3)}\sim \alpha_{\rm sl}^2\, H$.

%

\subsubsection{Contributions from reheating}

So far in this section we have focussed on the contribution from Horizon crossing. Eq.~(\ref{eq:reheating_redef}) tells us that there is a contribution to the $\zeta$ correlation functions coming directly from the reheating time. Even in this case, since the fermions appear quadratically in the expression for $\zeta$, the contribution to $\zeta$ correlation functions in Fourier space will consist of convolutions of correlation functions of Goldstino or gravitinos. Since these fields are not expected to obtain scale-invariant power spectra, the resulting contribution will not be scale invariant, and therefore will be negligible at the observed scales. If the power spectra were to be accidentally scale invariant, then there will be a scale invariant contribution, of the form of the so-called local shapes. For the three-point function, the size of this shape is usually quantified by $f_{\rm NL}^{\rm loc.}$.

 \section{Conclusions\label{sec:conclusions}}
     
     We have constructed the Supersymmetric Effective Field Theory of Inflation, {\it i.e.} the most general action for the fluctuations in an inflationary background in the case where supersymmetry happens to be a fundamental symmetry of Nature. The action was constructed with the following logic. Because of the structure of the algebra, the inflationary background, that we assume to be a quasi de Sitter expansion where time translations are spontaneously broken, spontaneously breaks SUSY. In the presence of gravity, supersymmetry is gauged to supergravity, which is therefore spontaneously broken as well. Constructing the Supersymmetric Effective Field Theory of Inflation in this case reduces therefore to constructing the most general Lagrangian that non-linearly realizes SUGRA and time diffs..  Several simplifications follow from the fact that SUSY is spontaneously broken. In particular, we can get rid of the auxiliary fields,  and write an action directly for the fluctuations, where no fields need to take a vev. In fact, the non-linear realization of time translations and SUSY, that implies the presence of a Goldstone boson and a Goldstino, allows us to choose a gauge where both these fields are set to zero. In this gauge, we write the  most general Lagrangian with the fields at our disposal, {\it i.e.} the graviton and the gravitino, compatible with the residual gauge invariance of the problem, which is time-dependent spatial diffs., and including only the operators that softly break time diffs.~and SUGRA. 
          
     Time diffs.~and SUGRA invariances are then introduced by performing a St\"uckelberg transformation. In particular, we can choose a field redefinition where the SUGRA transformation is such that, after St\"uckelbergization, the Goldstino appears only in terms that in unitary gauge contain a gravitino. Furthermore, the same field redefinition allows us to define a Goldstino that decouples at high energies from the gravitino. Though expected on general grounds, such decoupling follows from a non-trivial construction and allows for a major simplification of the theory at high energies. We have also discussed the formalism to include additional light fields and the parametrization for the effects coming from the reheating epoch. We stress that no additional light particle on top of the Goldstone and the Goldstino is required to non-linearly realize SUGRA: in inflation, the essential signatures of SUSY as a symmetry of Nature are associated to the gravitino and the Goldstino. 
     
     We find that, contrary to the case of the Goldstone boson of time translations, the Goldstino is mixed with the gravitino at energy scales of order and below the Hubble scale $H$. This means that to compute inflationary observables there is in general no sense in which one can neglect the gravitino for the Goldstino. We find that the Goldstino can have either a linear dispersion relation, $\omega\sim c_\lambda k$, with $c_\lambda-1$ up to order one, or a very non-relativistic quadratic dispersion relation $\omega\propto k^2$. Imposing subluminal propagation imposes some bounds on the parameter space. In particular the quadratic dispersion relation occurs only for an order one range of wavenumbers that are much larger than $H$.
  
     We then studied the observable consequences of our Lagrangian. Since the fundamental particles associated to non-linearly realized SUGRA, the gravitino and Goldstino, are fermions, they can affect the density perturbation of the universe only through loop effects. We estimated the size of these loop contributions. We found that  modes larger than $H$ running in the loops give an effect that is degenerate with the one induced by the operators of the standard bosonic EFTofI, parametrized by $\fnlrad$. This contribution is UV dependent and cannot be estimated within the EFT, but we showed that it can be as large as to induce $\fnlrad\sim 1$, and, for some non-minimal couplings, even $\fnlrad\gg1$. Instead, the contribution from modes of order $H$ is not degenerate with the one from the standard EFTofI, and detection of this effect would therefore represent a smoking gun for SUSY as a symmetry of Nature. Unfortunately, we find that the effect is expected to be very small.
          
          Several exploratory directions open up after our construction. For example, on the theory side, it would be interesting to relax the continuous shift symmetry of the Goldstone to a discrete one as well as to explore more the presence of additional fields. On the signature side, it would be interesting to better investigate the shape of the induced non-Gaussianities, in particular in the squeezed limit. More generally, our findings offer further motivation for improving our exploration of primordial non-Gaussianities, identifying again $\fnlrad\sim 1$ as an interesting threshold to reach~\cite{Alvarez:2014vva}.  Unfortunately, obtaining such a sensitivity requires a strong improvement in our capability of understanding the evolution of the primordial fluctuations in large scale structures. As it has done in the past, we hope that the cosmology community will be able to conquer this novel challenge.          
         
         \section*{Acknowledgments}

We thank all of those who wished to patiently share their precious time discussing with us what appeared to us to be very subtle points. In particular, we thank Nima Arkani-Hamed, Paolo Creminelli, Thomas Dumitrescu, Savas Dimopoulos, Sergei Dubovsky, Gia Dvali, Sergio Ferrara, Daniel Freedman, Sean Hartnoll, Luis Alverez-Gaume, Gian Giudice, Renata Kallosh, Mehrdad Mirbabayi, Massimo Porrati, Riccardo Rattazzi, Eva Silverstein, Giovanni Villadoro and Matias Zaldarriaga for many interesting conversations. We thank Sergio Ferrara and Zohar Komargodski for comments on the draft. 
The work of L.V.D. was
partially supported by the Swiss National Science Foundation. L.S. is partially supported by DOE Early Career Award DE-FG02-12ER41854.

 \appendix
 
 \section*{Appendix}
 \section{Goldstino dispersion relation in de Sitter space\label{app:Goldstino_dispersion}}
As an application of the action \eqref{dS_action_stu}, we will derive the dispersion relation of the helicity-1/2 mode in de Sitter space. As we will see, there exists a decoupling energy above which mixing of the Goldstino with the constrained helicity-1/2 modes of the gravitino can be ignored, and the Goldstino dispersion relation reduces to that of a massless Weyl fermion in Minkowski space, the spacetime that is relevant at those energies. As it is already quite evident from~\eqref{spelledout_fermbil_dS}, we will find that the decoupling energy is $E_{\rm mix} \gtrsim H$, with equality holding for $m_{3/2}=0$. Notice that the helicity-1/2 components of $\psi_\mu$ are non-dynamical, so that the dynamics of the Goldstino is not completely transparent from~\eqref{fermbil_dS}. The correct way to proceed is to fix the SUGRA gauge and integrate out the constrained variables in the action. It is useful to define $\wt\Lambda \equiv\Lambda + 3\mpl^2|m_{3/2}|^2$. The gravitino equation of motion $0=\delta S/\delta \bar\psi_\mu$ gives
\begin{equation}
0 = \mpl^2 \left( \varepsilon^{\mu\nu\rho\sigma}\bar\sigma_\nu D_\rho \psi_\sigma + 2 m_{3/2}^{*}\bar\sigma^{\mu\nu}\bar\psi_\nu\right) - \wt\Lambda(i\bar\sigma^\mu\lambda) \, .
\end{equation}
Contracting the equation of motion with $\sigma_\mu$ one has~\footnote{Here $\nabla_\mu$ is the full covariant derivative, which acts on the gravitino as $\nabla_\mu \psi_\nu = D_\mu \psi_\nu - \Gamma_{\mu\nu}^\rho \psi_\rho$. In eq.~\eqref{constr2}, for instance, the terms should be read as
\begin{equation*}
\nabla^i\psi_i = g^{i\mu}(D_\mu\psi_i - \Gamma_{\mu i}^\nu \psi_\nu) 
\qquad \hbox{and} \qquad
\nabla_i\psi_0 = D_i \psi_0 - \Gamma_{i0}^\mu \psi_\mu\, .
\end{equation*}
}
\begin{equation}\label{constr1}
{\mpl^2} \left(\cancel D \bar\sigma^\mu \psi_\mu + \nabla^\mu\psi_\mu - \frac{3i}{2}m_{3/2}^{ *}\sigma^\mu\bar\psi_\mu\right) = 2\wt \Lambda \lambda\, .
\end{equation}
The $\mu=0$ component of the equation of motion can be expressed as
\begin{equation}\label{constr2}
M_{\rm Pl}^2(\sigma^iD_i\bar\sigma^\nu\psi_\nu +  \nabla^i\psi_i + \sigma^0 \bar\sigma^i\nabla_i \psi_0  - i m_{3/2}^{ *}\sigma^i\bar\psi_i)
	= \tilde\Lambda  \lambda\, .
\end{equation}
After fixing the SUGRA gauge freedom to $\bar\sigma^i\psi_i = 0$, which eliminates one of the helicity-1/2 components of the gravitino, equations \eqref{constr1} and \eqref{constr2} can be combined to give
\begin{equation}
\mpl^2 \left(\sigma^0 \bar\sigma^i\nabla_i\psi_0  + \frac{3i}{2}m_{3/2}^{ *}\sigma^0\bar\psi_0\right)= -\wt \Lambda\lambda\, ,
\end{equation}
which eliminates the remaining helicity-1/2 component of the gravitino $\psi^0$. This constraint equation together with the Goldstino equation of motion $0=\delta S/\delta \bar \lambda$, which gives
\begin{equation}
2(\cancel D \lambda + 2i m_{3/2}^{ *}\bar\lambda) = \bar\sigma^\mu \psi_\mu\, ,
\end{equation}
leads to a dispersion relation for the helicity-1/2 degree of freedom of the form
\begin{equation}
2\mpl^2
\left(\sigma^0\bar\sigma^i D_i\sigma_0(\cancel D \lambda+ 2i m_{3/2}^{ *}\bar\lambda) -\frac{3i}{ 2}m_{3/2}^{ *}(\cancel D \bar\lambda+ 2i  m_{3/2}\lambda)\right)
 = \wt\Lambda \lambda\, .
\end{equation}
This expression can be simplified using the modified covariant derivative \eqref{cov_D_dS}:
\begin{equation}
2\mpl^2\sigma^0\bar\sigma^i \mathcal D_i\sigma_0\cancel{\mathcal  D} \lambda = \wt\Lambda \lambda\, .
\end{equation}

Now focusing for simplicity on the case $m_{3/2} = 0$, the solution is given by
\begin{equation}\label{Goldstino_dispersion}
\lambda(\tau,z) = 
	c_1\frac{e^{ik(z-\tau)}\tau^{5/2}}{3i+2k\tau}\left(\begin{array}{c} 1\\ 0\end{array}\right)		
	+ c_2\frac{e^{ik(z+\tau)}\tau^{5/2}}{3i-2k\tau}\left(\begin{array}{c} 0\\ 1\end{array}\right)\, ,
\end{equation}
where $c_1$ and $c_2$ are two (Grassmann) constants, $\tau=-1/(aH)$ is the conformal time and $\Lambda = 3\mpl^2 H^2$. This expression agrees with that found in Ref.s~\cite{Kallosh:1999jj,Giudice:1999yt} in a different gauge. Eq.~\eqref{Goldstino_dispersion} makes it clear that decoupling only happens for modes satisfying $\omega\sim k/a\gg H$. In this regime, after rescaling, the solution reduces to that of a free massless Weyl fermion in Minkowski space.

\section{Goldstino dispersion relation in the decoupling limit\label{app:Goldstino_prop}}

In this Appendix we study the Goldstino dispersion relation in the decoupling limit and perform a first study that the requirement of subluminal propagation has on the parameter space of the EFT.  For simplicity, we restrict to $m_\star=0$, whose implication for the dispersion relation are discussed in the main text. As also discussed in the main text, we are interested in the regime where 
\be
m_{3/2}\sim m_0\sim \tilde m_1\sim \tilde m_2 \quad\lesssim \quad H \ .
\ee
For energies and momenta $E,k \gg H$, the Goldstino decouples from the gravitino, and the mass term can be ignored ($\mathcal L _{\lambda \lambda}^{\rm kin}\gg \mathcal L _{\lambda \lambda}^{\rm mass}$). Here we are interested in the Goldstino dispersion in an FRW background
\begin{equation}
e_\mu^a = {\rm diag}\left[1,a(t),a(t),a(t)\right]\, ,
\end{equation}
where various tensors take the simple form
\begin{equation}
\Gamma^0_{\mu\nu} = H(g_{\mu\nu}+\delta_\mu^0\delta_\nu^0)\, , \qquad
R_{\mu\nu} = 2\dot H\delta_\mu^0\delta_\nu^0 - (\dot H + 3H^2)g_{\mu\nu} \, , \qquad
R = -6 \dot H - 12 H^2\, .
\end{equation}
The Goldstino kinetic term \eqref{kinkin3} can then be expressed as
\begin{equation}\label{kin_dec}
\frac{1}{\mpl^2 e}\mathcal L_{\lambda\lambda}^{\rm kin}
	= \frac{i}{2}\alpha (\bar\lambda \cancel D \lambda) + \frac{i}{2}(\alpha+ \beta) (\bar \lambda \bar\sigma^0 D^0 \lambda) + \gamma (\bar \lambda \bar\sigma^{0\mu}D_\mu\bar \lambda) + \hbox{c.c.}\, ,
\end{equation}
with 
\begin{subequations}\label{abc_coeffs}
\begin{align}
\frac{1}{2}\alpha(t)
	&=3|M|^2 +\wt m_1^2 - 2\dot{\wt m_1} + 2\Re(M m_0) - 4\wt m_1H + 2 \dot H + 3 H^2 \, ,\\
\frac{1}{2}(\alpha(t)+\beta(t))
	&= 2 \left[-\wt m_1^2 - \dot{\wt m_1} + \Re(M m_0) + \wt m_1 H + \dot H \right] \, ,\\
\gamma(t)
	&= 4 \left[-iM\wt m_2 + \wt m_1(M+m_0^* ) + \dot M - m_0^* H\right] \, .
\end{align}
\end{subequations}
In the decoupling limit one can take $D_\mu\to \d_\mu$, and the kinetic term becomes
\begin{equation}
\begin{split}
\frac{1}{\mpl^2e}\mathcal L_{\lambda\lambda}^{\rm kin}
	&\simeq i\alpha \bar\lambda \bar\sigma^i \d_i \lambda - i \beta \bar\lambda\bar\sigma^0\d_0\lambda + \left(\gamma\bar \lambda\bar\sigma^{0i}\d_i\bar\lambda+ \hbox{c.c.}\right)\\
	&= i\wt \alpha{\bar\chi} \bar\sigma^i \d_i \chi - i \wt\beta {\bar\chi}\bar\sigma^0\d_0\chi\, ,
\end{split}
\end{equation}
where in the second line we absorbed the $\gamma$ term with the field redefinition
\begin{equation}
\chi = \lambda + z  \sigma^0 \bar\lambda \, , \qquad \hbox{with}\quad  z= i e^{i {\rm Arg}(\gamma)}
\frac{|\gamma|}{\alpha + \sqrt{\alpha^2 + |\gamma|^2}}\, ,
\end{equation}
and $\alpha = \wt \alpha(1-|z|^2)$, $\beta = \wt \beta(1+|z|^2)$. $\chi$ enjoys a fairly simple dispersion relation, with
\begin{equation}
0 = \det [\wt\alpha k^i \sigma_i + \wt \beta\Omega \sigma_0] = \wt\beta^2\Omega^2 - a^2 \wt \alpha^2 k^2\,.
\end{equation}
Requiring that the speed of propagation is subluminal thus imposes
\begin{equation}\label{sub_c}
1\geq c_\lambda^2 = \frac{\wt \alpha^2}{\wt \beta^2} = \frac{\alpha^2}{\beta^2} \frac{1+|z|^2}{1-|z|^2}
= \frac{\alpha^2 + |\gamma|^2}{\beta^2} 
\quad \Leftrightarrow \quad \beta^2 \geq \alpha^2 + |\gamma|^2\, .
\end{equation}

It is interesting to see if we could have a value of $c_\lambda\ll1$.  Inspection of (\ref{sub_c}) tells us that a necessary condition to have $c_\lambda\ll1$ is to have $\beta\gg \alpha$. But from (\ref{abc_coeffs}) one easily see that if one pushes $\beta$ to be very large, then, unless we enforce some to-us unexpected cancellations among the parameters, $\alpha$ will be driven to $\alpha\to \beta$, and $c_\lambda\to 1$. We conclude that we expect $c_\lambda-1\sim {\cal O}(1)$.

Let us now return to \eqref{abc_coeffs} to see how the constraint \eqref{sub_c} affects the Lagrangian parameters. To get a better sense of what this condition implies, we consider some special cases.
\begin{itemize}
\item {Turn on only $m_{3/2}\neq 0$} --- in this case eq.~\eqref{sub_c} becomes
\begin{equation}
3\epsilon H^2 (H^2 + |m_{3/2}|^2)\geq |\dot m_{3/2}|^2\, ,
\end{equation}
where we wrote $\epsilon \equiv -\dot H/H^2$ and taken $\epsilon>0$ (as it is imposed by requiring that the kinetic term of $\pi$ has the healthy sign, in the case the $\pi$ sector is similar to slow-roll inflation). This is well satisfied for example if $\dot m_{3/2}/m_{3/2}^2\sim \dot H/H^2 \sim \epsilon$, which is a technically natural regime.

\item {Turn on only $\wt m_1 = {\rm const}\neq 0$} --- in this case eq.~\eqref{sub_c} becomes
\begin{equation}
(2H^2 - 3Hm_1 + m_1^2)(2H^2 \epsilon - H m_1 + m_1^2)\geq 0\, .
\end{equation}
Neither term in the brackets is positive definite. Any negative value for $m_1$ is allowed, however to leading order in $\epsilon\ll 1$ the LHS changes sign at $m_1/H=2\epsilon,\, 1-2\epsilon,\, 1,\, 2$. This leads to the forbidden (striped) regions of parameter space that are shown in Fig.~\ref{fig:forbidden2}.
\begin{figure}[h]
\centerline{
\includegraphics[width=0.6\linewidth, angle=0]{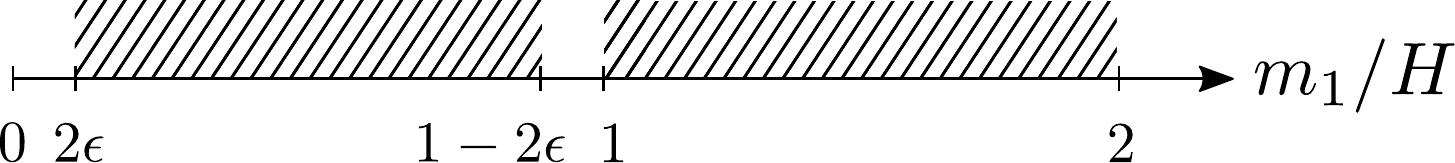}}
\caption{Striped regions are regions of the parameter space forbidden by superluminality in the case in which only the parameter $\tilde m_1={\rm const}$ is turned on. \label{fig:forbidden2}}
\end{figure}

\item {Turn on only $\wt m_2 = {\rm const}\neq 0$} --- $\wt m_2$ alone does not contribute to the speed of sound, and eq.~\eqref{sub_c} becomes
\begin{equation}
1\geq c_\lambda = 1+ \frac{4 \dot H}{|\beta|}\, ,
\end{equation}
with $\beta = -6H^2$. Thus $c_\lambda\leq 1$ as long as $\dot H \leq 0$, and to leading order in the slow-roll expansion $c_\lambda = 1 - \frac{2}{3}\epsilon + {\cal O}(\epsilon^2)$. In this case, the speed of sound of the Goldstino agrees with what found in \cite{Kahn:2015mla}. We stress however that this is particular to the case where the gravitino has no mass term in the unitary gauge (or just $\wt m_2$).

\item {Turn on only $m_0 = {\rm const}\neq 0$} --- assuming for simplicity that $M=0$ (i.e. $m_{3/2}=\frac{1}{2}m_0$) eq.~\eqref{sub_c} becomes
\begin{equation}
1\geq c_\lambda^2 = \left(1-\frac{2}{3}\epsilon\right)^2 + \frac{2}{3} \frac{|m_0|}{H} \, .
\end{equation}
This leads to the following constraint for $m_0$ from requiring subluminal propagation
\begin{equation}
|m_0|\leq 2\epsilon H + {\cal O}(\epsilon^2).
\end{equation}

\end{itemize}


\bibliographystyle{JHEP}
\bibliography{references}

%
%
%
%
%
%
%
%
%
%
%
%
\end{document}